\documentclass{aa}

\usepackage{graphicx}
\usepackage{txfonts}
\usepackage{url}
\usepackage{natbib}
\usepackage{color}
\usepackage{textcomp}
\usepackage{multirow}
\usepackage{array}
\usepackage{calc}
\usepackage{adjustbox}
\bibpunct{(}{)}{;}{a}{}{,}

\def\orcidlink#1{}

\begin{document}

\title{THEMIS 2.0: A self-consistent model for dust extinction, emission, and polarisation}

\author{N. Ysard\inst{\ref{inst1}, \ref{inst2}} \orcidlink{0000-0003-1037-4121}
\and A.P. Jones\inst{\ref{inst1}} \orcidlink{0000-0003-0577-6425}
\and V. Guillet\inst{\ref{inst1}, \ref{inst3}} \orcidlink{0000-0002-8881-3094}
\and K. Demyk\inst{\ref{inst2}} \orcidlink{0000-0002-5019-8700}
\and M. Decleir\inst{\ref{inst4}} \orcidlink{0000-0001-9462-5543}
\and L. Verstraete\inst{\ref{inst1}} \orcidlink{0000-0003-3482-4561}
\and I. Choubani\inst{\ref{inst2}}
\and M.-A. Miville-Desch{\^e}nes\inst{\ref{inst5}} \orcidlink{0000-0002-7351-6062}
\and L. Fanciullo\inst{\ref{inst6}} \orcidlink{0000-0001-9930-9240}
}

\institute{Universit{\'e} Paris-Saclay, CNRS, Institut d'Astrophysique Spatiale, 91405, Orsay, France\label{inst1}
\and IRAP, CNRS, Universit{\'e} de Toulouse, 9 avenue du Colonel Roche, 31028 Toulouse Cedex 4, France\label{inst2}
\and LUPM, Universit{\'e} de Montpellier, CNRS/IN2P3, CC 72, Place Eug{\`e}ne Bataillon, 34095 Montpellier Cedex 5, France\label{inst3}
\and Space Telescope Science Institute, 3700 San Martin Drive, Baltimore, Maryland, 21218, USA\label{inst4}
\and AIM, CEA, CNRS, Universit{\'e} Paris-Saclay, Universit{\'e} Paris Diderot, Sorbonne Paris Cit{\'e}, 91191 Gif-sur-Yvette, France\label{inst5}
\and National Chung Hsing University, 145 Xingda Rd., South Dist., Taichung City 402, Taiwan\label{inst6}\\
 \email{nathalie.ysard@cnrs.fr}}

\abstract
{Recent observational constraints in emission, extinction, and polarisation have at least partially invalidated most of the astronomical standard grain models for the diffuse interstellar medium. Moreover, laboratory measurements on interstellar silicate analogues have shown quite significant differences with the optical properties used in these standard models.}
{To address these issues, our objective is twofold: (i) to update the optical properties of silicates and (ii) to develop The Heterogeneous dust Evolution Model for Interstellar Solids (THEMIS) to allow the calculation of polarised extinction and emission.}
{Based on optical constants measured in the laboratory from $5~\mu$m to 1~mm for amorphous silicates and on observational constraints in mid-IR extinction and X-ray scattering, we defined new optical constants for the THEMIS silicates. Absorption and scattering efficiencies for spheroidal grains using these properties were subsequently derived with the discrete dipole approximation.}
{These new optical properties make it possible to explain the dust emission and extinction, both total and polarised. It is noteworthy that the model is not yet pushed to its limits since it does not require the perfect alignment of all grains to explain the observations and it therefore has the potential to accommodate the highest polarisation levels inferred from extinction measurements. Moreover, the dispersion of the optical properties of the different silicates measured in the laboratory naturally explain the variations in both the total and polarised emission and extinction observed in the diffuse interstellar medium.}
{A single, invariant model calibrated on one single set of observations is obsolete for explaining contemporary observations. We are proposing a completely flexible dust model based entirely on laboratory measurements that has the potential to make major advances in understanding the exact nature of interstellar grains and how they evolve as a function of their radiative and dynamic environment. Even if challenging, this is also relevant for future cosmic microwave background (CMB) missions that will aim to perform precise measurements of the CMB spectral distortions and polarisation.}

   \authorrunning{}
\titlerunning{THEMIS 2.0}
\maketitle
%

\section{Introduction}
\label{introduction}

Detailed knowledge of cosmic dust is crucial for understanding our Universe. It is almost impossible to avoid its effects and we should therefore never underestimate its impact on astronomical observations. Undoubtedly, the thermal emission from interstellar dust is prominent in all observations over the IR to sub-millimetre electromagnetic spectrum. The extinction of starlight is dominated by dust at all galaxy scales. Dust is therefore one of the best tracers of the processes that govern the evolution of cosmic matter in galaxies near and far, from their most diffuse interstellar clouds to the densest regions where stars and planets form. Yet dust is far from being just a tracer, but it is perhaps the major actor in the evolution of matter within the cosmos. For example, the dust opacity is a fundamental parameter for star formation as it determines whether a medium is optically thin or thick and hence how much of the local radiation field can be radiated away, which strongly affects the collapse process leading to new stars. Dust also heats the gas through the photoelectric effect and cools it through gas-grain collisions. It is thus a key player in determining the dynamics of the interstellar medium (ISM) through its interactions with the radiation field and with the gas. Dust is responsible for the diverse chemical complexity seen within galaxies. Its surfaces provide the sites for many chemical reactions that cannot occur in the gas phase, including the formation of H$_2$, and because it is very efficient at absorbing bond-dissociating UV photons it then protects those same molecules from destruction. Many of the most complex molecules, such as alcohols and sugars, exist only because they are formed in the ice layers that coat dust grains in dense star and planet-forming clouds. Further, ISM grains are the very precursors of the solid matter that makes the planets. All of the above processes depend on the exact grain size, structure, composition, and mass; and there is still a gap in our knowledge of these fundamental dust properties. Both the latest observations and laboratory measurements have cast doubt on the validity of the standard grain models for the diffuse ISM which we now consider in some detail.

From an observational point of view, several points are of concern. First of all, the silicate mid-IR vibrational bands are less prominent over the continuum extinction \citep[e.g.][]{Gordon2021} than those predicted by standard models \citep[e.g.][]{Desert1990, Draine2007, Compiegne2011, Jones2013, Guillet2018}, with most of them using optical properties derived from the astrosilicates defined by \citet{Draine1984}. Second, the Planck high frequency instrument (Planck-HFI) data revealed a number of inconsistencies between the model predictions for emission and extinction. This applies to both polarised and unpolarised extinction-to-emission ratios, with the deviation between observations and models being of the order of a factor of 2 to 3 \citep{Fanciullo2015, PlanckXXI2015, PlanckXXIX2016}. This second point is particularly problematic since it is directly related to the estimation of the mass of the ISM structures. Third, there is increasing evidence of variations in the grain properties in the diffuse ISM. Depending on the observed line of sight, the extinction curve has been shown to vary from the UV/optical \citep[e.g.][]{Gordon2009, Fitzpatrick2019, Massa2020} to the near-IR \citep[e.g.][]{Nishiyama2006, Nishiyama2009, Decleir2022} and the mid-IR \citep[e.g.][]{Gordon2021}. Regarding the dust emission, the Planck Collaboration began by showing that the grain sub-millimetre opacity varies on particularly tenuous lines of sight \citep[$N_H < 3 \times 10^{20}$~H/cm$^2$,][]{PlanckXI2014}. This pioneering study was followed by others pointing to variations in opacity in a variety of ways \citep{Reach2015, Reach2017} and showing that it was not accompanied by a significant change in the $E(B-V)/N_H$ ratio \citep[][]{Nguyen2018, Murray2018}. Moreover, the gas-to-dust mass ratio, which had been considered constant or even canonical until recently, varies from 20 to 60\% in relation to the historical measurement of \citet{Bohlin1978}, depending on the line of sight \citep[e.g.][]{Liszt2014, PlanckXI2014, Lenz2017, Murray2018, Nguyen2018, Remy2018, VanDePutte2023}. These variations are also corroborated by the spectro-polarimetric study led by \citet{Siebenmorgen2018}. All of these points lead to a fundamental problem with the dust models and not just to a marginal adjustment. The logical conclusion of this inventory is that there is a need to redefine the optical properties of the silicates responsible for the mid-IR extinction bands and a priori of a large part of the polarised and non-polarised thermal emissions from the far-IR to sub-millimetre. There are two ways of solving this problem: to create an empirical model from fitting of astronomical observations as performed by \citet{Hensley2023}\footnote{Their `astrodust' model is fully characterised in \citet{Draine2021b, Draine2021a, Draine2021c}.} or to start from laboratory measurements on interstellar grain analogues and check their agreement with the observations as in \citet{Siebenmorgen2022}. Both approaches are useful; however, in the logic of our previous work \citep[][and citations therein]{Jones2017} and in order to be able to characterise the chemical composition of the grains as well as possible and thus all of their physical, chemical, and dynamical properties, we followed the experimental rather than the empirical route.

The laboratory measurements performed by \citet{Demyk2017} are in this sense of paramount importance. They measured the mass absorption coefficients (MACs) of amorphous magnesium-rich glassy silicates, which are expected to be good analogues of diffuse ISM solid matter, between 5~$\mu$m and 1~mm. This was the first study to measure MACs of amorphous silicates over such a wide wavelength range, a particularly important point being that previous studies did not cover the vibrational bands at 9.7 and $18~\mu$m \citep{Agladze1996, Mennella1998, Boudet2005, Coupeaud2011}. Many of their results differ significantly from the silicates used in the standard models and are consistent with the direction needed to explain the observations: (i) the mid-IR vibrational bands are wider and less prominent with respect to the continuum; (ii) the far-IR to sub-millimetre opacity is higher, as well as wavelength and temperature dependent; and (iii) both opacity and bands are dependent on the chemical composition and structure at the nano-scale of the silicate samples. The optical constants corresponding to these laboratory measurements are now available \citep{Demyk2022} and our goal is to update the silicate component of our grain model THEMIS\footnote{Available here: \url{https://www.ias.u-psud.fr/themis/}.} \citep[The Heterogeneous dust Evolution Model for Interstellar Solids, ][and references therein]{Jones2017}. At the same time, we have moved from spherical to non-spherical grains to allow the calculation of polarisation with this new version of the model, referred to as THEMIS 2.0 afterwards. Solutions to the observational problems raised above have already been proposed by \citet{Siebenmorgen2022} and \citet{Hensley2023}, but our aim is to develop a model that will explain not only the `average values' observed in the diffuse ISM but also their dispersion.

This paper is organised as follows. Section~\ref{grain_composition} details the new optical constants used for the THEMIS 2.0 silicate component. The methodology of the optical property calculations is described in Sect.~\ref{methodology}. Section~\ref{grain_properties} then presents the variations in the optical properties depending on the grain shapes and sizes. The observational constraints are discussed in Sect.~\ref{observations} and the THEMIS 2.0 model is subsequently defined in Sect.~\ref{themis_ii}. Section~\ref{conclusion} finally summarises our results.

\section{THEMIS 2.0 grain chemical compositions}
\label{grain_composition}

Our starting point is the original version of THEMIS \citep{Jones2013, Koehler2014, Jones2017} in which the grains consist of (sub-)nanometric aromatic-rich a-C carbon grains ($0.4 \leqslant a \leqslant 20~$nm), larger amorphous carbonaceous grains with an H-rich a-C:H core and a 20~nm thick aromatic-rich a-C mantle and amorphous magnesium-rich silicates with metallic iron and iron sulphide nano-inclusions coated with a 5~nm thick aromatic-rich a-C carbon mantle\footnote{This core/mantle structure is in agreement with observations of primitive chondrites \citep{Ishii2018} and GEMS (glass with embedded metal and sulfides) inside interplanetary dust particles \citep{Davidson2015}.}. For the large amorphous carbonaceous grains, the core material is taken to be that typical of the aliphatic-rich, wide band gap, material formed around C-rich evolved stars \citep[e.g.][]{Goto2000, Goto2007}, while the outer mantle is assumed to have been UV photo-processed in the low density ISM \citep[see][for a more comprehensive explanation]{Jones2016RSOS}. For the amorphous silicate component, THEMIS consists of two chemical compositions: one with a stoichiometry similar to that of enstatite and one to that of forsterite\footnote{Forsterite is a mineral of the silicate group and the mesosilicate subgroup. With a composition of Mg$_2$SiO$_4$, it is the pure magnesian end member of olivine. Enstatite is a mineral species in the silicate subgroup of inosilicates. It is a member of the pyroxene family, with the formula MgSiO$_3$. When we talk of forsterite, olivine, enstatite or pyroxene in an amorphous state what we implicitly imply is silicates that have these stoichiometries and not the crystalline silicates to which these mineralogical terms apply.} \citep{Scott1996}, with both optical constants quite similar to those of the `astrosilicates' defined by \citet{Draine1984} for $\lambda \gtrsim 18~\mu$m. As stated in Sect.~\ref{introduction}, the complex refractive index ordinarily assumed for the silicate grains can no longer be considered as the most reliable data set and needs to be replaced by the more consistent measurements performed by \citet{Demyk2017}.

\subsection{Silicate grains}
\label{silicate_grains}

\citet{Demyk2017} performed measurements on interstellar silicate analogues from the mid-IR to sub-millimetre at 10, 100, 200, and 300~K and subsequently derived the associated complex refractive indices ($m = n + ik$) as described in \citet{Demyk2022}. The analogues were amorphous magnesium-rich glassy silicates with the stoichiometry of forsterite \citep[X35 sample in][]{Demyk2017} and enstatite (X50a and X50b samples). A third sample had a stoichiometry in-between forsterite and enstatite (X40 sample). For clarity and also because it was measured only below $\lambda_{\rm FIR} = 500~\mu$m, we choose to discard this sample and rather mix the X35 and X50 samples (see Sect.~\ref{silicate_nk}). The X50a and X50b samples were measured up to $\lambda_{\rm FIR} = 0.75$ and 1~mm, respectively. Above these threshold values, \citet{Demyk2022} extrapolated the imaginary part of the refractive index and computed its real part according to the Kramers-Kronig relations. The optical constants used in THEMIS 2.0 are those measured at 10~K, the temperature closest to that expected in the diffuse ISM, and it should be noted that \citet{Demyk2017} found no significant variation up to 30~K.

As stated later in Sect.~\ref{methodology}, computing the optical properties of core/mantle spheroidal grains is an enormous time-consuming process. This means that it is impossible to test all of the silicate chemical composition combinations in the spheroidal case. A first step in the definition of THEMIS 2.0 is therefore to constrain a plausible silicate chemical composition based on previous studies.

\subsubsection{Observational constraints for silicate composition}
\label{observational_constraints}

The majority of the silicates in the diffuse ISM appear to be amorphous \citep[e.g.][]{Kemper2004, Zeegers2019}, which makes it difficult to characterise their chemical composition. Indeed, contrary to crystalline solids, the 9.7 and 18~$\mu$m absorption bands of amorphous silicates are broad and featureless. A second issue concerns the nature of the iron within the grains -- in the matrix or in metallic form.

Several studies have compared the mid-IR silicate bands measured in the ISM to those measured on analogues in the laboratory. Although the proportions vary depending on the line of sight considered, most of these studies concluded that the composition is a mixture of olivine and pyroxene stoichiometry mineralogies \citep[e.g.][]{Chiar2006, VanBreemen2011}. Observing the Galactic Centre in the 9.7~$\mu$m feature, \citet{Kemper2004} pointed towards a predominance of olivine over pyroxene in proportions of $\sim 85$ to 15\%. Combining both features at 9.7 and 18~$\mu$m, \citet{Min2007} further showed that interstellar silicates have to be highly magnesium-rich (Mg/(Mg+Fe) $> 0.9$) thus leading to the conclusion that iron is incorporated in silicates in the form of metallic inclusions rather than in the matrix. This is consistent with two laboratory studies. The first showed that amorphous Fe-rich silicates that are annealed in the presence of carbon leads to the reduction of the iron and the formation of Fe nano-particles embedded within a matrix of Mg-rich amorphous silicate \citep{Davoisne2006, Djouadi2007}. The second found that H$^+$ irradiation of amorphous silicates with iron in the matrix leads to selective oxygen sputtering and thus again to the reduction of Fe$^{2+}$ to metallic Fe \citep{Jager2016}.

This is also confirmed by observations of dust X-ray absorption and scattering at the Si, Mg, O, and Fe K-edges. Modelling of these data with silicate laboratory analogue optical properties lead to the conclusions that: (i) the Si, Mg, and O K-edge spectral shapes exhibit a major contribution from olivine and pyroxene type silicates \citep{Costantini2005, Zeegers2017} ; (ii) the Fe K-edge structure implies Mg-rich silicates with metallic iron and troilite (FeS) inclusions to be dominant \citep{Xiang2011, Costantini2012}. Regarding the relative contributions of olivine and pyroxene, two recent X-ray studies concluded that olivine are dominant over pyroxene with proportions of up to $\sim 70$ to 30\% \citep{Rogantini2019} and $\sim 80$ to 20\% \citep{Zeegers2019}. This result is in agreement with analyses of pre-solar silicates \citep{Hoppe2018, Hoppe2022}. These proportions may however vary depending on local environment \citep{Demyk2001} and even be reversed \citep{Psaradaki2022, Psaradaki2023}. 

\begin{figure*}[!t]
\centerline{\begin{tabular}{cc}
\includegraphics[width=0.5\textwidth]{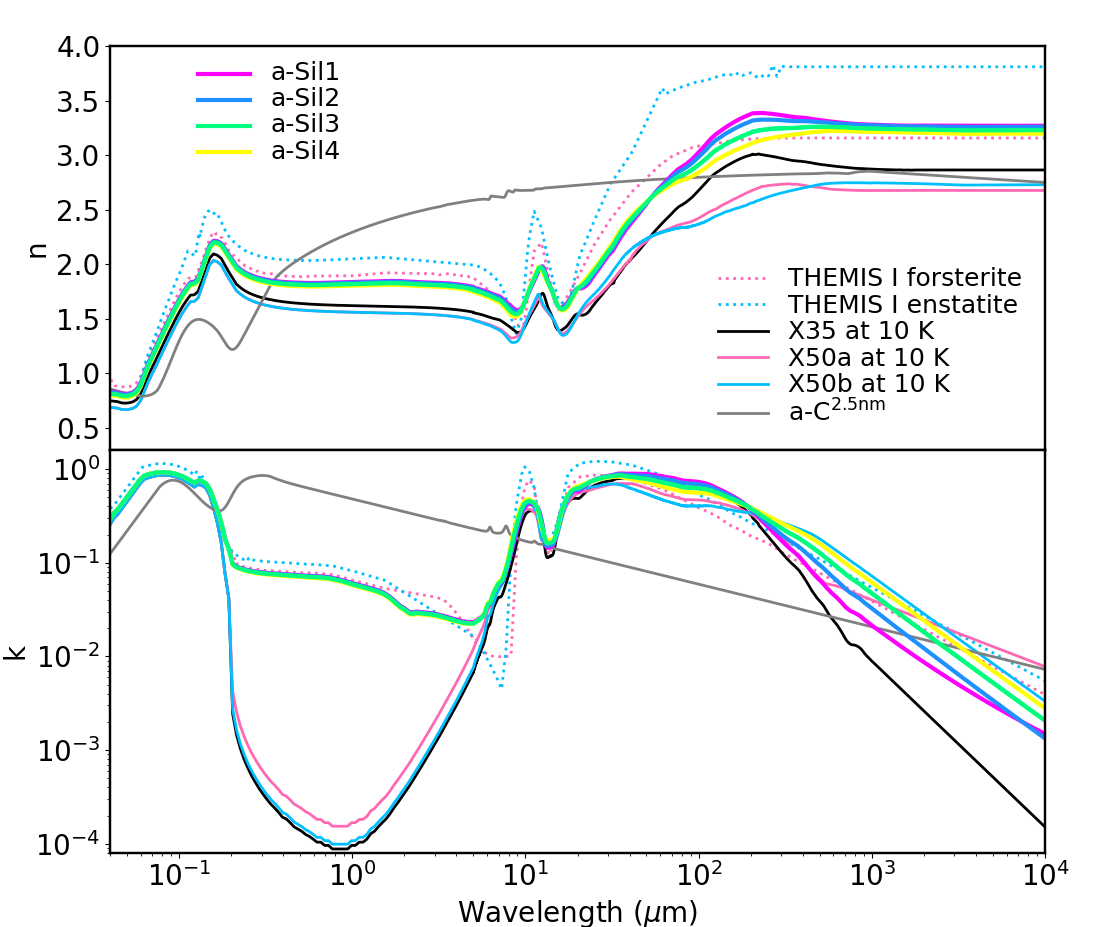} & \includegraphics[width=0.5\textwidth]{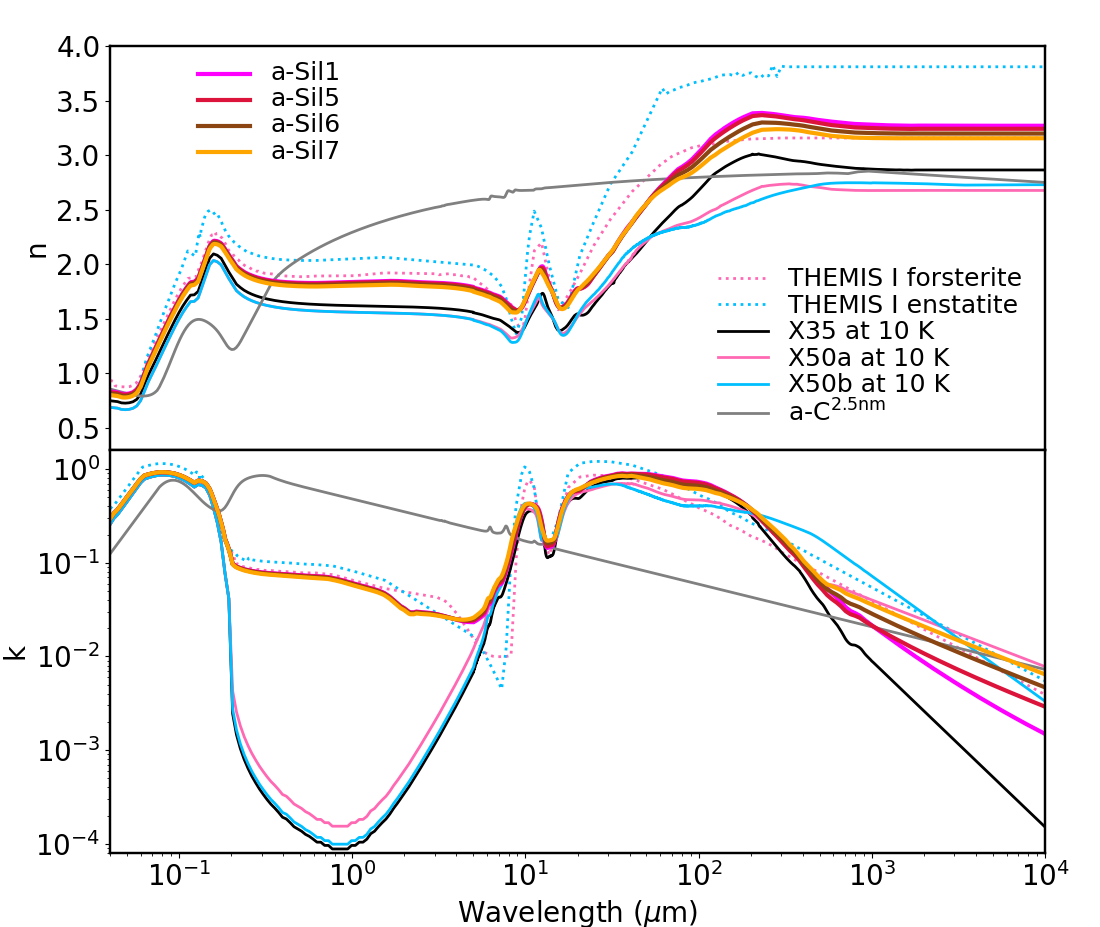}
\end{tabular}}
\caption{Complex refractive indices, $m = n + ik$, of the silicate materials. {\it Top:} Real part. {\it Bottom:} Imaginary part. {\it Left:} Complex refractive indices for the calculation of the silicate grain optical properties. \citet{Demyk2017} silicate samples measured at 10~K are shown: forsterite X35 sample (black), enstatite X50a sample (light pink), enstatite X50b sample (light blue) along with the aSil-1, aSil-2, aSil-3, and aSil-4 mixtures as pink, blue, green, and yellow thick lines, respectively (see Sect.~\ref{silicate_nk} and Table~\ref{table_composition} for details). We note that 10\% of the silicate volume is occupied by metallic nano-inclusions of Fe and FeS (70\% and 30\% of the total volume of the inclusions, respectively), hence the increase in the visible to mid-IR imaginary part of the refractive index. For comparison, the optical constants of the two silicate materials used in the original THEMIS are shown with dotted lines.  The optical constants used for the 5~nm-thick aromatic-rich carbon mantles, a-C$^{2.5{\rm nm}}$, are also plotted (grey, see Sect.~\ref{carbonaceous_grains} for details). {\it Right:} Same for the aSil-5, a-Sil-6, and aSil-7 mixtures as red, brown, and orange thick lines, respectively.}
\label{refractive_indices_silicates} 
\end{figure*}

\subsubsection{Silicate refractive index}
\label{silicate_nk}

Even if there are still some remaining discrepancies concerning the exact silicate composition and the nature of iron in dust, previous studies seem to indicate that silicates are mostly amorphous Mg-rich with metallic Fe and possibly FeS inclusions\footnote{For a complete presentation of the different samples measured by \citet{Demyk2017}, including iron-rich silicates, see \citet{Demyk2022}. See also the study by \citet{Siebenmorgen2022} for their effects when used in dust models.}, with the majority of them belonging to the olivine mineralogical family and the others to the pyroxene family.

\citet{Demyk2022} determined the refractive indices of the magnesium-rich end members of the amorphous silicates with stoichiometries typical of olivine and pyroxene series, forsterite and enstatite, respectively. They considered two samples with the composition of enstatite, X50a and X50b (see Fig.~\ref{refractive_indices_silicates}). The main differences are in the far-IR and sub-millimetre range for the imaginary part of the refractive index. Beyond the band at 18~$\mu$m, the X50a sample shows a much steeper decay than X50b followed by a flattening for $\lambda \gtrsim 700~\mu$m. Sample X50b does not show a break in the slope. We also note that the band at 10~$\mu$m is more contrasted for X50b than X50a. In the following and to test the influence of the choice of the initial laboratory analogue sample, we consider three possible compositions for the silicates. In the first case, we model silicates as being made up of 80\% X35 forsterite sample and 20\% enstatite, half of which consists of the X50a sample and half of the X50b sample (see Table~\ref{table_composition} and Fig.~\ref{refractive_indices_silicates}). In the second (third) [fourth] case, silicates consist of 70\% (50\%) [30\%] X35 forsterite and 30\% (50\%) [70\%] X50b enstatite. In the fifth (sixth) [seventh] case, silicates consist of 70\% (50\%) [30\%] X35 forsterite and 30\% (50\%) [70\%] X50a enstatite. We use the Bruggeman effective medium theory to derive the refractive index of such silicate mixtures \citep{BohrenHuffman}. Then, true to the original concept of THEMIS, we use the Maxwell Garnett mixing rule to derive the refractive index of Fe/FeS nano-inclusions: Fe and FeS \citep{Ordal1985, Ordal1988, Pollack1994} occupy 70 and 30\% of the volume of the inclusions, respectively \citep{Koehler2014}. The Maxwell Garnett rule is then used to derive the refractive index of a mixture 90\% amorphous silicate plus 10\% Fe/FeS inclusions. These optical constants, presented in Fig.~\ref{refractive_indices_silicates}, are those of the materials called `aSil-n' afterwards (n = 1 to 7) for cases 1 to 7 silicate mixtures, respectively. For the material mixtures containing the X50b sample, the main difference between the three mixtures is an increase in the imaginary part of the spectral index in the sub-millimetre range by a factor of about 1.5 (2.2) [2.9] when moving from aSil-1 to aSil-2 (aSil-3) [aSil-4]. This increase will be reflected in the emissivity of the grains, as will the variations in spectral index. The material mixtures containing the X50a sample present a major break in the slope around 600~$\mu$m, the severity of this break being dependent on the X35-to-X50a ratio. The different silicate mixtures are summarised in Table~\ref{table_composition}. For the sake of comparison, we also perform calculations for the original THEMIS optical constants using a 50\%-50\% mixture of the amorphous silicates with the normative compositions of forsterite and enstatite as defined in \citet{Koehler2014}.

\begin{table}[t]
\centering
\caption{Silicate core compositions with corresponding line colours in the figures.} 
\label{table_composition}
\begin{tabular}{l|cccc}
\hline
Name   & X35  & X50a & X50b & Colour/linestyle  \\
\hline
aSil-1 & 80\% & 10\% & 10\% & magenta \\
aSil-2 & 70\% & --   & 30\% & blue    \\
aSil-3 & 50\% & --   & 50\% & green   \\
aSil-4 & 30\% & --   & 70\% & yellow  \\
aSil-5 & 70\% & 30\% & --   & red     \\ 
aSil-6 & 50\% & 50\% & --   & brown   \\
aSil-7 & 30\% & 70\% & --   & orange   \\
\hline
THEMIS I & \multicolumn{3}{c}{as in \citet{Koehler2014}} & dotted lines \\
\hline
\end{tabular}
\end{table}

\begin{figure}[!t]
\centerline{\includegraphics[width=0.45\textwidth]{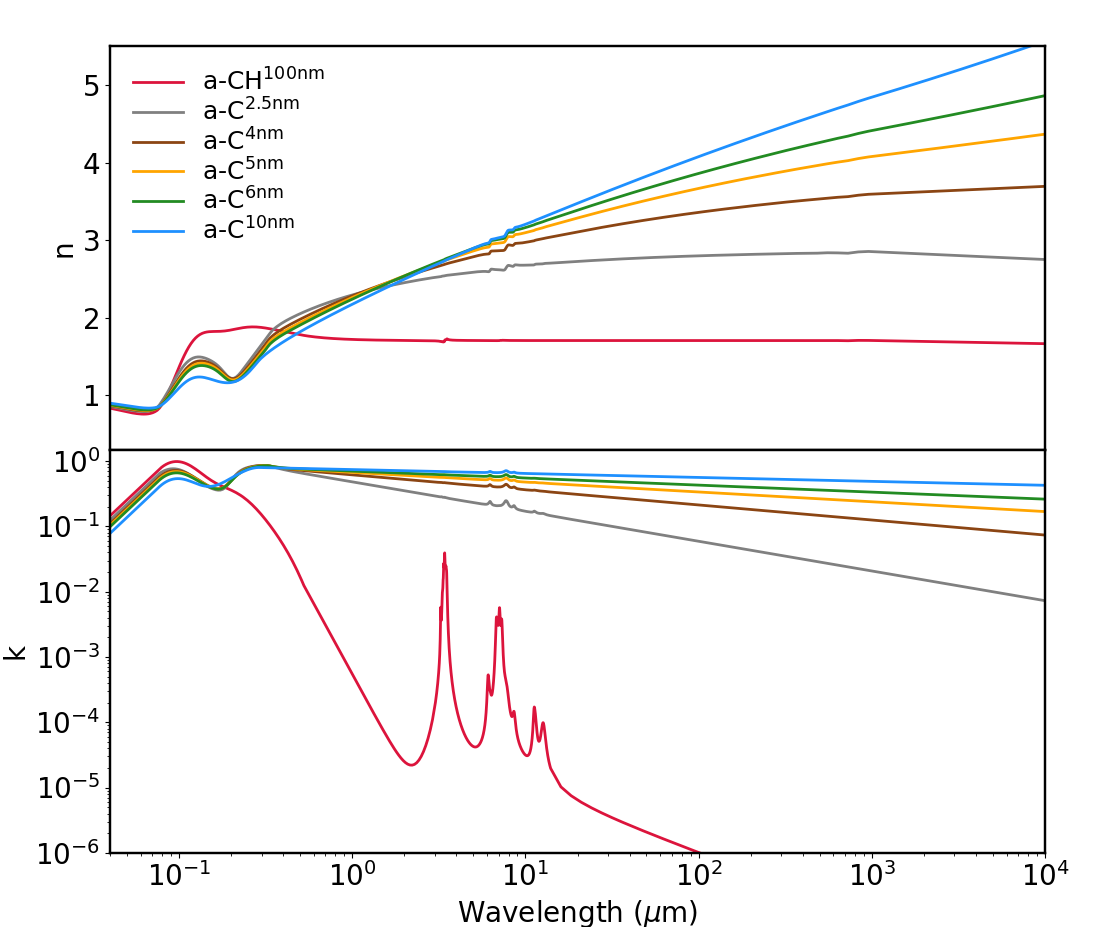}}
\caption{Complex refractive indices, $m = n + ik$, for the calculation of the carbonaceous grain optical properties with those used for the mantle in grey (2.5~nm a-C), brown (4~nm a-C), orange (5~nm a-C), green (6~nm a-C), and blue (10~nm a-C) and the example of a 100~nm a-CH core in red. {\it Top:} Real part. {\it Bottom:} Imaginary part.}
\label{refractive_indices_carbons} 
\end{figure}

\subsection{Carbonaceous grains}
\label{carbonaceous_grains}

For the semi-conducting hydrocarbon grains and mantles, we do not modify the optical constants already defined in the original THEMIS framework \citep{Jones2012a, Jones2012b, Jones2012c}. However, these being grain size- and hydrogen content-dependent, a size and a band gap have to be chosen in order to define the carbonaceous grain optical properties. For the smallest aromatic-rich a-C grains with $0.4 \leqslant a \leqslant 20$~nm, we keep the definition given by \citet{Jones2013} with $E_g = 0.1$~eV $\sim 4.3 X_{\rm H}$ \citep[see][]{Tamor1990}. 

For the larger grains, we depart from the original THEMIS model. For the cores, we keep the original definition of THEMIS by using the actual aliphatic-rich core size to define its optical constants and a band gap of $E_g = 2.5$~eV. For the mantles, the original definition corresponds to a perfectly homogeneous 20~nm-thick mantle having a band gap of $E_g = 0.1$~eV and an equivalent radius equal to half the mantle thickness, 10~nm (called `characteristic size' afterwards and by definition always smaller than or equal to half the mantle thickness). Since the choice of this characteristic size completely determines the emissivity and spectral index in the far-IR and sub-millimetre range, we test several characteristic sizes for the mantle: 2.5, 4, 5, 6, and 10~nm but keep $E_g = 0.1$~eV and the mantle thickness equal to 20~nm. Schematically, the smallest characteristic size reflects the case where the 20~nm-thick mantle comes mainly from the accretion of aromatic-rich small nano-grains and the largest characteristic size where the 20~nm-thick mantle comes from the aromatisation by stellar UV photons of a parent aliphatic grain in an isotropic and homogeneous way \citep{Jones2013}. In the following, the different carbonaceous mantle types are labelled a-C$^{\rm 2.5nm}$, a-C$^{\rm 4nm}$, a-C$^{\rm 5nm}$, a-C$^{\rm 6nm}$, and a-C$^{\rm 10nm}$ for increasing characteristic size. We insist that only the characteristic size describing the mantle formation varies whereas the mantle thickness is fixed at 20~nm. Figure~\ref{refractive_indices_carbons} shows that, at long wavelengths, the refractive index increases with characteristic size -- leading to stronger emissivities -- while the spectral index of its imaginary part decreases.

\section{Methodology of the optical property calculation}
\label{methodology}

In contrast to total emission and extinction, the dust polarised emission and extinction depend not only on its optical properties, size and shape, but also on the efficiency of its alignment with the local magnetic field and on the orientation of the latter with respect to the line of sight. In this section, we describe the methodology of our calculations for the orientated optical properties (Sect.~\ref{DDA}) which depend on the way the polarised emission and extinction are computed (Sect.~\ref{methodology_orientation}).

\subsection{Maximum polarisation fraction for spheroids}
\label{methodology_orientation}

The simplest and most commonly used grain shape in diffuse ISM models is the spheroid \citep[e.g.][among others]{Lee1985, Draine2009, Siebenmorgen2014, Voshchinnikov2016, Guillet2018}, which is obtained by the rotation of an ellipse along one of its main axes. If the rotation is made around the short axis, we refer to it as an oblate grain and if it is around the long axis, as a prolate grain. Calling $a$ the semi-major axis of revolution and $b$ the two other semi-major axes, the ratio $b/a$ is greater than one for oblates and smaller than one for prolates. The effective radius of the sphere of the same volume is then $a_{eff} = (ab^2)^{1/3}$. In the following, the total and polarised emissions and extinctions are calculated with the DustEM\footnote{Available here: \url{https://www.ias.u-psud.fr/DUSTEM/}.} numerical tool \citep{Compiegne2011}, following the formalism detailed in \citet{Guillet2018}, itself based on the approaches of \citet{Hong1980} and \citet{Das2010}, that we only briefly summarise below.

For the grains producing the polarised extinction and emission, that is those aligned with the magnetic field, DustEM needs absorption and scattering efficiencies for two orientations of the \textbf{E} component of an incident electromagnetic wave: (i) $Q_1^{{\rm abs, sca}}$ with \textbf{E} parallel to the projection of the magnetic field onto the plane of the sky and (ii) $Q_2^{{\rm abs, sca}}$ with \textbf{E} perpendicular to it. Here we only seek to calculate the maximum in polarised extinction and emission and therefore assume that the grains are in perfect spinning alignment, no precession or nutation of the grain spin axes around the magnetic field, that is assumed to be in the plane of the sky. It follows that the absorption and scattering efficiencies depend only on the inclination angle $\psi$ of the grain symmetry axis with respect to the plane of the sky \citep[see Fig. B1 in][]{Guillet2018}. Spheroids rotating around their axis of greatest moment of inertia, itself aligned with the magnetic field, implies that their absorption and scattering efficiencies change periodically and thus need to be time-averaged over the grain spinning dynamics. Calling \textbf{E} and \textbf{H} the components of an incident electromagnetic wave in the plane containing the symmetry axis of the grain and the line of sight, the efficiencies can be written as:
\begin{eqnarray}
\label{equations_for_oblates}
Q_1^{\rm oblates} = Q^E(\psi = \pi /2)\\
Q_2^{\rm oblates} = Q^H(\psi = \pi /2)
\end{eqnarray}
for oblate grains since their axes of rotation and symmetry are identical and parallel to the magnetic field. And:
\begin{eqnarray}
Q_1^{\rm prolates} = \frac{2}{\pi} \int_0^{\pi /2} Q^H(\psi) d\psi \\
Q_2^{\rm prolates} = \frac{2}{\pi} \int_0^{\pi /2} Q^E(\psi) d\psi
\label{equations_for_prolates}
\end{eqnarray}
for prolate grains, the spin axes of which are perpendicular to the magnetic field, implying the need to integrate their absorption and scattering efficiencies over the grain spinning dynamics around their minor axes.

Our alignment model is based on polarisation cross-sections that are calculated for grains in perfect spinning alignment with a magnetic field in the plane of the sky. To model the possible imperfect alignement of grains, the polarisation cross-sections are weighted by a parametric function of the grain size as in \citet{Guillet2018}:
\begin{equation}
\label{alignment_fraction}
f(a_{{\rm eff}}) = \frac{1}{2} f_{{\rm max}} \left[ 1 + \tanh \left( \frac{\ln{a_{{\rm eff}}/a_{{\rm thresh}}}}{p_{{\rm stiff}}} \right) \right]
\end{equation}
where the grain alignment fraction increases with the grain size, with a threshold size $a_{{\rm thresh}}$ for which $f(a_{{\rm thresh}}) = 1/2 f_{{\rm max}}$ and the parameter $p_{{\rm stiff}}$ setting the stiffness of the transition from the not-aligned to best-aligned regimes. These parameters can be adjusted from the observations. As stated by \citet{Guillet2018}, this function is sufficient for our purpose of modelling the maximum polarisation fraction of spheroids. This function correctly reproduces the size dependence of the grain alignment by radiative torques \citep[RATs,][]{Hoang2016} or by superparamagnetic Davis-Greenstein alignment \citep{Voshchinnikov2016}.

\subsection{The discrete dipole approximation} 
\label{DDA} 

The grain absorption and scattering efficiencies are computed according to the discrete dipole approximation \citep[DDA,][]{Purcell1973}, using the 7.3.3 version of the \texttt{ddscat} routine described in \citet{DDA1, DDA2} and \citet{DDA3}. In DDA, the grain is assumed to be well represented by an assembly of point-like electric dipole oscillators. \citet{DDAmanual} advise that the dipole size, $\delta$, has to be chosen according to the following criterion: $|m| 2 \pi \delta / \lambda < 1/2$. This criterion is met by all grains and at all wavelengths used in our calculations.

As stated in Sect.~\ref{methodology_orientation} and in Eqs.~\ref{equations_for_oblates} to \ref{equations_for_prolates}, the absorption and scattering efficiencies have to be averaged over the grain spinning dynamics. The \texttt{ddscat} routine calculates the absorption and scattering efficiencies while varying the orientation of the grain relative to the incident electromagnetic wave. We define and orientate the grain in the \texttt{ddscat} input parameter files for the $\psi$ angle of Eqs.~\ref{equations_for_oblates} to \ref{equations_for_prolates} to match the $\Phi$ angle defined in \citet[][see their Fig.~7 where $\widehat{x}$ is the direction of propagation of the incident wave and where we fix the grain axis of symmetry along $\widehat{a_2}$]{DDAmanual}. The grain is then rotated in 10 steps of $10^\circ$ around the $\Phi \equiv \psi$ angle from 0 to $\pi/2$. This allows relatively fast calculations and gives sufficient accuracy after integration \citep[see][]{Koehler2012, Mishchenko2017}.

We perform the DDA calculations for compact grains\footnote{As per the original version, THEMIS 2.0 is constructed to model the dust in the low density, diffuse ISM, we therefore do not consider porous grains (vacuum inclusions).} of $a_{eff} \leqslant 0.7~\mu$m size and core/mantle structures, with the mantle being 5~nm thick for silicates and 20~nm thick for carbon grains. For the amorphous silicates, the cores have the refractive indices of aSil-1, 2, 3, 4, 5, 6, 7, and THEMIS I materials, and the mantles of a 2.5~nm a-C grain with $E_g = 0.1$~eV. The carbonaceous grain materials are as used in THEMIS I, that is cores with the refractive index of an a-C:H with $E_g = 2.5$~eV but having mantles of a 2.5, 4, 5, 6, or 10~nm a-C material with $E_g = 0.1$~eV. Two different elongations are considered for both compact oblate and prolate silicate and carbonaceous spheroids: $e = 1.3$ and 2.

\subsection{Influence of the core-mantle structure: Approximation for the largest grains}
\label{core_mantle_structure}

The larger the grains, the longer the computation time: to allow THEMIS 2.0 to be defined in a reasonable time, we have performed a series of tests to assess the carbon mantle influence on the silicate dust optical properties, extinction curve and SED as a function of the grain effective size $a_{eff}$ and mantle thickness. The mantle volume accounts for $\sim 50$\% of the total volume of a spheroidal grain with $a_{eff} = 25$~nm and a 5~nm thick mantle and for less than 10\% when $a_{eff}$ is greater than 150 to 175~nm depending on the grain elongation. When $a_{eff} \gtrsim 275$~nm, the mantle volume amounts then to less than $\sim 5$\% of the total. It is therefore legitimate to question the extent to which accurate DDA core-mantle calculations are required for the largest silicate grain sizes.

For the absorption and scattering efficiencies, both total and polarised, in order to reach an accuracy better than 5\% at all wavelengths, the exact description of the core-mantle structure must be made up to $a_{eff} = 175$~nm in the case of a silicate with a 5~nm-thick carbon mantle. Above this size, effective medium theory (EMT) can be used to estimate the complex refractive indices of equivalent homogeneous spheroids. If the approximation is used at smaller sizes, the absorption efficiency is underestimated in the UV and visible -- leading to an underestimation of the grain temperature -- and the features at 10 and 20~$\mu$m are altered in two ways: shifting of the peak by a few tenths of a micron and a decrease in the band-to-continuum ratio. For example, for a 100~nm oblate grain with $e = 1.3$, the features at 10 and 20~$\mu$m are redshifted by 0.1~$\mu$m and blueshifted by 0.3~$\mu$m, respectively, and the integrated intensity of the features underestimated by about 5 and 15\%, respectively.

Such a simplification cannot be made in the case of grains with an a-C:H core ($E_g = 2.5$~eV). When illuminated by the interstellar radiation field of \citet[][ISRF at $D_G = 10$~kpc, labelled Solar neighbourhood]{Mathis1983}, their very low far-IR emissivity yields temperatures higher by about 4~K when the a-C component is mixed within the a-C:H core instead of being located at the grain surface \citep[see for instance][]{Ysard2019, Ysard2018} which strongly modifies the resulting SED. Exact core/mantle calculations are thus performed for all sizes for the carbonaceous grains.

\section{Spheroidal grain properties}
\label{grain_properties}

Based on the chemical compositions and methodology presented above, we now detail the properties of the related compact grains. We first present the grain optical properties and then their volume densities as a function of size.

\subsection{Absorption and scattering efficiencies}
\label{efficiencies}

\begin{figure*}[!th]
\centerline{\begin{tabular}{cc}
\includegraphics[width=0.4\textwidth]{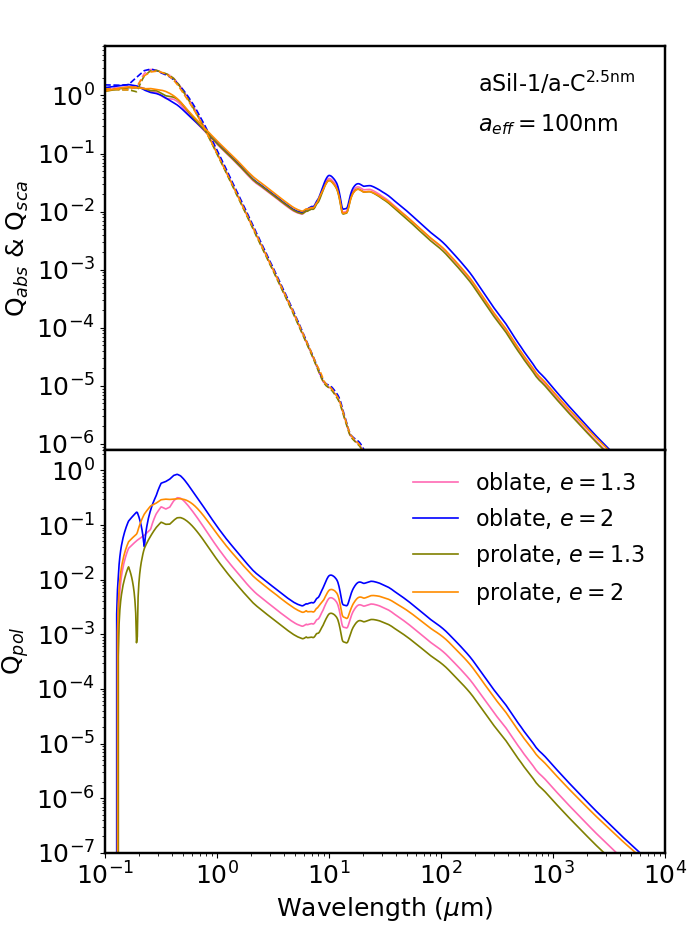} & \includegraphics[width=0.4\textwidth]{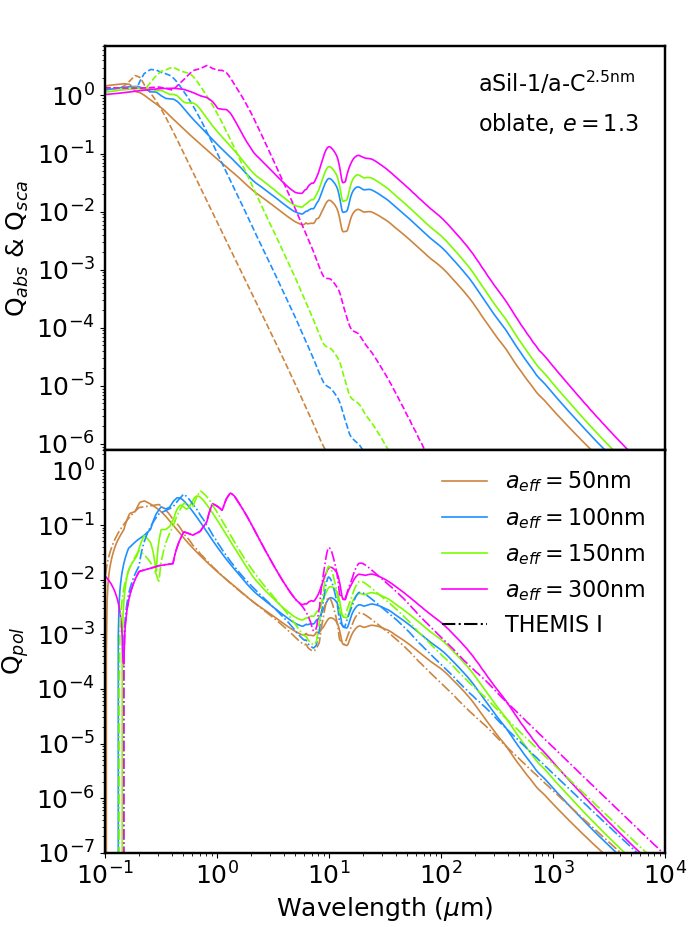} \\ \includegraphics[width=0.4\textwidth]{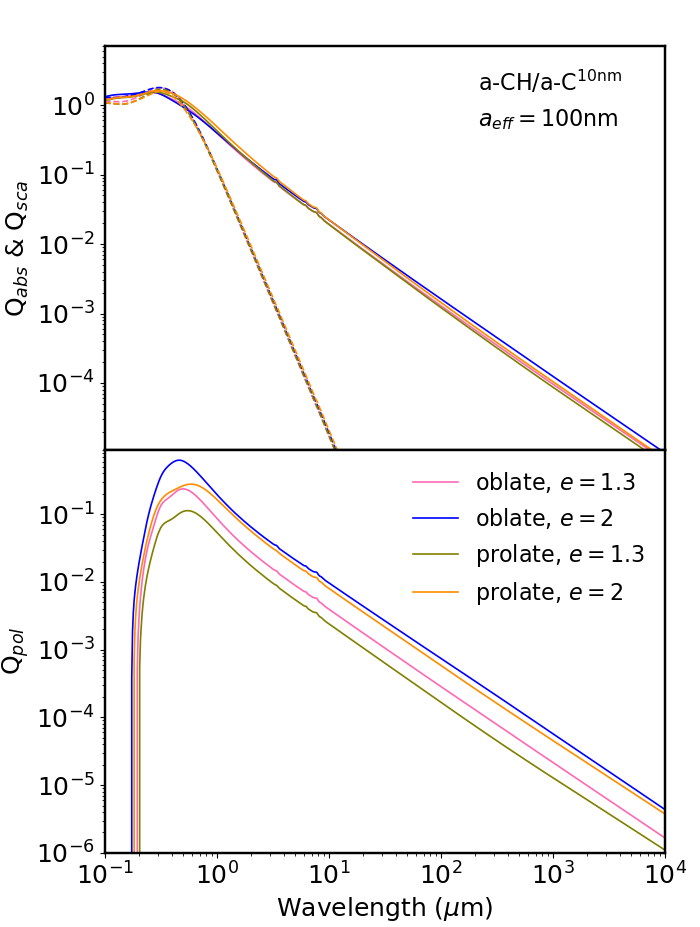} & \includegraphics[width=0.4\textwidth]{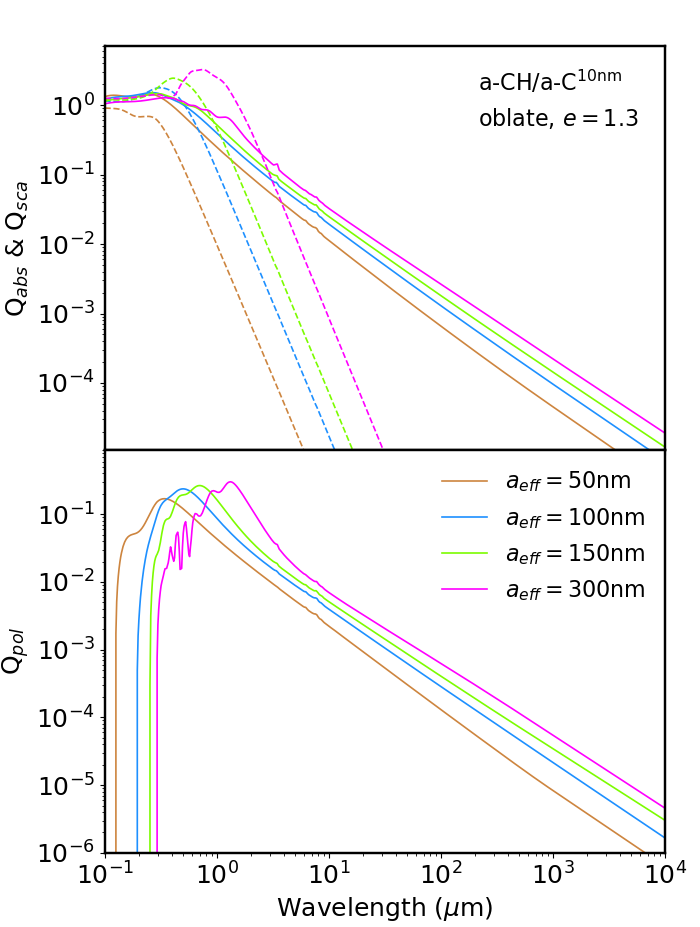}
\end{tabular}}
\caption{Dependence of optical properties on grain shape (left) and size (right) for silicate grains made of aSil-1/a-C$^{\rm 2.5nm}$ materials (top) and carbonaceous grains with a-CH/a-C$^{\rm 10nm}$ composition (bottom). Each block of figures presents the total absorption and scattering efficiencies as solid and dashed lines (upper figure), respectively, and the polarisation efficiency $Q_{pol} = (Q_{2,ext}-Q_{1,ext})/2$ (lower figure). For comparison, $Q_{pol}$ for THEMIS I oblate silicates with $e = 1.3$ and variable size are shown in the lower figure of the top right block.}
\label{efficiencies_size_shape} 
\end{figure*}

\begin{figure}[!th]
\centerline{\includegraphics[width=0.4\textwidth]{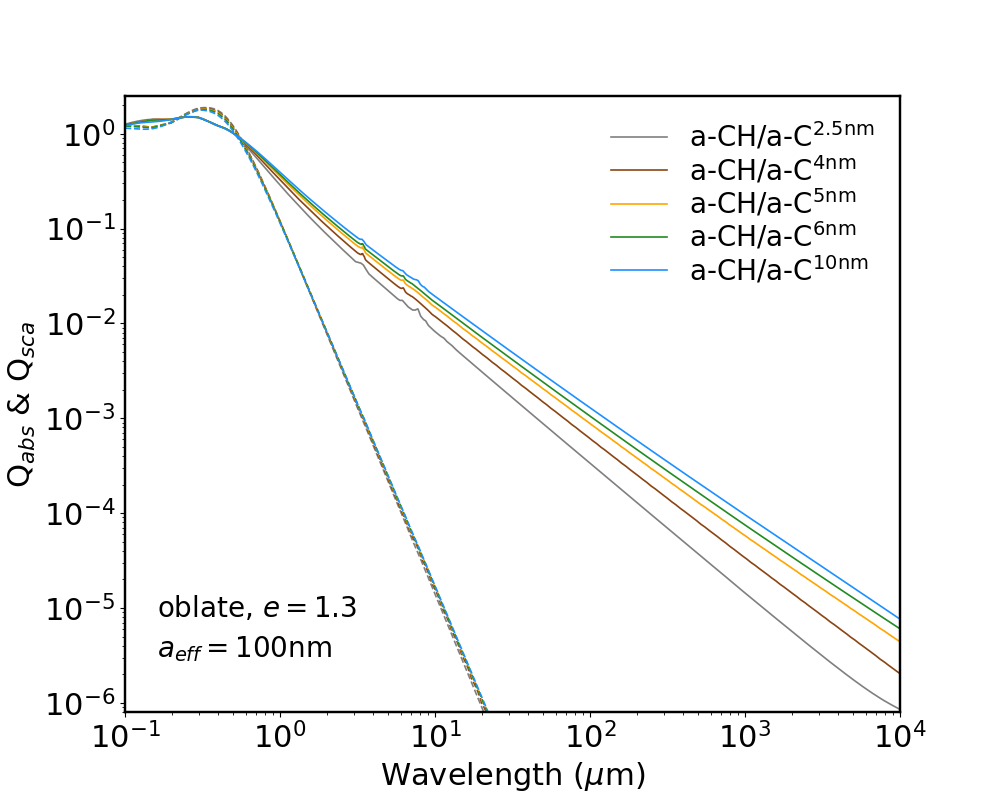}}
\caption{Dependence of optical properties on grain composition for carbonaceous grains with a-CH/a-C$^{\rm 2.5nm}$, a-CH/a-C$^{\rm 4nm}$, a-CH/a-C$^{\rm 5nm}$, a-CH/a-C$^{\rm 6nm}$, and a-CH/a-C$^{\rm 10nm}$ compositions in grey, brown, orange, green, and blue, respetively  ($a_{eff} = 100$~nm, oblate with $e = 1.3$). Solid lines show $Q_{abs}$ and dashed lines $Q_{sca}$.}
\label{efficiencies_carbons} 
\end{figure}

Figure~\ref{efficiencies_size_shape} shows the variations in the optical properties as a function of grain shape and size for two grain types: aSil-1/a-C$^{\rm 2.5nm}$ and a-CH/a-C$^{\rm 10nm}$. The optical properties vary as expected with higher efficiencies for oblates than prolates as well as for larger elongations \citep[see for instance][and references therein]{Ysard2018}. Mie theory predicts that in the far-IR the absorption and scattering efficiencies vary as a function of $a$ and $a^4$, respectively, which is found for both grain types. However, the break in the slope of the imaginary part of the refractive index of aSil-1 material around 1~mm (Fig.~\ref{refractive_indices_silicates}) is observed for the aSil-1/a-C$^{\rm 2.5nm}$ grains, at approximately the same wavelength. The amplitude of this break decreases with the grain size, reflecting the decreasing mantle-to-core volume ratio. For a given shape, the change in elongation from $e = 1.3$ to 2 leads to a change in equilibrium temperature by about 0.5~K (for a grain illuminated by the ISRF with $G_0 = 1$) for both carbonaceous and silicate grains. For a given elongation, prolate grains are slightly hotter than oblate grains due to their lower absorption efficiency at long wavelengths (and hence lower emissivity) while the differences in the absorption efficiency in the UV-visible range remain modest.

Figure~\ref{efficiencies_carbons} shows the variations in the optical properties as a function of grain composition for a carbonaceous oblate grain with $e = 1.3$ and $a_{eff} = 100$~nm. As expected from the variations in their refractive index (Fig.~\ref{refractive_indices_carbons}), the absorption efficiency of the carbonaceous grains increases with the characteristic size of the a-C representing their 20~nm-thick mantles by a factor of about 4 at 1~mm from a-C$^{\rm 2.5nm}$ to a-C$^{\rm 10nm}$. In the meantime, the spectral index in the far-IR/sub-millimetre range decreases from $\beta = 1.35$ for a-C$^{\rm 2.5nm}$ to 1.12 for a-C$^{\rm 10nm}$ where $Q_{abs} \propto \lambda^{-\beta}$. For a grain with $a_{eff} = 100$~nm illuminated by the ISRF with $G_0 = 1$, this leads to a decrease in the equilibrium temperature by about 5~K from 23.3 to 18~K (decrease by about 4 and 6.5~K for $a_{eff} = 25$ and 700~nm, respectively).

\begin{figure*}[!th]
\centerline{\begin{tabular}{cc}
\includegraphics[width=0.4\textwidth]{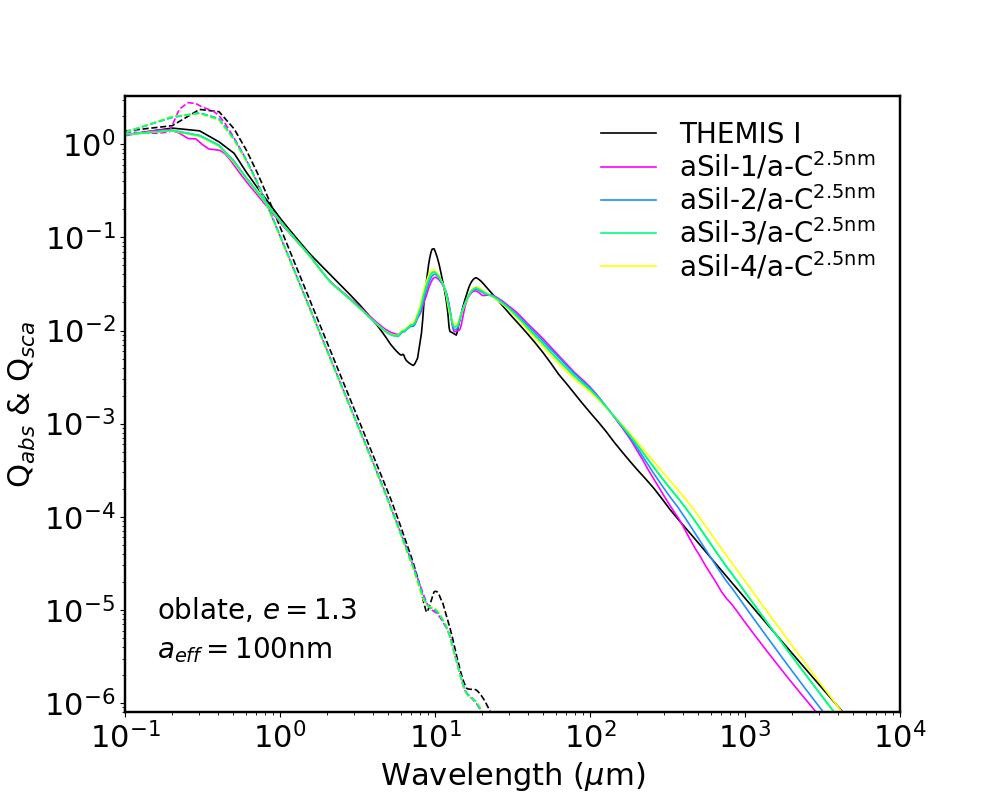} & \includegraphics[width=0.4\textwidth]{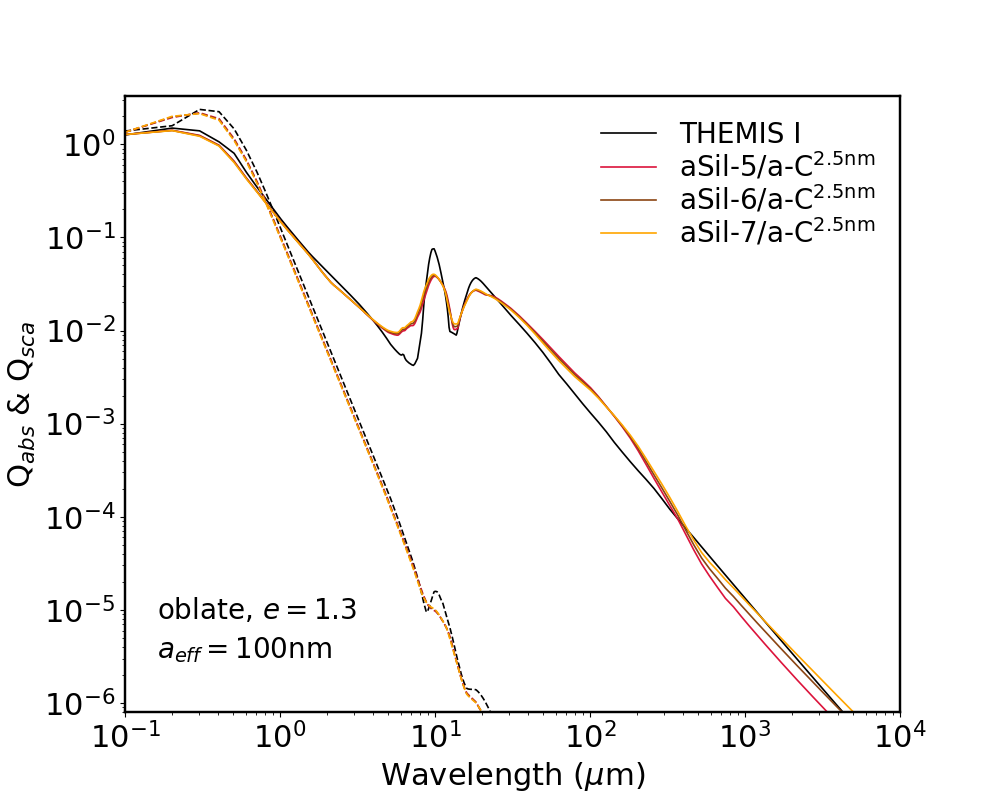} \\ 
\includegraphics[width=0.4\textwidth]{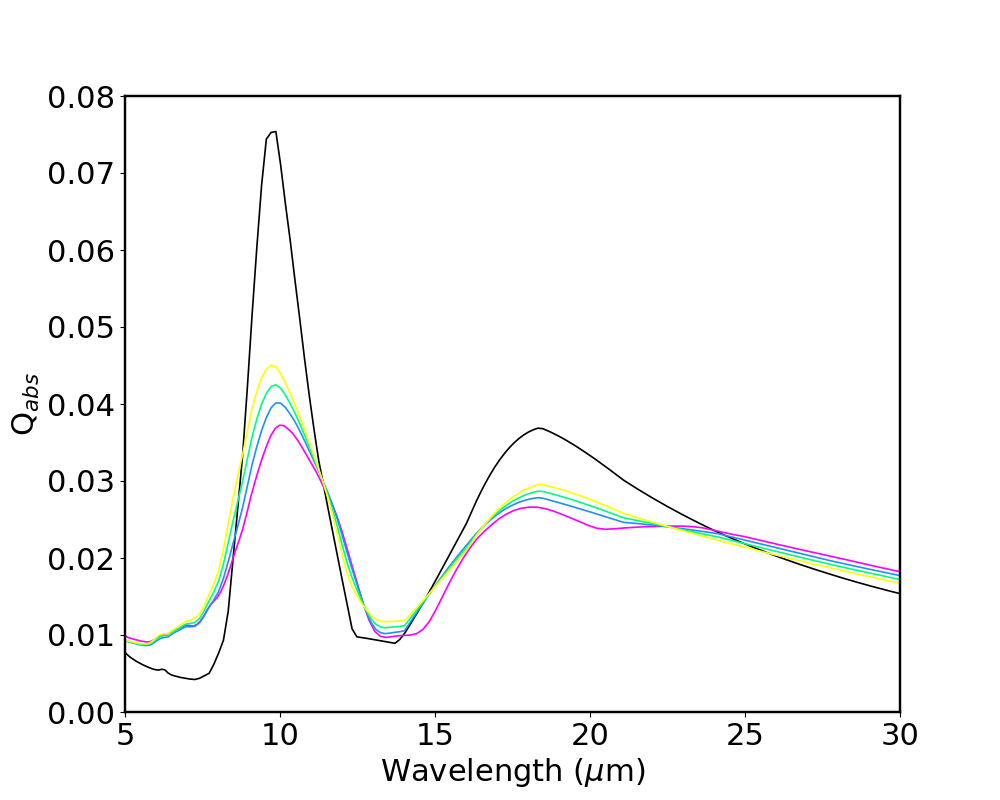} & \includegraphics[width=0.4\textwidth]{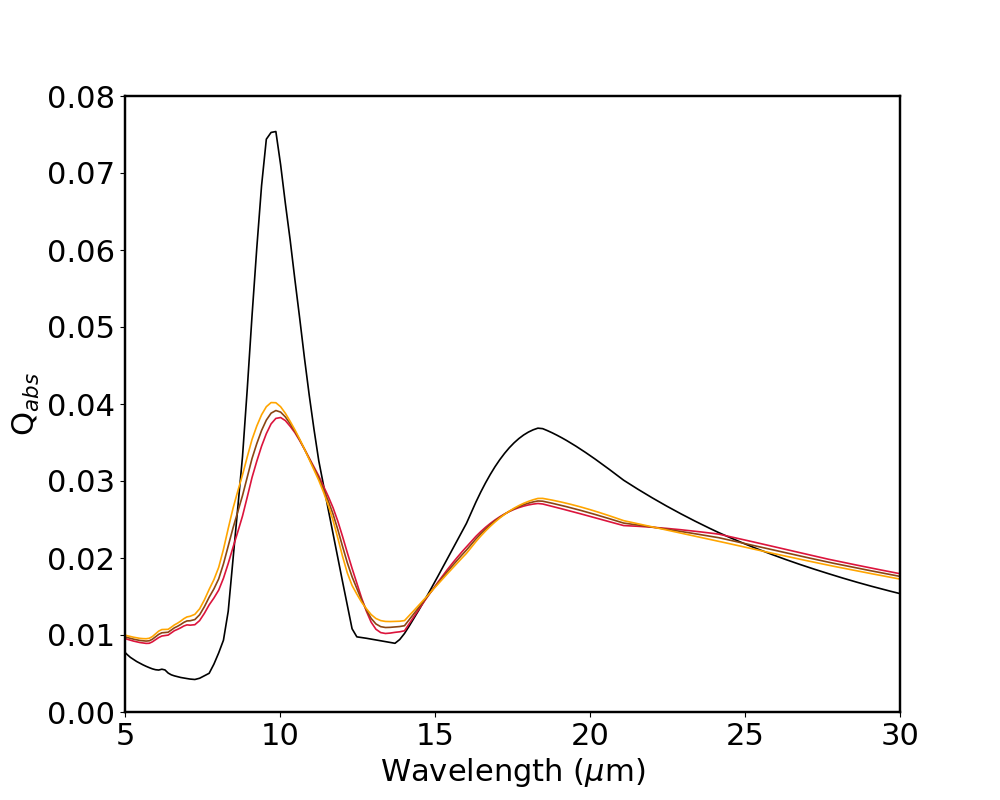} \\ 
\includegraphics[width=0.4\textwidth]{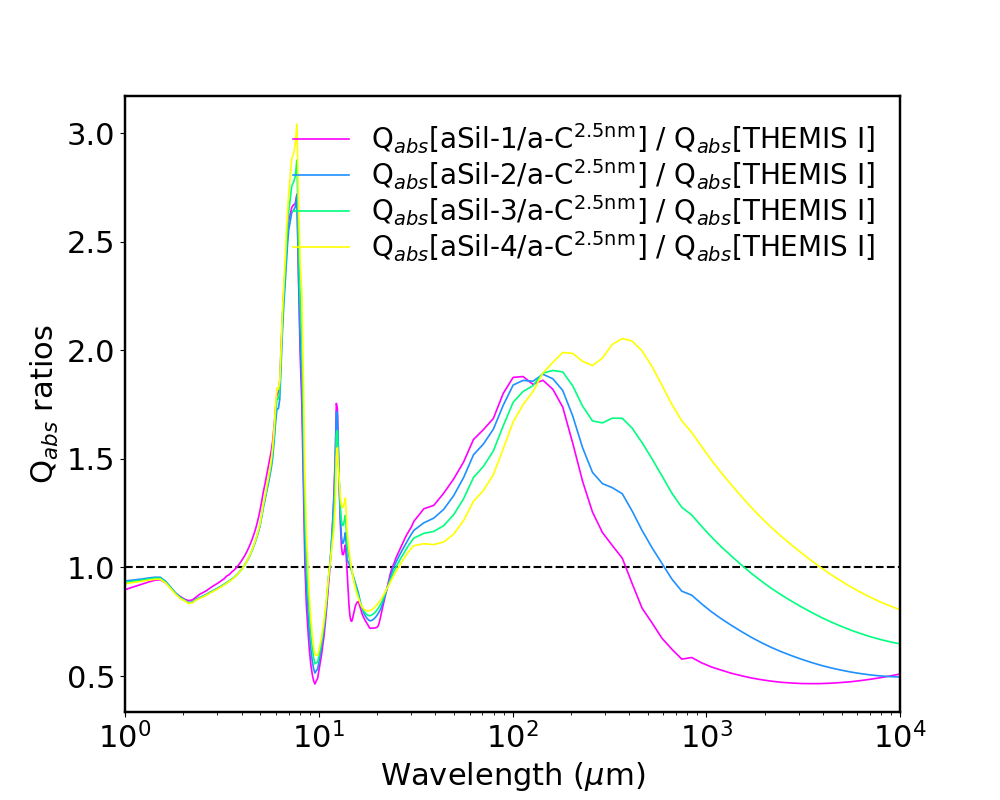} & \includegraphics[width=0.4\textwidth]{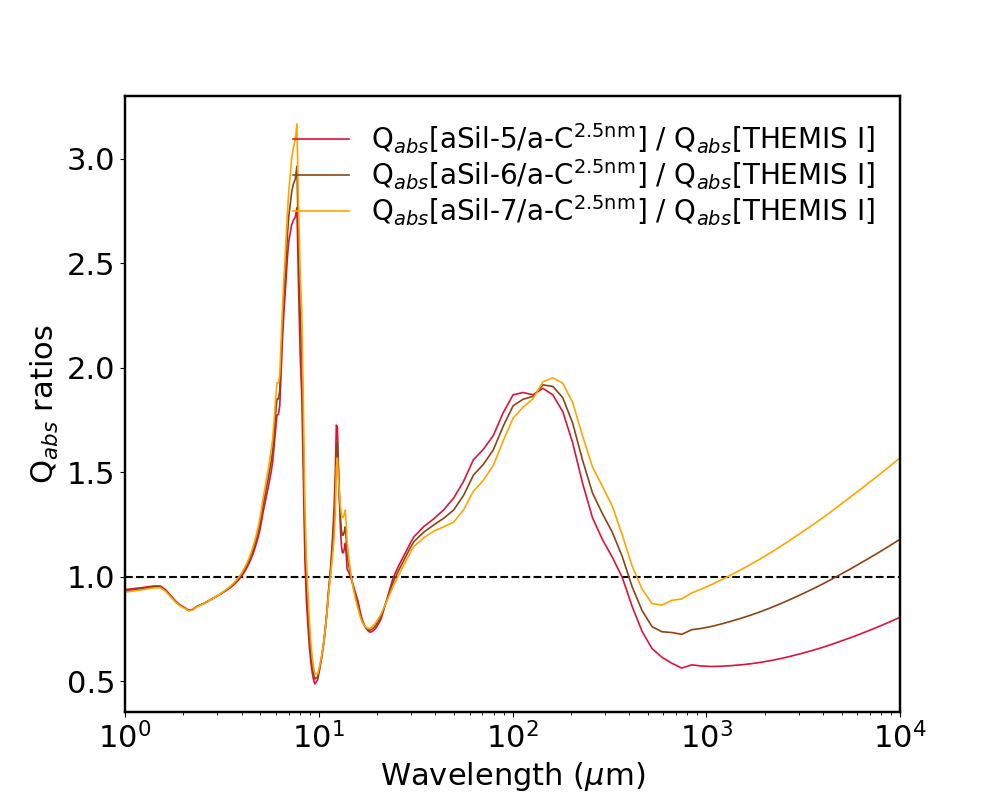} \\ 
\end{tabular}}
\caption{Dependence of optical properties on grain composition for silicate grains with cores made of aSil-1, aSil-2, aSil-3, and aSil-4 materials in pink, blue, green, and yellow, respectively ($a_{eff} = 100$~nm, oblate with $e = 1.3$) on the left and cores made of aSil-5, aSil-6, and aSil-7 in red, brown, and orange, respectively, on the right. Top: Solid lines show $Q_{abs}$ and dashed lines $Q_{sca}$. Middle: Zoom in on the silicate mid-IR features. Bottom: Ratios of $Q_{abs}$ to THEMIS I.}
\label{efficiencies_silicates} 
\end{figure*}

\begin{table}[!t]
\centering
\caption{Characteristics of the absorption efficiencies of the silicate grains (Fig.~\ref{efficiencies_silicates}). The first column indicates the grain composition. The second and third columns give the far-IR ($200 \leqslant \lambda \leqslant 500~\mu$m) and millimetre ($1 \leqslant \lambda \leqslant 3$~mm) spectral indices, respectively. The fourth and fifth columns give the ratios of the aSil-$x$/a-C$^{\rm 2.5nm}$ to THEMIS I absorption efficiencies at 850~$\mu$m (353~GHz) and 2~mm (143~GHz), respectively. The numerical values are given for an oblate grain with $e = 1.3$ and $a_{eff} = 100$~nm.} 
\label{table_Qabs}
\begin{tabular}{l|cccc}
\hline
Composition & $\beta_{200-500\mu{\rm m}}$ & $\beta_{1-3{\rm mm}}$ & $r_{850\mu{\rm m}}$ & $r_{2{\rm mm}}$ \\
\hline
THEMIS I                 & 2.00 & 1.94 & 1.00 & 1.00 \\
aSil-1/a-C$^{\rm 2.5nm}$ & 2.80 & 2.09 & 0.58 & 0.48 \\
aSil-2/a-C$^{\rm 2.5nm}$ & 2.47 & 2.25 & 0.87 & 0.64 \\
aSil-3/a-C$^{\rm 2.5nm}$ & 2.20 & 2.25 & 1.24 & 0.91 \\
aSil-4/a-C$^{\rm 2.5nm}$ & 2.02 & 2.26 & 1.62 & 1.20 \\
aSil-5/a-C$^{\rm 2.5nm}$ & 2.97 & 1.85 & 0.61 & 0.60\\
aSil-6/a-C$^{\rm 2.5nm}$ & 2.88 & 1.77 & 0.76 & 0.85 \\
aSil-7/a-C$^{\rm 2.5nm}$ & 2.79 & 1.73 & 0.92 & 1.11 \\
\hline
\end{tabular}
\end{table}

The differences in the optical properties of the four types of silicates containing the X50b enstatite sample properties mostly show in the far-IR and sub-millimetre range (Fig.~\ref{efficiencies_silicates}). The differences in optical properties between the different silicate mixtures are explained by the break in slope of the imaginary part of the refractive index of the enstatite sample X50a, present only for aSil-1, and the various weights of X50b for aSil-2, aSil-3 and aSil-4, this sample being much more emissive than the forsterite sample X35 (Fig.~\ref{refractive_indices_silicates}). This break leads to a variable spectral index from the far-IR/sub-millimetre to the millimetre range (see Table~\ref{table_Qabs}). For grains illuminated by the ISRF with $G_0 = 1$, the difference in temperature is small between grains of the four compositions. Compared to aSil-1, the increase in the emissivity for $\lambda \gtrsim 200~\mu$m is indeed compensated by a decreased emissivity at shorter wavelengths ($20 \lesssim \lambda \lesssim 200~\mu$m) by about 5, 12, and 19\% for aSil-2/a-C$^{\rm 2.5nm}$, aSil-3/a-C$^{\rm 2.5nm}$, and aSil-4/a-C$^{\rm 2.5nm}$, respectively, in parallel with a slight increase in the UV/visible absorption efficiency. The difference in temperature between THEMIS I silicates and these four grain types is $2 \leqslant \Delta T \leqslant 2.5$~K. The silicate mid-IR features also depend on the exact grain composition. In the case of an oblate grain with $a_{eff} = 100$~nm and $e = 1.3$, the first feature shifts from $\sim 9.7$ to 10.0, 9.9, 9.8, and 9.7~$\mu$m from THEMIS I to aSil-1/a-C$^{\rm 2.5nm}$, aSil-2/a-C$^{\rm 2.5nm}$, aSil-3/a-C$^{\rm 2.5nm}$, and aSil-4/a-C$^{\rm 2.5nm}$, with a decrease in the band strength of a factor of $\sim 2.0$, 1.9, 1.8, and 1.7 when comparing aSil-1/a-C$^{\rm 2.5nm}$, aSil-2/a-C$^{\rm 2.5nm}$, aSil-3/a-C$^{\rm 2.5nm}$, and aSil-4/a-C$^{\rm 2.5nm}$ to THEMIS I, respectively. The peak of the second feature shifts from $\sim 18.4$ to 18.0, 18.3, 18.3, and 18.4~$\mu$m with a decrease in the band strength of a factor of $\sim 1.4$, 1.3, 1.3, and 1.2 when comparing aSil-1/a-C$^{\rm 2.5nm}$, aSil-2/a-C$^{\rm 2.5nm}$, aSil-3/a-C$^{\rm 2.5nm}$, and aSil-4/a-C$^{\rm 2.5nm}$ to THEMIS I, respectively. The spectral profile of this feature is also modified due to the decreasing influence of the forsterite X35 sample from aSil-1 to aSil-4 which gives rise to a shoulder around 23~$\mu$m.

The differences in the optical properties of the four types of silicates containing the X50a enstatite sample again mostly show in the far-IR and sub-millimetre range (see Fig.~\ref{efficiencies_silicates} and Table~\ref{table_Qabs}). The inclusion in the silicates of the X50a rather than the X50b sample leads to stronger wavelength-dependent variations in the spectral index. The general pattern of variations in the mid-IR features is the same whether the enstatite is represented by sample X50a or X50b, with the X50a case showing less significant variations. In the case of an oblate grain with $a_{eff} = 100$~nm and $e = 1.3$, the first feature shifts from $\sim 9.7$ to 9.95, 9.85, and 9.75~$\mu$m from THEMIS I to aSil-5/a-C$^{\rm 2.5nm}$, aSil-6/a-C$^{\rm 2.5nm}$, and aSil-7/a-C$^{\rm 2.5nm}$, with a decrease in the band strength of $\sim 2.0$, 1.9, and 1.9 when comparing aSil-5/a-C$^{\rm 2.5nm}$, aSil-6/a-C$^{\rm 2.5nm}$, and aSil-7/a-C$^{\rm 2.5nm}$ to THEMIS I, respectively. The peak of the second feature shifts from $\sim 18.4$ to 18.3, 18.3, and 18.4~$\mu$m with a decrease in the band strength of a factor of about 1.4 for the three silicate types compared to THEMIS I. 

The use of laboratory data therefore leads to variations in the optical properties of the grains, both in terms of spectral index and opacity. These variations are visible from one sample to another, but also in relation to THEMIS I, and depend on the wavelength considered, with an increase or decrease in opacity and a lower or higher spectral index depending on the sample and the spectral range considered.

\subsection{Grain densities}
\label{grain_densities}

\begin{figure}[!t]
\centerline{\includegraphics[width=0.45\textwidth]{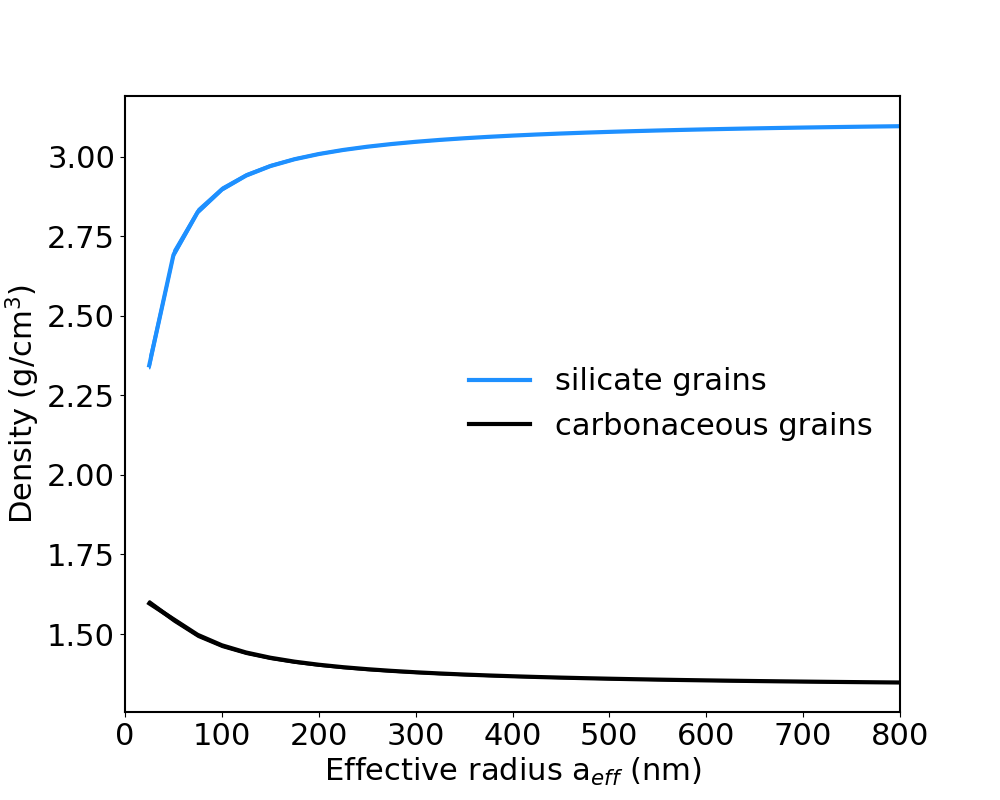}}
\caption{core/mantle spheroidal grain volume densities. The blue line shows the case of grains with silicate cores with 5~nm thick a-C mantles and the black line of a-C:H cores with 20~nm thick a-C mantles. The thickness of the lines shows the little dispersion of the density as a function of the grain shape, oblate or prolate, and elongation.}
\label{densities} 
\end{figure}

Figure~\ref{densities} shows the volume densities of spheroidal core/mantle grains with either carbon or silicate cores for $1.3 \leqslant e \leqslant 2$. The shape and elongation have little influence on density with less than 1.5\% and 0.7\% variations in the case of silicate and carbonaceous grains, respectively. However, the large variation in the core-to-mantle volume ratio as a function of size leads to a variation in the volume density of about +30\% for silicates and -20\% for carbonaceous grains when $a_{eff}$ increases from 25 to 700~nm. The dust SED being inversely proportional to the grain density, the aforementioned variations will be accounted for when producing both total and polarised SEDs.

\section{Observational constraints}
\label{observations}

A wide range of observational constraints are used to define the state of the THEMIS 2.0 chemical composition, size distribution and dust-to-gas mass ratio. In addition to the constraints already presented in Sect.~\ref{observational_constraints}, the defining model parameters are tested against the spectral energy distribution from the mid-IR to the millimetre range and the extinction from the UV to the mid-IR, total and polarised quantities in both cases. One of the most important points is to have as much coherent data as possible that is representative of the same type of weakly irradiated diffuse medium with low column density. To define the dust-to-gas mass ratios of each dust population, one must know how the extinction scales with the gas column density (assuming a perfect mixing of gas and dust along the line of sight). This is usually described by the quantity $N_{\rm H} / E(B-V)$ which was recently shown to vary with increasing value from the Galactic plane to regions at higher latitude \citep[i.e. from high to low column density regions, see for instance references in][]{Ysard2020, Hensley2021, Siebenmorgen2022}. The range of measured values is fairly wide, ranging from around 4 to $9\times 10^{21}$~cm$^{-2}$/mag \citep[see for instance][]{Bohlin1978, Liszt2014, PlanckXI2014, Lenz2017, Murray2018, Nguyen2018, Remy2018, VanDePutte2023}. In agreement with and to be comparable with the models of \citet{Hensley2023} and \citet{Siebenmorgen2022}, we adopt the value measured by \citet{Lenz2017}, $N_{\rm H}/E(B-V) = 8.8 \times 10^{21}$~cm$^{-2}$/mag, as representative of low column density high latitude lines of sight. However, to test the impact of the uncertainty of this parameter, we also consider the Bohlin's ratio, $N_{\rm H}/E(B-V) = 5.8 \times 10^{21}$~cm$^{-2}$/mag (see details in Sect.~\ref{obs_polEXT}).

\citet{Hensley2021} recently presented a review of the observational constraints available in the literature, the aim being to bring them together in a coherent way to make a reference dataset on which to test dust models for the diffuse ISM. In the following, we use part of this dataset but sometimes depart from it for reasons that are detailed in the next sections.

\subsection{Element abundances in the solid phase}
\label{element_abundance}

\citet{Hensley2021} proposed numerical values derived from the literature. It should be noted, however, that estimating the abundance of elements is a complicated exercise and that variations of greater or lesser magnitude depending on the elements and the lines of sight have been observed \citep[e.g.][]{Jenkins2009}. \citet{deCia2021} recently highlighted spatial variations in the metallicity in our Galaxy by a factor greater than 10. Moreover, as pointed out by \citet{Compiegne2011}, the existence of a stellar standard representative of the diffuse ISM is questionable. Should we start from the solar abundance, the enhanced solar abundance following a Galactic Chemical Enrichment model \citep[GCE, ][]{Bedell2018}, or from F,G-type stars? This choice already leads to a difference in the total abundance of up to 90~ppm for carbon or to more than 200~ppm for oxygen \citep[c.f.][]{Compiegne2011, Hensley2021}. As shown by \citet{Mishra2017}, the various methods used to determine [Si or C/H]$_{\rm dust}$ also lead to different estimates (dust model-dependent or -independent methods).

Carbon is an element that appears to cycle rapidly between the gas phase and the solid phase \citep{Jones2014}, which is highlighted by depletion measurements. \citet{Sofia2011} made the first estimates of the abundance of carbon in the gas phase using the strong transition at 1334~$\mathring{A}$ rather than the weak feature at 2325~$\mathring{A}$. This resulted in more reliable estimates of the abundance of carbon in the gas that are systematically lower than estimates based on the 2325~$\mathring{A}$ feature. On average, they found an additional 80~ppm available for the solid phase. Using this method, \citet{Parvathi2012} measured in the diffuse medium $100 \lesssim [{\rm C/H}]_{\rm dust} \lesssim 290$~ppm.

The amount of silicon included in the silicate grains is also uncertain. The measurements range from $[{\rm Si/H}]_{\rm dust} \sim 15$ to 40~ppm depending on the studies \citep[e.g.][]{Sofia2001, Compiegne2011, Nieva2012, Hensley2021}. \citet{Voshchinnikov2010} have also shown a variation in the amount of silicon locked in grains correlated with the Galactic latitude.

Insofar as elemental abundances appear to vary from one line of sight to another, we impose no other limits on the models than to respect the following three criteria based on the articles cited above: $100 \lesssim [{\rm C/H}]_{\rm dust} \lesssim 290$~ppm, $15 \lesssim [{\rm Si/H}]_{\rm dust} \lesssim 40$~ppm and $[{\rm Mg/H}]_{\rm dust} \lesssim 60$~ppm. This allows all the other abundances to be in line with the different values presented in the articles cited above and the references therein.

\subsection{Total SED}
\label{obs_SED}

From far-IR to sub-millimetre, \citet{Hensley2021} tabulated colour-corrected estimates with uncertainties based on Planck Collaboration results \citep{PlanckXVII2014, PlanckXXII2015, PlanckXI2020}. This SED is an average SED for lines of sight with $N_{\rm HI} \sim 3 \times 10^{20}$~H/cm$^2$ and we adopt it in the following (see their Table~3). We further include the measurements made by \citet{Bianchi2017} at 250, 350, and 500~$\mu$m, by correlating the Herschel Virgo Cluster survey data (HeViCs) with HI observations from the Arecibo Legacy Fast ALFA (ALFALFA). As shown by \citet{Hensley2021}, the two datasets agree perfectly.

In the mid-IR, \citet{Hensley2021} use a Spitzer IRS spectrum observed by \citet{Ingalls2011} in the direction of the translucent cloud DCld~300.2-16.9. This cloud, which has a column density of the order of $4 \times 10^{21}$~H/cm$^2$, that is, 10 times higher than the regions used for the far-IR to sub-millimetre part of the spectrum, was affected by supernovae explosions 2 to $6 \times 10^5$ years ago. Shortwards of $5~\mu$m, an average spectrum of star-forming galaxies is added from \citet{Lai2020}. \citet{Hensley2021} then normalise these two spectra to the DIRBE photometric measurements of \citet{Dwek1997}. Since the size distribution of sub-nanometre sized carbon grains, whether PAHs or amorphous hydrocarbons, is strongly affected by shocks and intense radiation fields, the two aforementioned spectra may not be representative of the weakly irradiated diffuse ISM used for the far-IR to sub-millimetre part of the SED. We thus choose not to include any spectrospic data as none exists to date for the diffuse medium at high latitude\footnote{As no changes are made to the properties of the carbonaceous nano-grains, the spectroscopic predictions presented in our previous studies for denser/brighter regions remain valid \citep[e.g.][]{Jones2012c, Jones2013, Jones2016}. Similarly, the microwave spinning dust emission is not affected by the update that we make here and the conclusions of \citet{Ysard2022} still hold for THEMIS 2.0.}. In conclusion, for the mid-IR part, we follow \citet{Compiegne2011} who selected regions of the sky at latitudes $|b| > 15^\circ$ and $N_{\rm H} < 5.5 \times 10^{20}$~H/cm$^2$, avoiding point sources, nearby molecular clouds and the Magellanic clouds. This leads to the inclusion of only four mid-IR data points in the DIRBE filters at 3.53, 4.9, 12, and 25~$\mu$m.

\subsection{Polarised SED}
\label{obs_polSED}

We adopt the polarised far-IR to sub-millimetre SED described by \citet{Hensley2021} which is based on the results of \citet{PlanckXXII2015} and \cite{PlanckXI2020}. The dust model must be able to reproduce all cases and we therefore adopt the maximum polarisation of 19.6\% at 353~GHz derived by \citet{Hensley2021} which leads to a polarised emission per hydrogen of $2.51 \times 10^{-28}$~erg/s/sr/H at 353~GHz.

Comparing the results of the Planck Collaboration with those of BLASTPol \citep[at 250, 350, and 500~$\mu$m, ][]{Ashton2018}, \citet{Hensley2021} conclude that the polarisation fraction from 250~$\mu$m to 3~mm is essentially constant. To reach this conclusion, \citet{Hensley2021} use the Planck SED for the diffuse medium and $\lambda \geqslant 850~\mu$m. The BLASTPol data come from observations in the direction of the translucent molecular cloud Vela C detected in CO, which lies in the Galactic plane and for which the column density is estimated to be around $5\times 10^{21}$~H/cm$^2$, that is, more than one order of magnitude higher than those lines of sight used to produce the Planck SEDs. Studies of comparable translucent clouds have shown that dust is already expected to have significantly grown by coagulation at such column densities \citep[e.g.][]{Ysard2013, Fanciullo2017}. All grains are then mixed and can easily explain why such a ratio would be flat at all wavelengths.

The nature of the regions observed is therefore sufficiently different to warrant at least a cautious warning when comparing the polarisation fraction wavelength-dependency with models designed for the high-latitude diffuse ISM. We therefore do not include this ratio in our fits, as we do not consider it to be binding. We only present the comparison with the results obtained to allow comparison with other dust models \citep{Hensley2023, Siebenmorgen2022}.

\subsection{Total extinction}
\label{obs_EXT}

The extinction in the UV has been extensively investigated and \citet{Hensley2021} present a review of the various studies made since the 1980s. Most of the measurements are consistent and the dispersion over the lines of sight is relatively limited. In the following, we include the observations presented by \citet{Cardelli1989}, \citet{Gordon2009} and \citet{Fitzpatrick2019} for $\lambda \leqslant 0.4~\mu$m. In the visible ($0.4 \leqslant \lambda \leqslant 1~\mu$m), we also follow the prescription of \citet{Hensley2021}, based on the results of \citet{Schlafly2016} and \citet{Wang2019}. Most dust models try to reproduce the standard value $R_V = A(V)/E(B-V) = 3.1$ which corresponds to the average value measured in the Milky Way \citep[see for instance][and references therein]{Fitzpatrick2019}. It is important to note, however, that $R_V$, like most of the parameters used to define grains in the diffuse ISM, can vary significantly from one line of sight to another. As described by \citet{Siebenmorgen2023}, the OB stars often used to measure the reddening curve are generally at a great distance and the lines of sight therefore pass through several individual interstellar clouds, simulating a more or less constant -- average -- value of $R_V$. Selecting 53 well characterised single-cloud lines of sight (distances, spectral types) they find variations between $2 \lesssim R_V \lesssim 4.3$ and an average of $3.1 \pm 0.4$. This is similar to the results of \citet{Decleir2022} who find values between $2.4 \lesssim R_V \lesssim 5.3$ for lines of sight with $0.8 \lesssim A(V) \lesssim 3$, while respecting the 3.1 average value.

The extinction curve in the near-IR ($1 \lesssim \lambda \lesssim 4~\mu$m) is usually represented by a power law $A(\lambda) \propto \lambda^{-\alpha}$. Recent photometric measurements agree on a value of $\alpha$ around 1.7-1.8 in the diffuse ISM \citep[e.g.][]{Chen2018, Nagatomo2019, Wang2019}, which was confirmed by the first spectroscopic extinction curve measurements made by \citet{Decleir2022} between 0.8 and 5~$\mu$m. For a range of measured values between 1.4 to 2.2, they found an average for the diffuse ISM of $\alpha = 1.71 \pm 0.01$. These results are in agreement with older determinations such as those of \citet{Rieke1985}, \citet{Cardelli1989} and the lines of sight with the lowest extinction in \citet{McClure2009}. \citet{Hensley2021} deviate from this and choose to set $\alpha = 1.55$, which gives a much flatter extinction. We do not follow them and therefore use the recent extinction curves referred to above. Their choice appears to be motivated by the shape of the mid-IR extinction that they adopt. 

In the mid-IR ($4 \lesssim \lambda \lesssim 35~\mu$m), \citet{Hensley2021} built on the work of \citet{Hensley2020} for the determination of the silicate band profiles at $\sim 10$ and 18~$\mu$m. Their observations were made in the direction of Cyg OB2-12 which has an $A(V)$ of $\sim 10$. This line of sight shows a definite flattening for $4 \lesssim \lambda \lesssim 8~\mu$m. This result is in agreement with other studies towards regions with similar relatively high $A(V)$ either in the direction of the Galactic plane or of dark clouds such as the Coalsack nebula \citep[e.g.][]{Indebetouw2005, Gao2009, Wang2013, Wang2015, Xue2016} and which can be modelled with grain models including significant grain growth compared to dust in the diffuse ISM \citep{Weingartner2001}. These lines of sight are a priori significantly different from the high latitude lines of sight used to determine total and polarised SEDs with column densities below a few $10^{20}$~H/cm$^2$. However, more recent measurements have since been reported by \citet{Gordon2021}, based on Spitzer IRS, IRAC, and MIPS data: some of their sightlines have quite high $A(V)$ as in previous studies but nine of them have $A(V) < 3$. When looking at their Fig.~5 in which the extinction curves are ranked according to the dust column density, we see two things: (i) no strong flattening shortwards of 8~$\mu$m; (ii) the band-to-continuum ratio decreases with decreasing column density. The spectral shape shortwards the 10~$\mu$m silicate feature appears to be consistent with previous studies and the $10~\mu$m silicate feature itself is compatible with those presented in \citet[][sightlines with the lowest extinction]{Chapman2009} and \citet{VanBreemen2011}. Since the extinction curves measured by \citet{Gordon2021} with $A(V) < 3$ are more representative of the high latitude low density regions used to characterise the total and polarised SEDs, we use them in the following. To combine these mid-IR extinction data with those in the near-IR by \citet{Decleir2022}, it would be best to perform a joint analysis using the same reference stars for both datasets in order to get a more reliable slope from near- to mid-IR. Similarly, the extrapolation to infinite wavelengths that is used to estimate $A(V)$ would benefit from being re-done fitting all near- to mid-IR data simultaneously\footnote{\citet{Gordon2023} carried out such a study from the UV to mid-IR using the samples of \citet{Gordon2009},  \citet{Fitzpatrick2019}, \citet{Gordon2021} and \citet{Decleir2022}. However, to produce an average extinction curve, they use observations of lines of sight with $A(V) \gtrsim 4$ which results in a higher mid-IR band-to-continuum ratio.}. However this is beyond the scope of this paper and in the following, we simply normalise the mid-IR results of \citet{Gordon2021} to the near-IR results of \citet{Decleir2022} at $\lambda = 5~\mu$m.

\subsection{Polarised extinction}
\label{obs_polEXT}

In accordance with the literature presented by \citet{Hensley2021}, the UV to near-IR extinction is represented by a Serkowski law \citep{Serkowski1975}: 
\begin{equation}
p/p_{max} = \exp\left[ -K \ln^2(\lambda_{max}/\lambda)\right],
\end{equation}
with $K = 0.87$ and $\lambda_{max} = 0.55~\mu$m from 0.12 to 1.38~$\mu$m ; a power law with $\beta = 1.6$ is adopted for $1.38 \leqslant \lambda \leqslant 4~\mu$m. It is worth noting, however, that measurements of $\lambda_{max}$ along diffuse to translucent lines of sight, out of the Galactic Plane, have shown that it increases with the visual extinction $A(V)$. For instance, \citet{Vaillancourt2020} show variations from $\lambda_{max} = 0.43$ to 0.72~$\mu$m for $A(V) \leqslant 4$, which can be approximated by the linear relationship $\lambda_{max} \sim 0.51 + 0.02 A(V)$. 

As for the polarised emission, the model should be able to reproduce the maximum polarisation, which is usually expressed in the V band per unit reddening $p_V / E(B-V)$, assuming that $R_V = 3.1$. Compared to the classical value of 9\% per mag \citep[e.g.][]{Serkowski1975}, this maximum has recently been revised upwards in two studies. For the low column density lines of sight used to estimate the total and polarised SEDs, \citet{PlanckXII2020} found a maximum value of about $p_V / E(B-V) \sim 13$\%. \citet{Panopoulou2019} found that 13\% is only a lower limit. They could linearly fit their data with a value of $15.9 \pm 0.4$\% and found that all their lines of sight had $p_V / E(B-V) < 20$\%. Similar results were also found by \citet{Angarita2023}.

Given the uncertainty in the measurement of the polarised extinction, we make different fits in the following, one with a normalisation at $p_V / E(B-V) = 13$\% and one using the low classical value of 9\%. As stated in Sect.~\ref{observations}, we choose to first normalise our model to \citet{Lenz2017}. If instead we normalise to the lower value previoulsy measured by \citet{Bohlin1978}, a maximum optical polarisation of 13\% (9\%) would be equivalent to a maximum of 19.7\% (13.6\%) in the normalisation frame of \citet{Lenz2017}. \citet{Bohlin1978} and \citet{Lenz2017} thus roughly encompass the dispersion in $N_{\rm H}/E(B-V)$ measured until now\footnote{We can also illustrate this in terms of $R_V$: the Bohlin normalisation is equivalent to the Lenz normalisation if we decrease the value of $R_V$ from 3.1 to 2.1, for example. This exploration should therefore lead to a dispersion of the value of $R_V$ in the models in line with the measurements presented in Sect. ~\ref{obs_EXT}.} \citep[e.g.][]{Liszt2014, PlanckXI2014, Nguyen2018, Remy2018, Zuo2021, VanDePutte2023}.  

A handful of polarisation measurements in the 3.4~$\mu$m hydrocarbon band were also made towards the Galactic Centre \citep{Adamson1999, Chiar2006}, three young stellar objects \citep{Ishii2002} and a Seyfert 2 galaxy \citep{Mason2007}. All the studies either only measured upper limits or very low values, compatible with a non-polarised band. For instance, \citet{Chiar2006} observed two Galactic Centre sources, GCS-3II and GCS-3IV, for which they measured $\Delta p_{3.4\mu{\rm m}} = 0.06 \pm 0.13$\% and $0.15 \pm 0.31$\%, respectively. We do not include these values in our fitting procedure but compare them a posteriori with our results to allow comparison with other models \citep[e.g.][]{Siebenmorgen2022, Hensley2023}. The observed lines of sight are indeed too different from those used to produce the total and polarised SEDs to be used as binding constraints. A measurement of the polarisation in the 10~$\mu$m-silicate band has also been made by \citet{Telesco2022} in the direction of Cyg OB2-12: they find $p_{10.2\mu{\rm m}} = (1.24 \pm 0.28)$\%. Combining this measure with that of \citet{McMillan1977}, \citet{Hensley2021} find $p_{10.2\mu{\rm m}}/p_{0.55\mu{\rm m}} = 0.014 \pm 0.03$. This gives an interesting anchor point for dust models but again, due to the disparity with the lines of sight used to build the total and polarised SEDs, we do not use it as a binding constraint in our fitting routine but only compare it with our results afterwards.

\subsection{Sub-millimetre to visible observational ratios}
\label{obs_ratios}

\citet{PlanckXII2020} also estimated two interesting ratios to test dust models: $R_{P/p} = P_{353{\rm GHz}}/p_V$, the ratio of the polarised intensity in the sub-millimetre to the degree of polarisation in the visible, and $R_{S/V} = (P/I)_{353{\rm GHz}} / (p/\tau)_V$, the ratio of the submillimetric to visible polarisation fractions. They found $R_{P/p} = 5.4 \pm 0.5$~MJy/sr/mag and $R_{S/V} = 4.2 \pm 0.5$. In constrast to the all-sky study performed by \citet{PlanckXII2020}, \citet{Panopoulou2019} restricted themselves to a sky area with a high submillimetric polarisation fraction of $\sim 20$\%, similar to that of the polarised SED defined in Sect.~\ref{obs_polSED}, and found $R_{P/p} = 4.1 \pm 0.1$~MJy/sr/mag. In the following, these ratio values are not included in our fitting routine, but we compare our results against them retrospectively.

\subsection{Interstellar radiation field}
\label{obs_ISRF}

With fixed grain properties, the SED then only depends on the radiation field. The choice of the radiation spectral shape and intensity therefore affects the final estimation of the gas-to-dust mass ratio. We fix its spectral shape to that of \citet{Mathis1983} at a galactocentric distance of 10~kpc, scaled by the $G_0$ factor\footnote{Scaling factor for the radiation field integrated between 6 and 13.6~eV. The standard ISRF corresponds to $G_0 = 1$ and to a flux of $1.6 \times 10^{-3}$~erg/s/cm$^2$ \citep{Parravano2003}.}. \citet{Fanciullo2015} showed that the ISRF intensity varies in the diffuse ISM as at most $0.5 \lesssim G_0 \lesssim 1.8$.

\section{THEMIS 2.0 definition}
\label{themis_ii}

From the laboratory and astronomical data presented above, the model can now be defined. We first present the methodology of our adjustments, the associated results, and finally define the new version of THEMIS.

\subsection{Grain alignment}
\label{grain_alignment}

The polarisation of radiation comes from the alignment of aspherical grains with the magnetic field, this alignment being made possible by the magnetic properties of the grains. The detailed explanation of grain alignment, which is beyond the scope of this paper, has made great progress in the last decade both theoretically and observationally \citep[e.g.][]{Andersson2015, Hoang2016, Lazarian2020, Hoang2023}. 

It is widely accepted that interstellar silicates contain iron, so these grains are paramagnetic and can align themselves effectively with the magnetic field. In most models of interstellar grains, the population of large carbons is usually represented by graphite, a diamagnetic material, which cannot align effectively with the magnetic field. Indeed, graphite paramagnetic properties only arise from the rotation of charged grains or hydrogen attachment to the surface, both resulting in weak magnetic moments \citep[see for instance][]{Hoang2016}. However, THEMIS does not assume graphite but instead hydrogenated amorphous carbons. A paramagnetic defect in a disordered hydrocarbon lattice can be defined by an electronic configuration with an unpaired spin that gives rise to an electronic state close to the Fermi level \citep{Delpoux2009}. For a lattice composed entirely of $sp^3$ carbons, defects are $sp^3$ sites with dangling bonds. In the case of a material containing both $sp^2$ and $sp^3$ carbons, two types of defects can be observed. The first is an isolated dangling bond. The second, any cluster with an odd number of $sp^2$ sites generates a Fermi state. The magnetic moment of an a-C:H grain therefore depends on the concentration of paramagnetic defects in the material \citep[e.g.][]{Esquinazi2005, Akhukov2013, Sakai2018}. This concentration has been estimated experimentally for amorphous hydrocarbons in meteorites \citep{Binet2002, Binet2004} for which it is comparable to the paramagnetic defects in silicates \citep{Delpoux2009}.

The theoretical study carried out by \citet{Hoang2023} shows that if a-C:H grains larger than 50~nm are present in the diffuse ISM, they can produce significant polarisation due to efficient internal alignment and B-RAT alignment (see their Sect.~7). They further suggest to test this finding against measurement of the polarisation in the 3.4~$\mu$m hydrocarbon feature. 

While we do not consider the possibility of iron inclusions in a-C(:H) materials, we point out that their doping with sulphur atoms can induce magnetic properties, as highlighted by \citet{JonesAlone2013}. Thus, there appears to be no reason why a-C(:H) grains should not be aligned. In what follows, we therefore assume that the two populations of large grains can align with the magnetic field but remind the reader that the concentrations of free radicals and impurities with uncompensated electron spins as well as the possibility of iron inclusions are still uncertain for the diffuse ISM carbonaceous dust \citep[see for instance][]{Lazarian2020}.

\subsection{Methodology}
\label{definition_methodology}

We have seven types of silicates and five types of large core/mantle carbonaceous grains, each for two oblate and two prolate shapes. As we have no prior knowledge, we test the association of all carbonaceous and silicate grains, the populations being allowed to have identical or different shapes. In the same way, we test the cases of a universal grain alignment law or a differentiated alignment between amorphous hydrocarbons and silicates. We consider three free parameters for the analytical alignment function defined by Eq.~\ref{alignment_fraction}: $f_{max}$, $a_{thresh}$ and $p_{stiff}$. The radiation field is allowed to vary within the limits defined in Sect.~\ref{obs_ISRF}.

To perform the fits, we use the DustEMWrap\footnote{Available here: \url{http://dustemwrap.irap.omp.eu/Home.html}.} IDL extension of the DustEM code. DustEMWrap allows for iterative fitting of SEDs and extinction curves, including the fit of linear polarisation (Stokes parameters I, Q, and U). The total and polarised SEDs are fitted together with the Serkowski law. The model total extinction is only a posteriori qualitatively compared to the observations which are too widely spread to be included in the fitting routine. Our focus points at that stage are mainly the slope in the near-IR and the shape and intensity of the two mid-IR silicate features. The polarisation fraction is not included in the fit either, but simply compared a posteriori with the results for the reasons given in Sect.~\ref{obs_polSED}. Finally, elemental abundances are calculated after the fitting procedure and we exclude models that do not comply with the limits given in Sect~\ref{element_abundance}.

\subsection{Fitting results}
\label{results}

\begin{figure*}[!th]
\centerline{\begin{tabular}{ccc}
\includegraphics[width=0.32\textwidth]{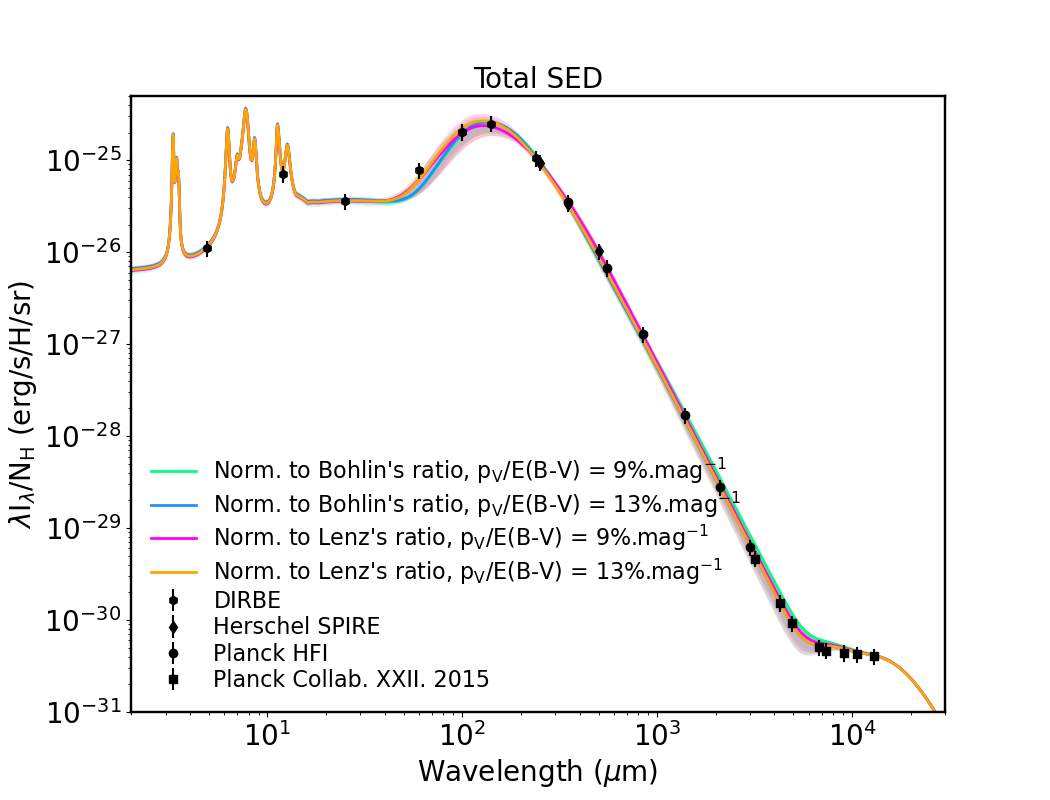} & \includegraphics[width=0.32\textwidth]{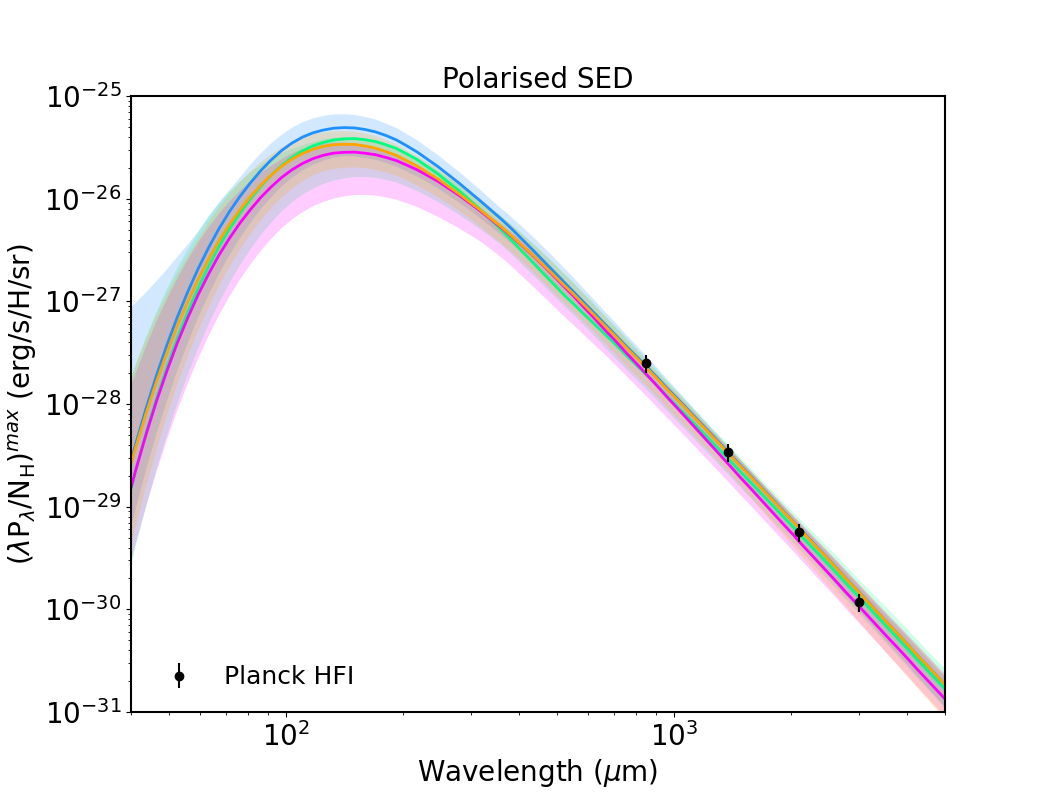} & \includegraphics[width=0.32\textwidth]{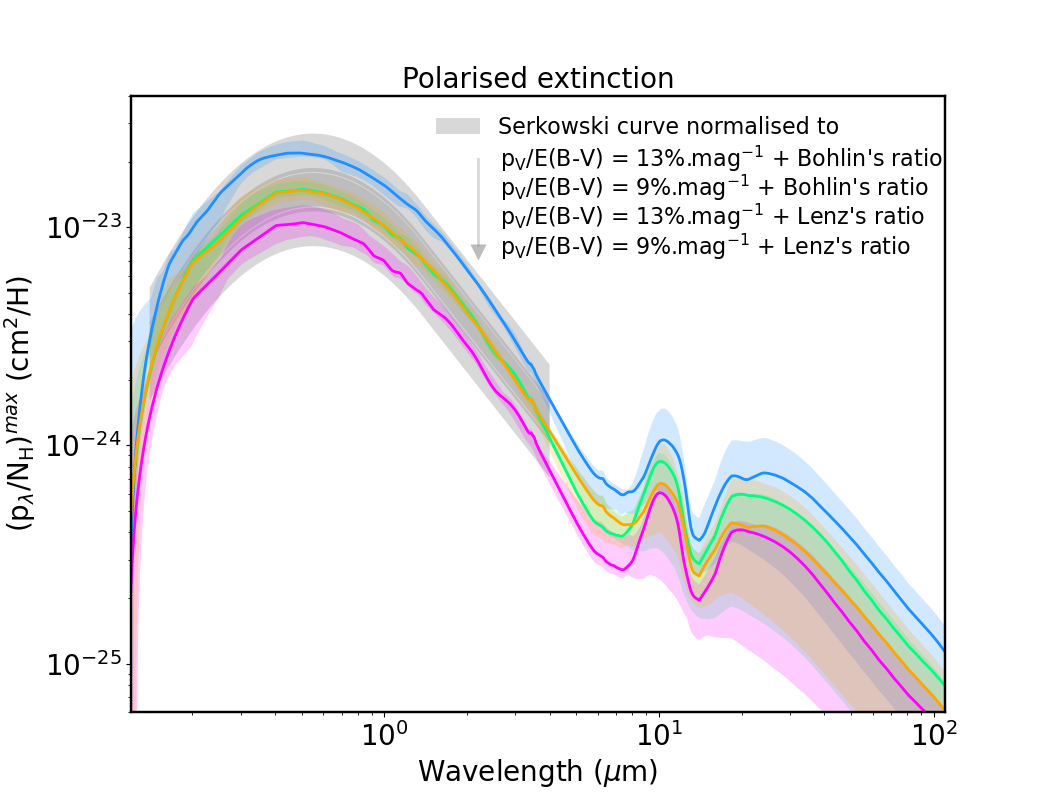} \\ 
\includegraphics[width=0.32\textwidth]{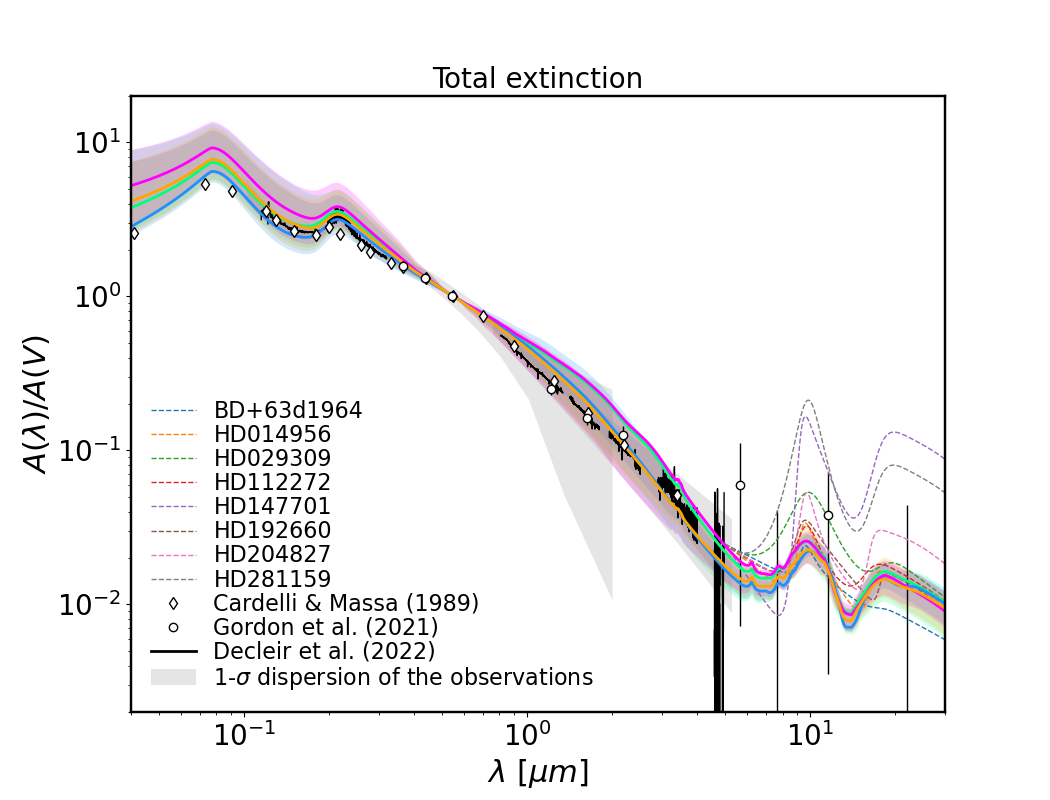} & \includegraphics[width=0.32\textwidth]{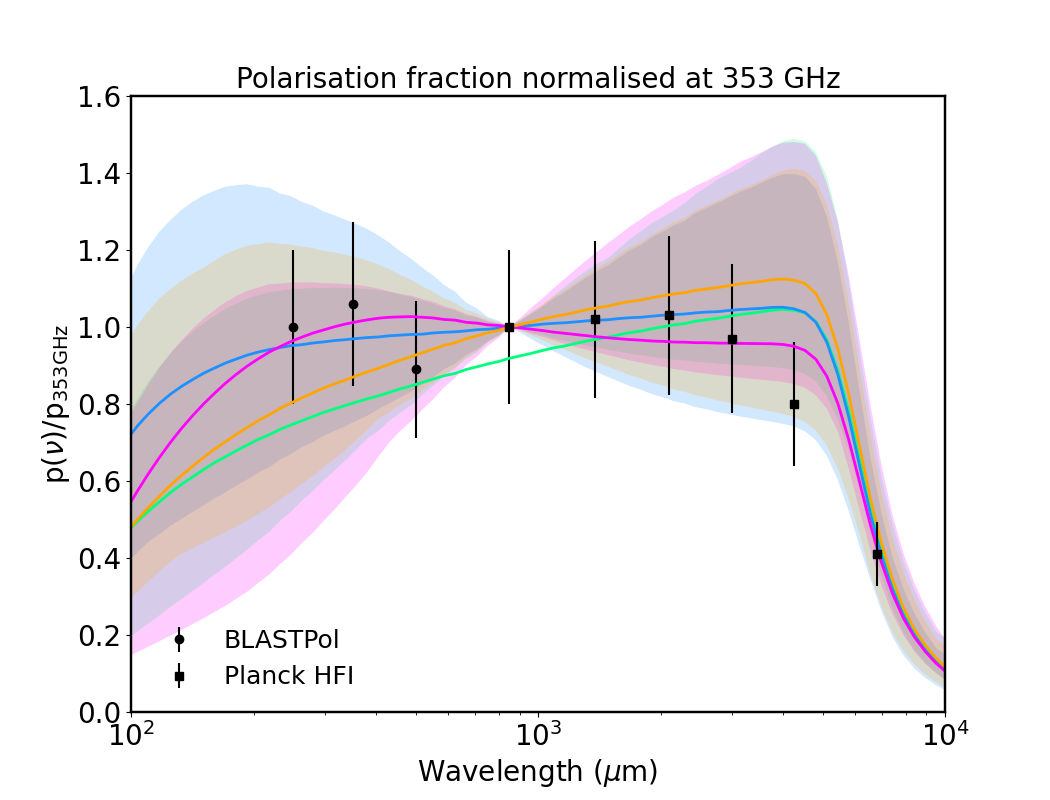} & \includegraphics[width=0.32\textwidth]{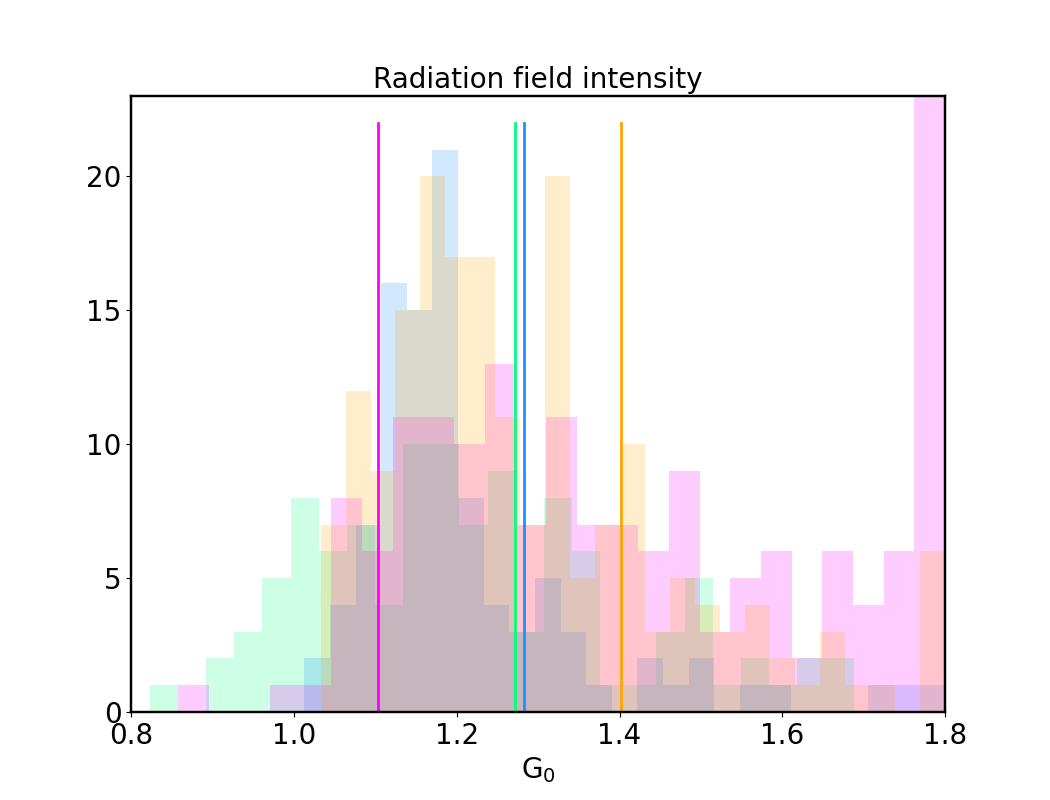} \\
\includegraphics[width=0.32\textwidth]{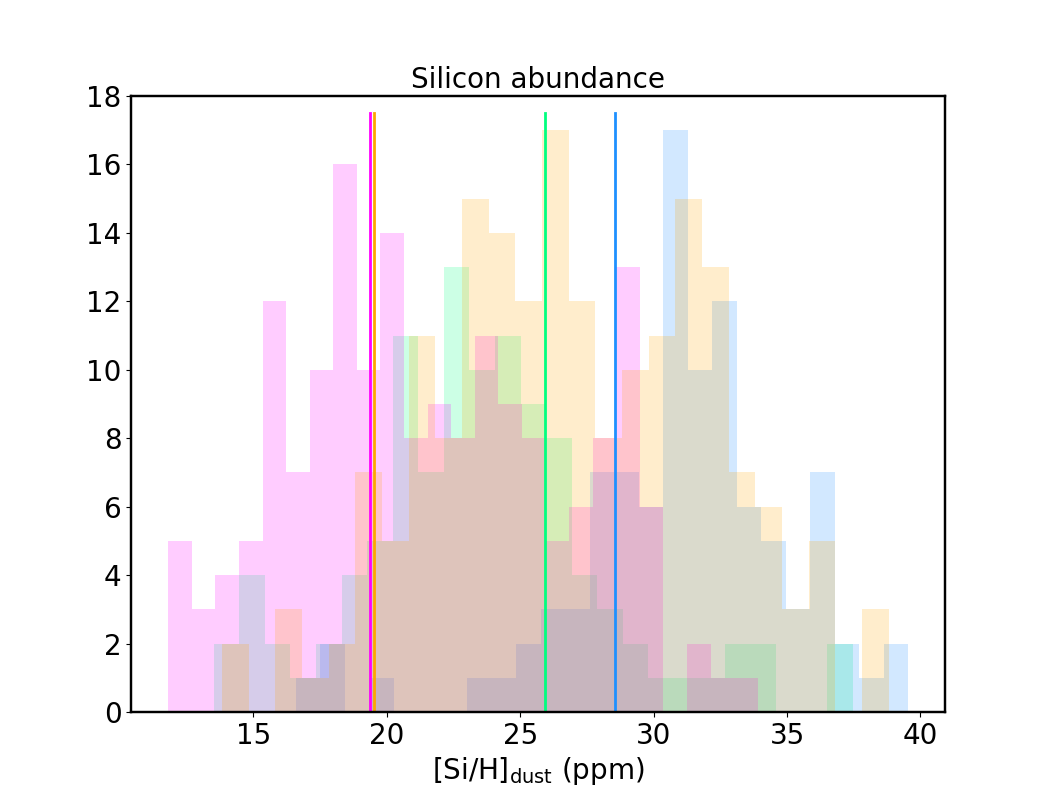} & \includegraphics[width=0.32\textwidth]{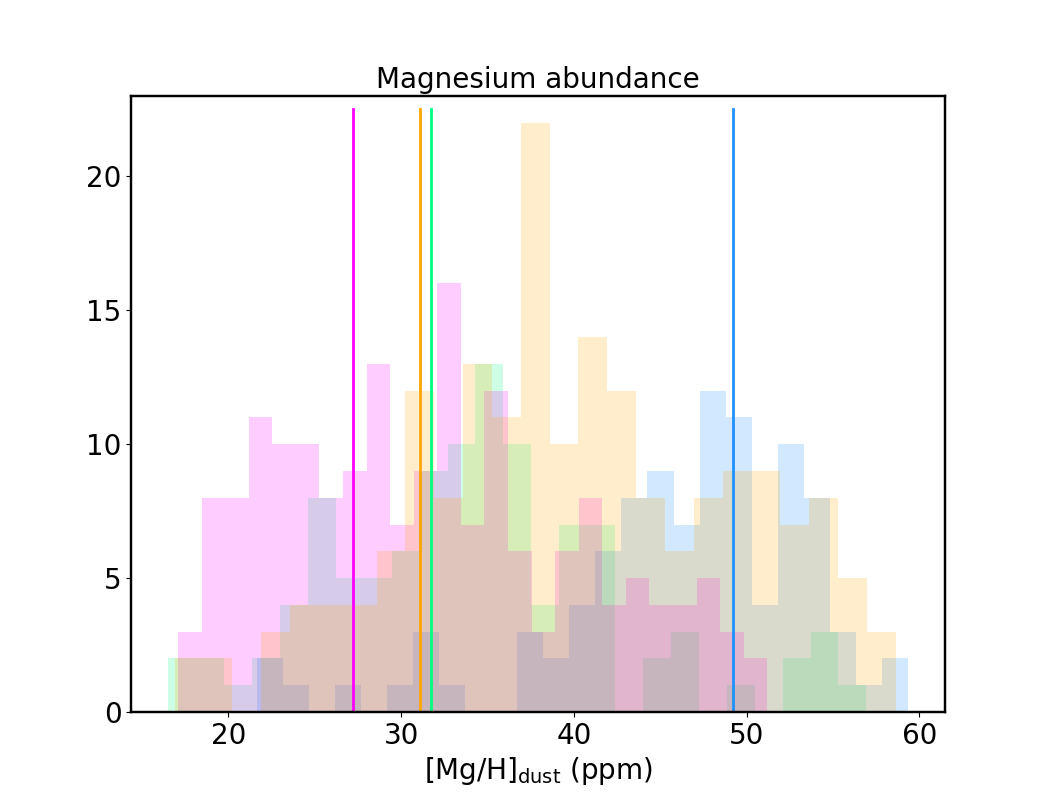} & \includegraphics[width=0.32\textwidth]{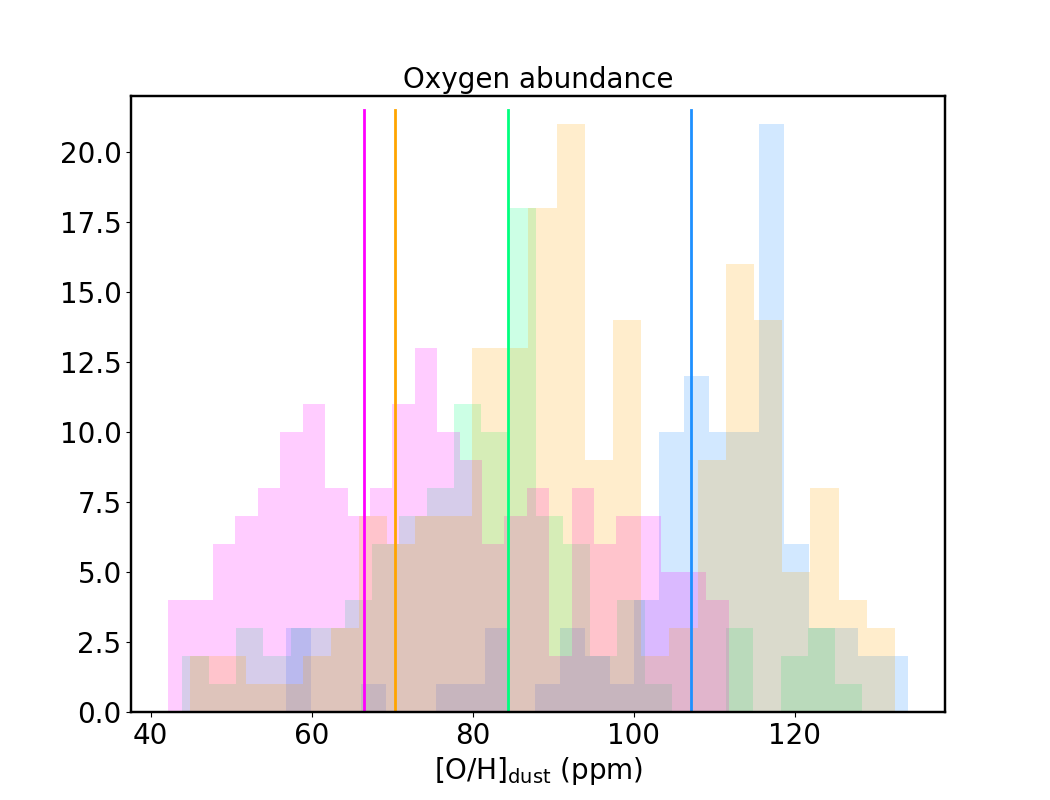} \\
\includegraphics[width=0.32\textwidth]{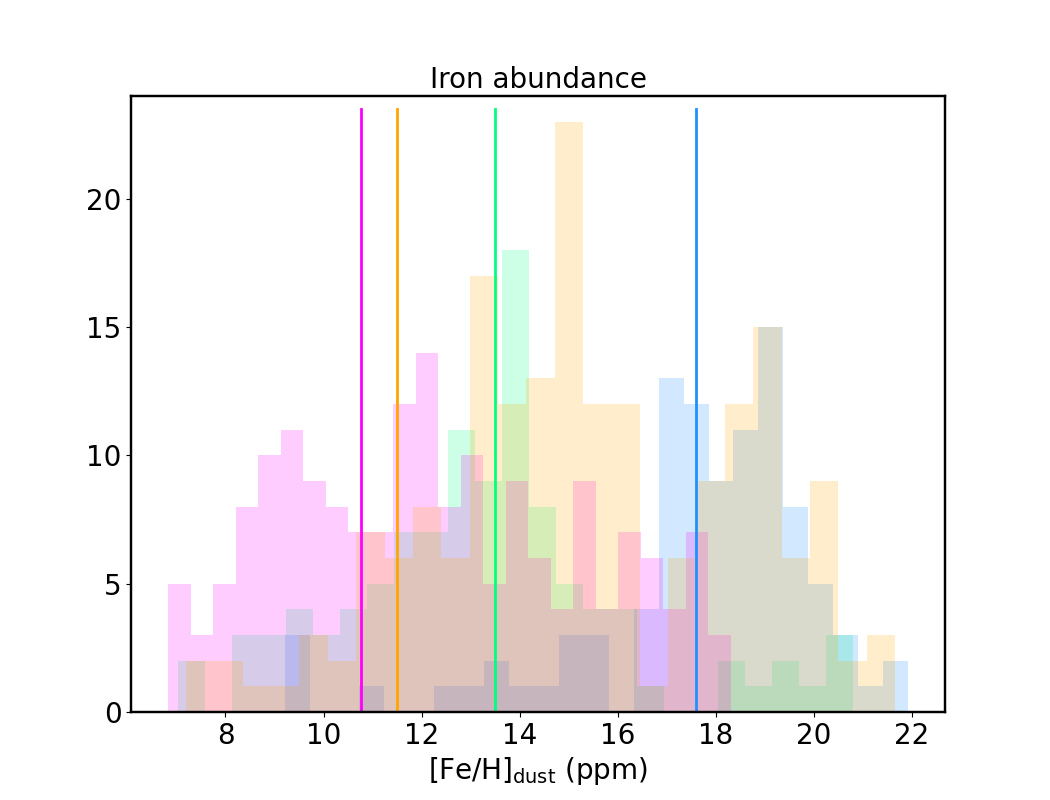} & \includegraphics[width=0.32\textwidth]{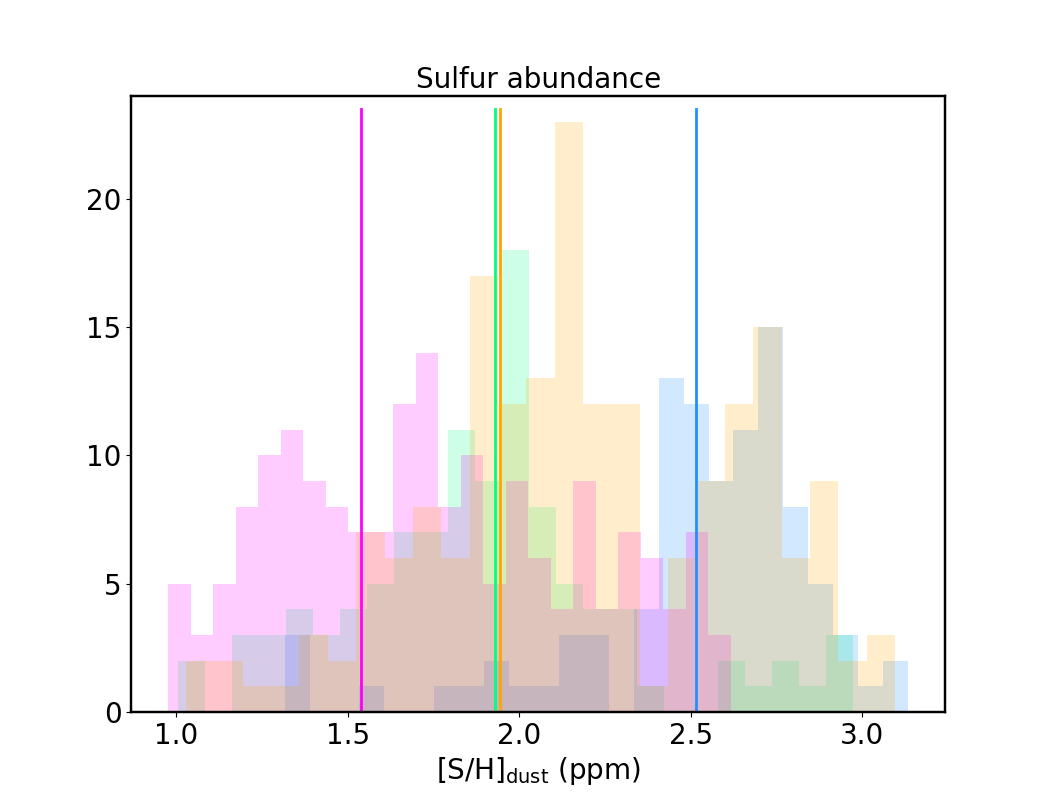} & \includegraphics[width=0.32\textwidth]{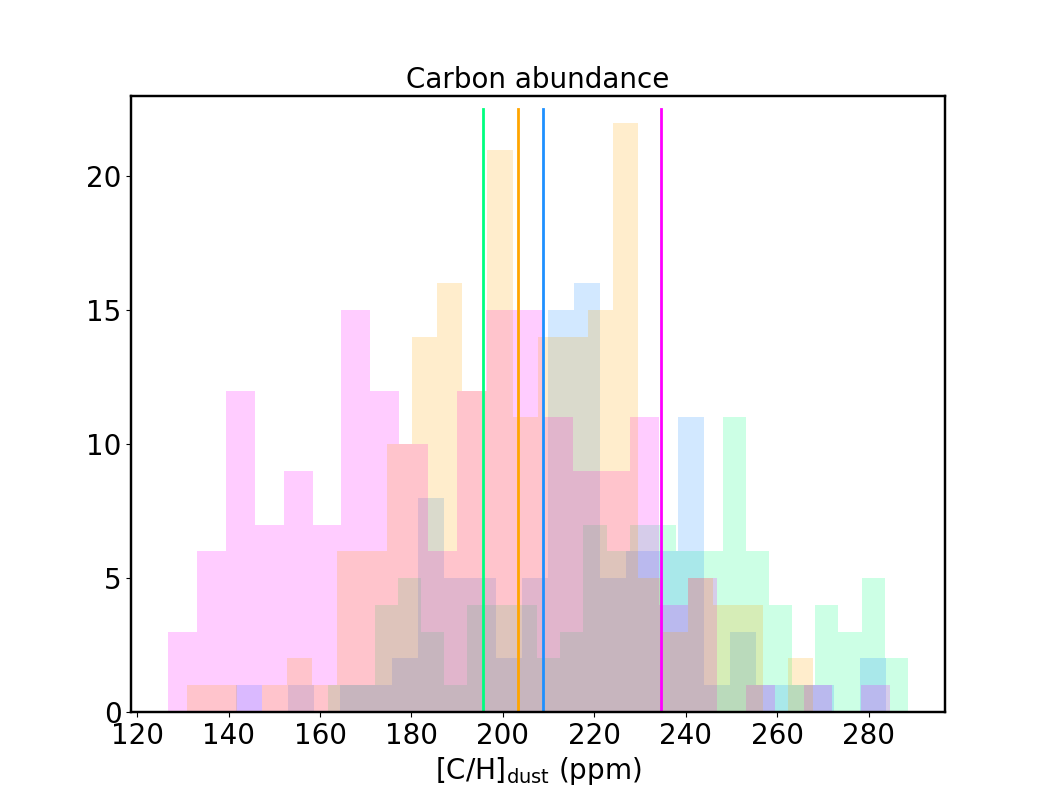} \\
\end{tabular}}
\caption{Data are those presented in Sect.~\ref{observations}. From top to bottom and left to right, the figures first show the models against the observations: total SED, polarised SED, polarised extinction, total extinction ; and then some model parameters: radiation field, silicon abundance, magnesium abundance, oxygen abundance, iron abundance, sulphur abundance, and carbon abundance. The light coloured areas represent the dispersion of the models in agreement with the observations and the matching thick solid lines show the best fits (see Sect.~\ref{results} for details). As detailed in the text, two gas-to-dust mass ratios and two maximum optical polarisation values are considered to normalise the models. The models normalised to the ratio of \citet{Lenz2017} are shown in magenta and orange for $p_{\rm V}/E(B-V) = 9$\% and 13\%, respectively. Those normalised to the ratio of \citet{Bohlin1978} are shown in green and blue for $p_{\rm V}/E(B-V) = 9$\% and 13\%, respectively.}
\label{fit_results} 
\end{figure*}

\begin{figure}[!t]
\centerline{\includegraphics[width=0.45\textwidth]{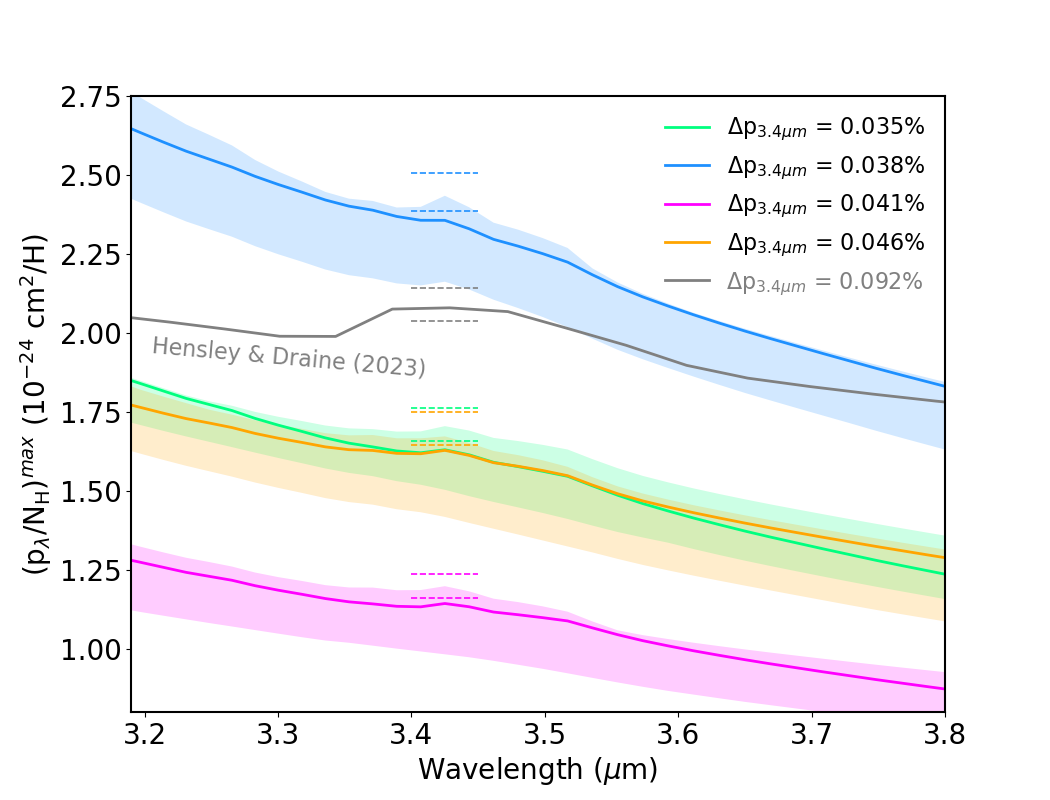}}
\caption{Same colour scheme as in Fig.~\ref{fit_results}. Zoom on the 3.4~$\mu$m hydrocarbon feature seen in polarised extinction. The dashed lines show the upper limits measured by \citet{Chiar2006} towards GCS 3-II and GCS 3-IV, lowest and highest lines, respectively, above the continuum of our four best-fit models (see Sect.~\ref{results} for details). For comparison, we also show in grey the model of \citet{Hensley2023} which complies with the upper limit measured towards GCS 3-IV but not with the lower one towards GCS 3-II. \citet{Hensley2023} gives indeed $\Delta p_{3.4\mu{\rm m}}$ 2 to 2.6 times as large as our best models.}
\label{3p4um} 
\end{figure}

\begin{figure}[!t]
\centerline{\includegraphics[width=0.45\textwidth]{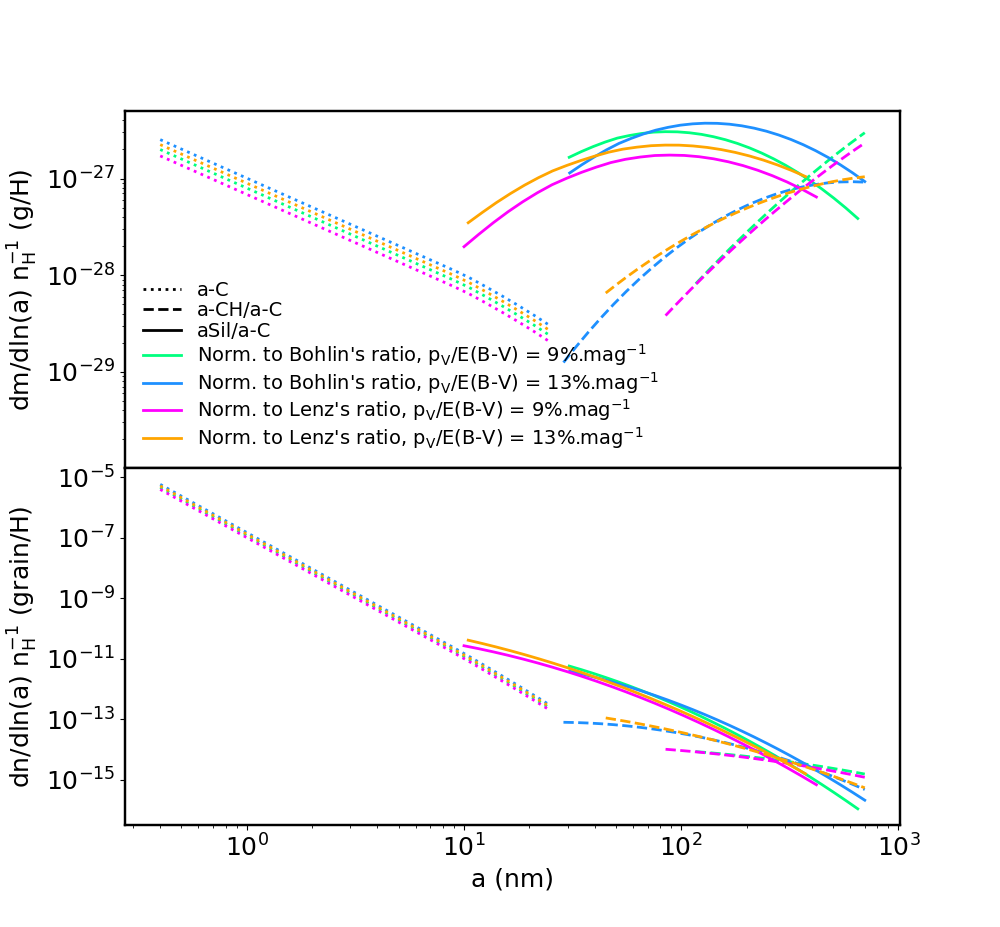}}
\caption{Same colour scheme as in Fig.~\ref{fit_results}. {\it Top:} Grain mass distributions of the best-fit models shown in Fig.~\ref{fit_results}. {\it Bottom:} Equivalent grain number distributions. In both figures, the distributions for the amorphous nano-carbon grains, a-C, are shown as dotted lines, for the a-C:H/a-C grains as dashed lines and for the aSil/a-C grains as solid lines.}
\label{size_distribution} 
\end{figure}

\begin{figure}[!t]
\centerline{\includegraphics[width=0.45\textwidth]{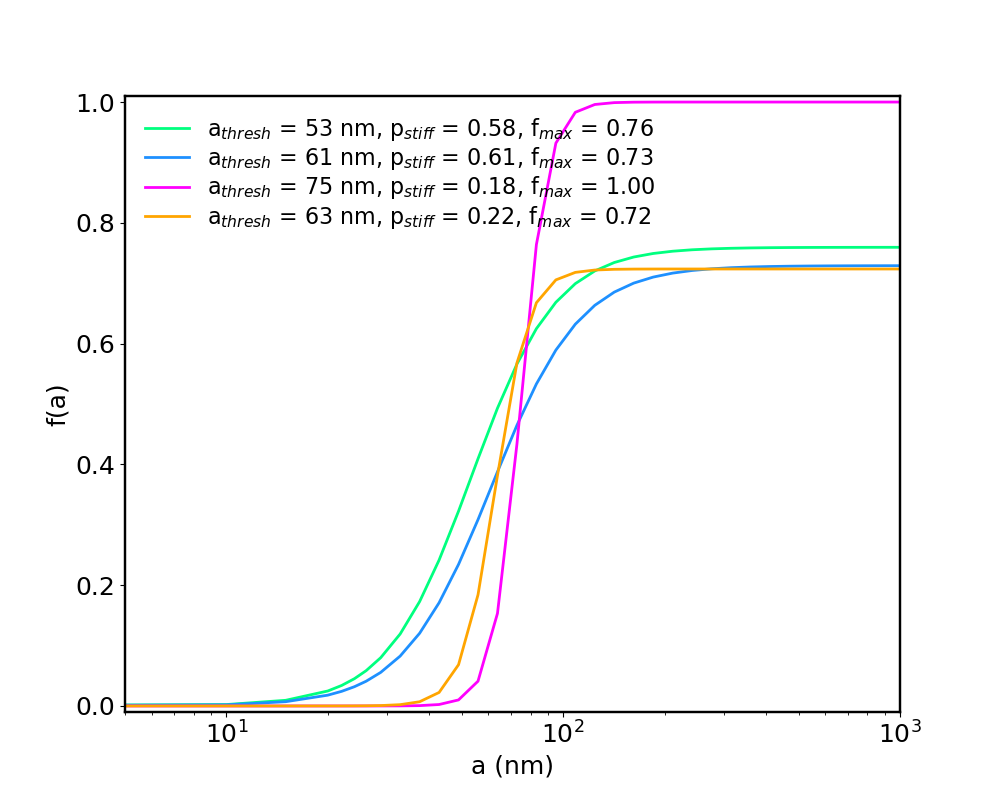}}
\caption{Alignment functions of the best fits presented in Fig.~\ref{fit_results} and Sect.~\ref{results}. The colour scheme is the same as in Fig.~\ref{fit_results}. Models normalised to the ratio of \citet{Lenz2017} are shown by magenta and orange lines for $p_{\rm V}/E(B-V) = 9$\% and 13\%, respectively. Both models are for prolates with $e = 2$. Those normalised to the ratio of \citet{Bohlin1978} are shown by green and blue lines for $p_{\rm V}/E(B-V) = 9$\% and 13\%, respectively. The model plotted in green is for oblates with $e = 1.3$ and the one in blue for prolates with $e = 2$.}
\label{alignment} 
\end{figure}

The fitting results and the associated parameters for defining the dust model are shown in Figs.~\ref{fit_results}, \ref{3p4um}, \ref{size_distribution} and \ref{alignment}. As expected from Fig.~\ref{efficiencies_silicates}, the mid-IR silicate features are not very strong but agree with half of the lines of sight with $A(V) < 3$ presented by \citet{Gordon2021}. This illustrates just how crucial an out-of-Galactic plane diffuse extinction curve ($A(V) < 1$) would be for calibrating grain models\footnote{The Planck-HFI data showed that the dust is probably different in the Galactic plane, where a higher spectral index is observed than at higher latitudes, and that this spectral index is correlated with temperature, unlike in the high-latitude sky where these two quantities are anti-correlated \citep[see for instance][and references therein]{PlanckXI2014}.}. As a complement, Appendix~\ref{appendix_contributions} shows the contributions of the different grain populations to total and polarised emission and extinction (Figs.~\ref{appendix_pol_ext}, \ref{appendix_seds} and \ref{appendix_extinction}). For comparison, fitting results obtained with THEMIS I silicates are presented in Appendix~\ref{appendix_THEMIS_I} (Figs.~\ref{3p4um_old}, \ref{size_distribution_old}, \ref{alignment_old} and \ref{fit_results_old}).

We can see that the new silicates derived from laboratory measurements make it possible to explain the UV to cm observations without violating the constraints on the abundance of the various elements making up the grains. Two comments can be made about grain composition, independent on the data normalisation: (i) all the silicate mixtures can explain the observations; (ii) all the mantle descriptions for a-C:H/a-C lead to acceptable fits, with the exception of a-C$^{\rm 2.5nm}$ (same result if THEMIS I silicates are used instead of lab-derived silicates). The first comment agrees with the constraints derived from the X-ray and mid-IR observations presented in Sect.~\ref{observational_constraints} showing that silicates are a mixture of olivine and pyroxene stoichiometry mineralogies. This makes it difficult to conclude on the proportion of one type of silicate to the other from this study. This is not particularly surprising, since this proportion is bound to vary according to local physical and chemical conditions \citep{Demyk2001, Psaradaki2022, Psaradaki2023}. The second comment gives indications on the evolution of carbons formed around evolved stars in the a-C:H form and then evolving to become a-C:H/a-C. The exclusion of the a-C$^{\rm 2.5nm}$ shows that the formation of aromatic mantles cannot originate entirely from nano-grain accretion but must be, at least partly, due to the photo-processing of the grain surfaces in agreement with what was presented in \citet[][and references therein]{Jones2013}.

Whether we force the two populations of carbonaceous and silicate grains to have the same shape, or allow the two populations to have different shapes, then the four shapes tested here lead to reasonable fits matching the observations at $\sim \pm 20$\% at almost all wavelengths (see Sect.~\ref{pm20percent} and Fig.~\ref{appendix_deviation}). Not being restrictive on this point seems reasonable, since we can see no physical reason why all the grains in the ISM should have exactly the same shape. Similarly, the alignment function depends on the optical properties and shape of the grains under consideration to obtain the same polarisation levels. We thus find solutions with a universal alignment function for both populations as well as different alignment functions, whether or not the grains are the same shape, owing to the flexibility of our parametric alignment function (see Eq.~\ref{alignment_fraction} and Figs.~\ref{alignment_fraction}, \ref{appendix_fmax}). This is again independent on the choice made for the data normalisation. Furthermore, the vast majority of acceptable models do not require perfect grain alignment, but rather $0.6 \lesssim f_{max} \lesssim 0.9$. This means that whatever the $N_{\rm H}/E(B-V)$ normalisation chosen, our models are able to reproduce the high levels of polarisation found by \citet{Panopoulou2019} and \citet{Angarita2023} since there is still room to increase $f_{max}$ for almost all combinations of compositions and shapes.

Although this constraint was not used in our fitting routine (see Sect.~\ref{obs_polSED}), as already shown by \citet{Siebenmorgen2022}, two polarising grain populations can explain the relative constancy of the polarisation fraction from the far-IR to sub-millimetre (Fig.~\ref{fit_results}). Depending on the mass ratio between the silicates and the carbonaceous grains, as well as on their alignment functions, the models can generate a flat, slightly increasing or slightly decreasing tendency. THEMIS I silicates are also able to reproduce this trend (Fig.~\ref{fit_results_old}). To be able to really understand the differences, or lack of them, in the spectral variations of total and polarised emissions, detailed studies of diffuse regions will have to be carried out rather than working on average SEDs as here.

Finally, Fig.~\ref{3p4um} zooms in on the spectral profile of the hydrocarbon band at 3.4~$\mu$m. In all cases and whatever the chosen normalisation, the polarisation excess above the continuum is below the upper limits measured by \citet[][see Sect.~\ref{obs_polEXT}]{Chiar2006} with $\Delta p_{3.4\mu{\rm m}} \leqslant 0.07$ and 0.08\% (0.09 and 0.13\%) for a normalisation at $p_V/E(B-V) = 9$ and 13\%.mag$^{-1}$, respectively, with the Lenz's ratio (with the Bohlin's ratio). In the case of a normalisation to the Lenz's ratio, about 20 to 30\% of the acceptable models shown in Fig.~\ref{fit_results} are even compatible with no polarisation at 3.4~$\mu$m. This proportion falls to between $\sim 5$ to 16\% for a normalisation to the Bohlin's ratio.

\subsection{Variations in the diffuse ISM}
\label{comparison_PCXI}

\begin{figure*}[!th]
\centerline{\begin{tabular}{cc}
\includegraphics[width=0.45\textwidth]{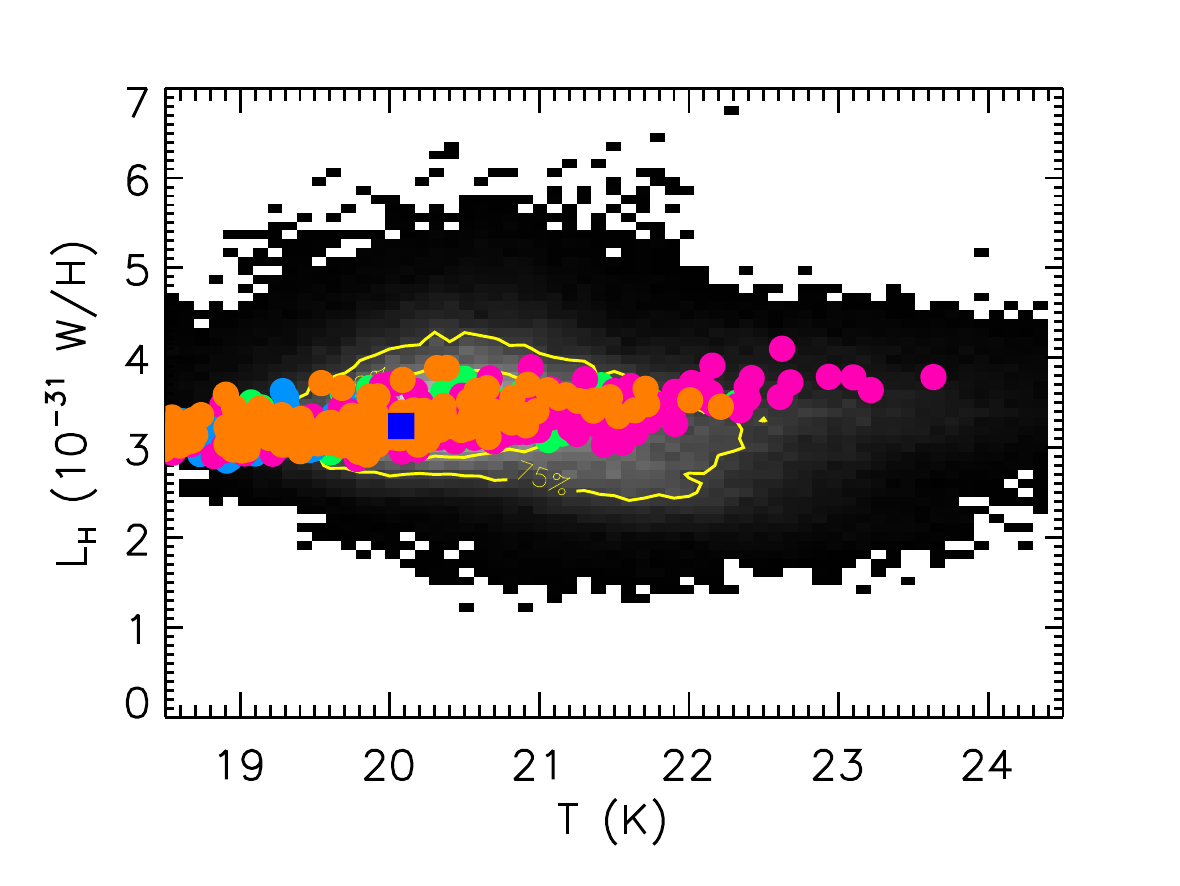} & \includegraphics[width=0.45\textwidth]{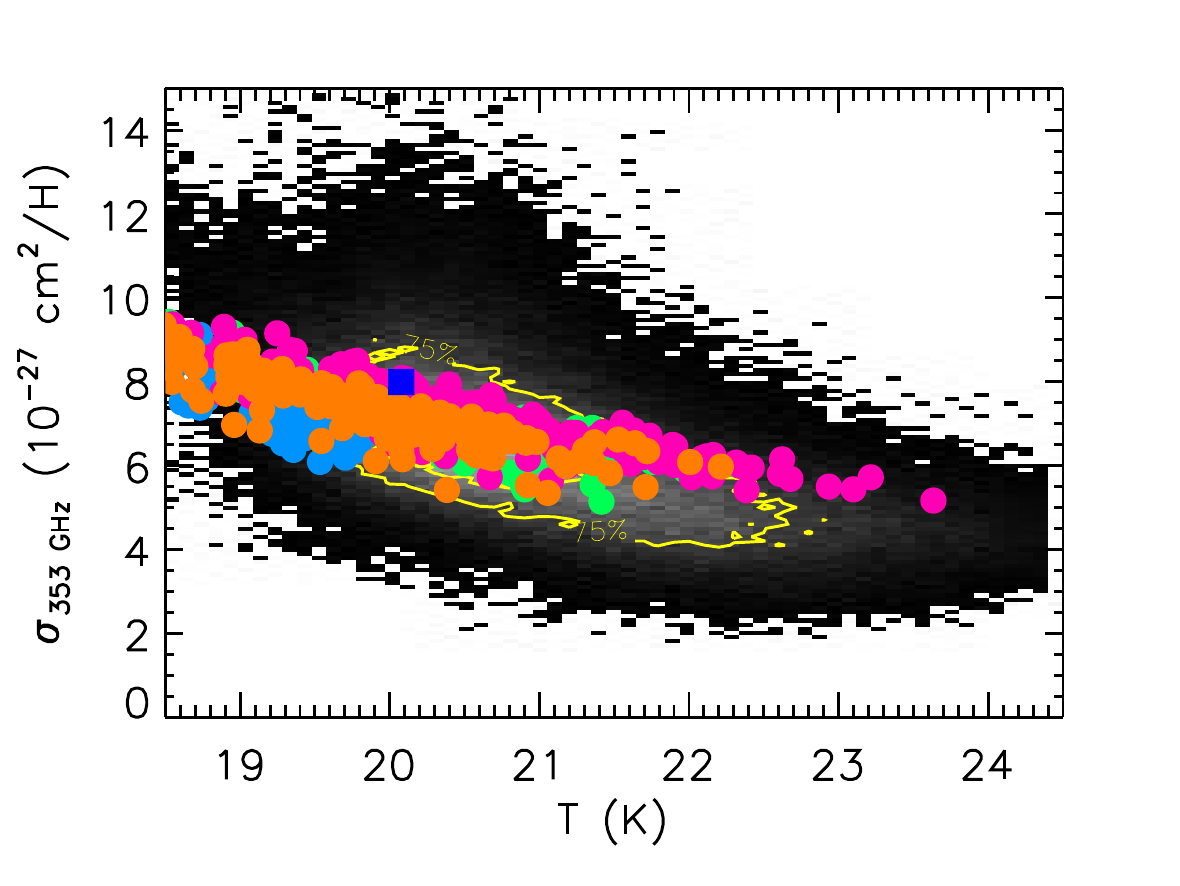} \\ 
\includegraphics[width=0.45\textwidth]{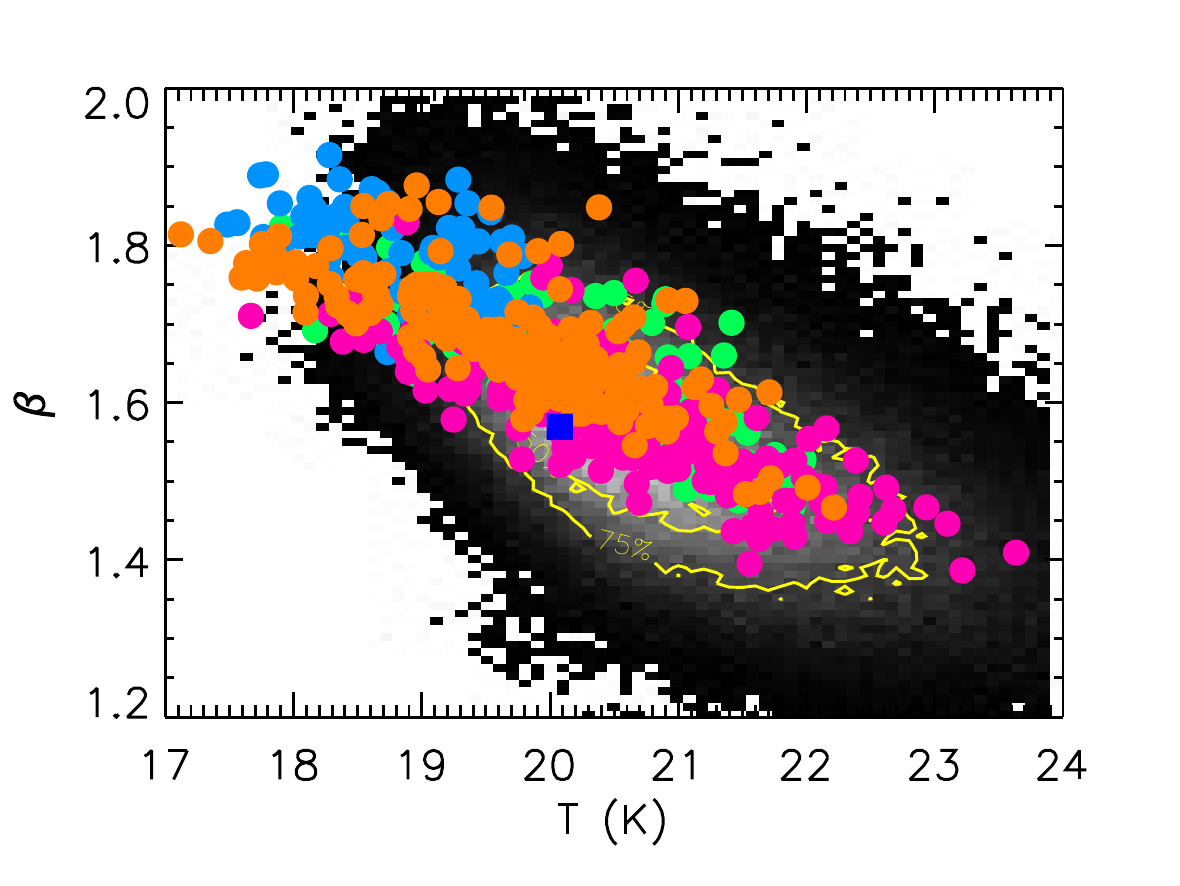} & \includegraphics[width=0.45\textwidth]{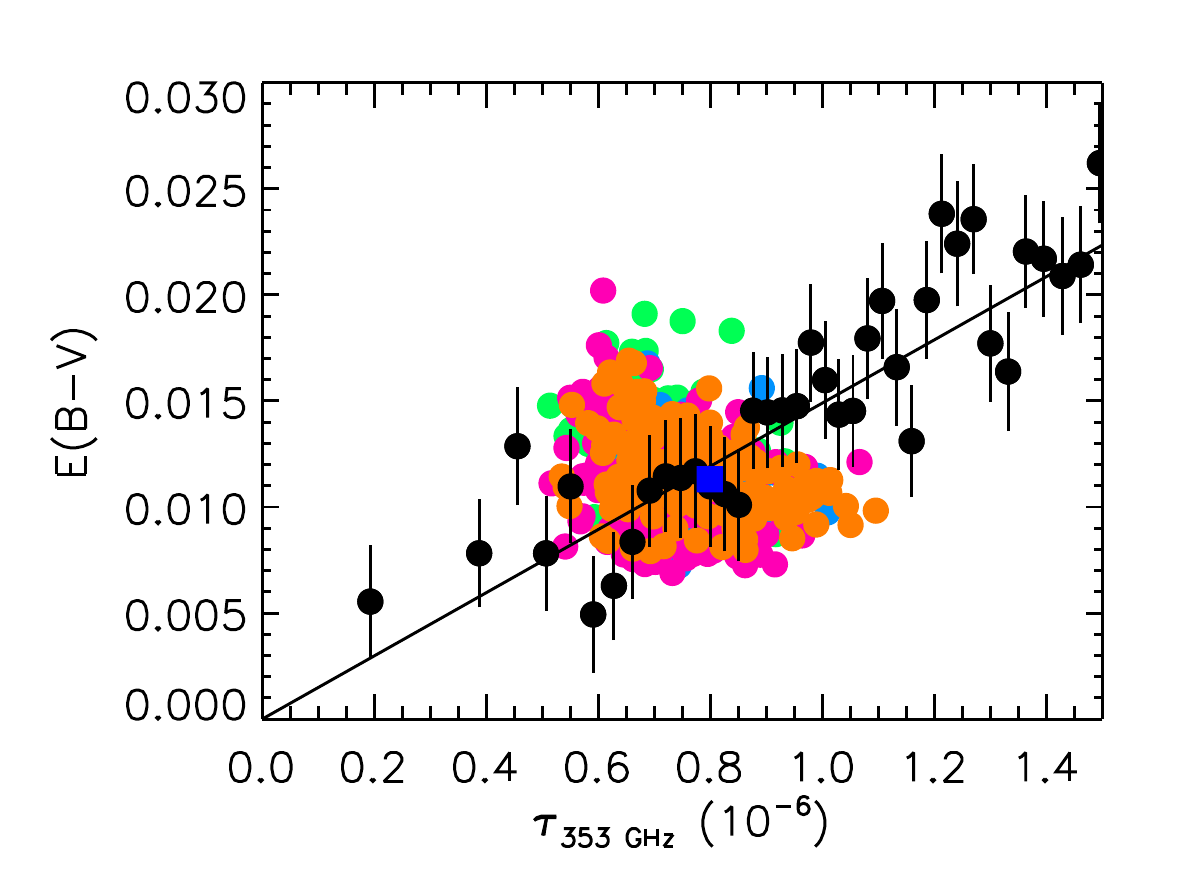}
\end{tabular}}
\caption{Variations in the dust parameters presented by \citet{PlanckXI2014}, based on a pixel-by-pixel modified blackbody $\chi^2$ fit (see Eq.~\ref{MBB} and Sect.~\ref{comparison_PCXI} for details). In the top left and right, and the bottom left figures, the observational results are the density of points maps, on which we overplot yellow contours: the central contour encloses 50\% of the observed pixels and the external contour 75\%. In the bottom right figure, the observational results are the black dots with error bars. In the four figures, the model results are the coloured dots, with the same colour scheme as in Fig.~\ref{fit_results}. Models normalised to the ratio of \citet{Lenz2017} are shown by magenta and orange dots for $p_{\rm V}/E(B-V) = 9$\% and 13\%, respectively. Those normalised to the ratio of \citet{Bohlin1978} are shown by green and blue dots for $p_{\rm V}/E(B-V) = 9$\% and 13\%, respectively. For comparison, the dark blue squares show the parameters derived from the model of \citet{Hensley2023}. {\it Top left:} luminosity vs. temperature. {\it Top right:} opacity at 353~GHz vs. temperature. {\it Bottom left:} spectral index vs. temperature. {\it Bottom right:} E(B-V) vs. optical depth at 353~GHz, models shown for $N_H = 10^{20}$~H/cm$^2$.}
\label{fit_Planck} 
\end{figure*}

\begin{figure*}[!th]
\centerline{\begin{tabular}{*2{>{\centering\arraybackslash}p{.5\textwidth-2\tabcolsep}}}
\includegraphics[width=0.45\textwidth]{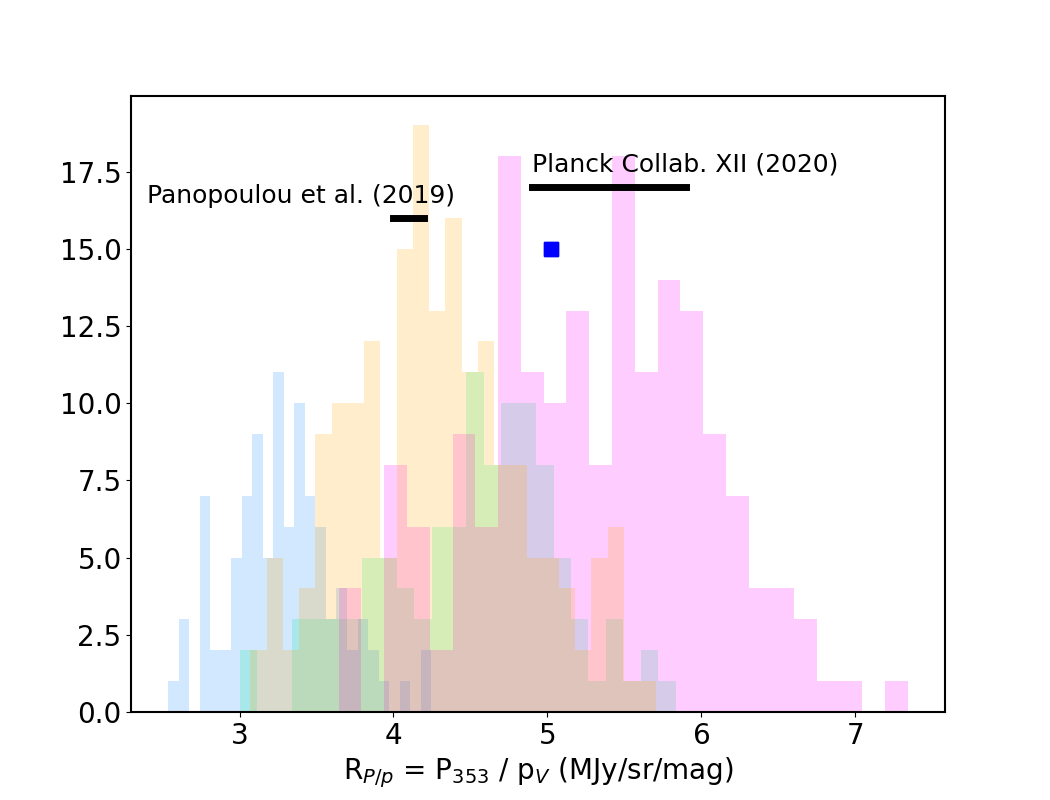} & \includegraphics[width=0.45\textwidth]{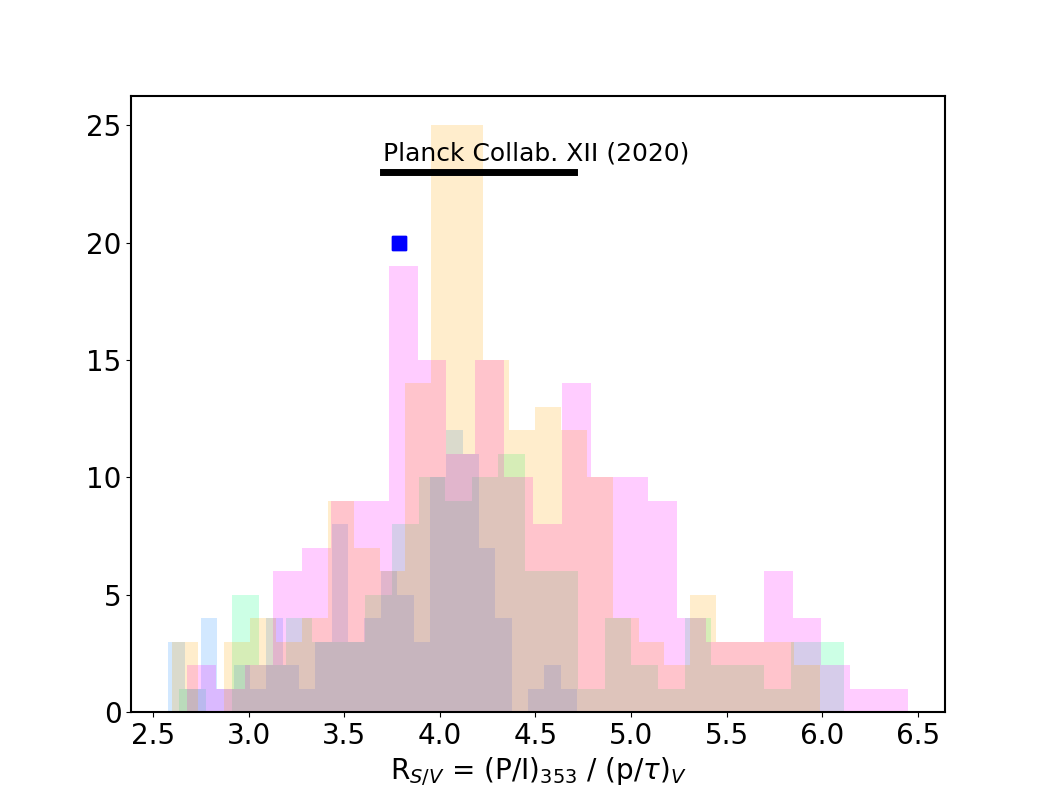} \\ 
\includegraphics[width=0.45\textwidth]{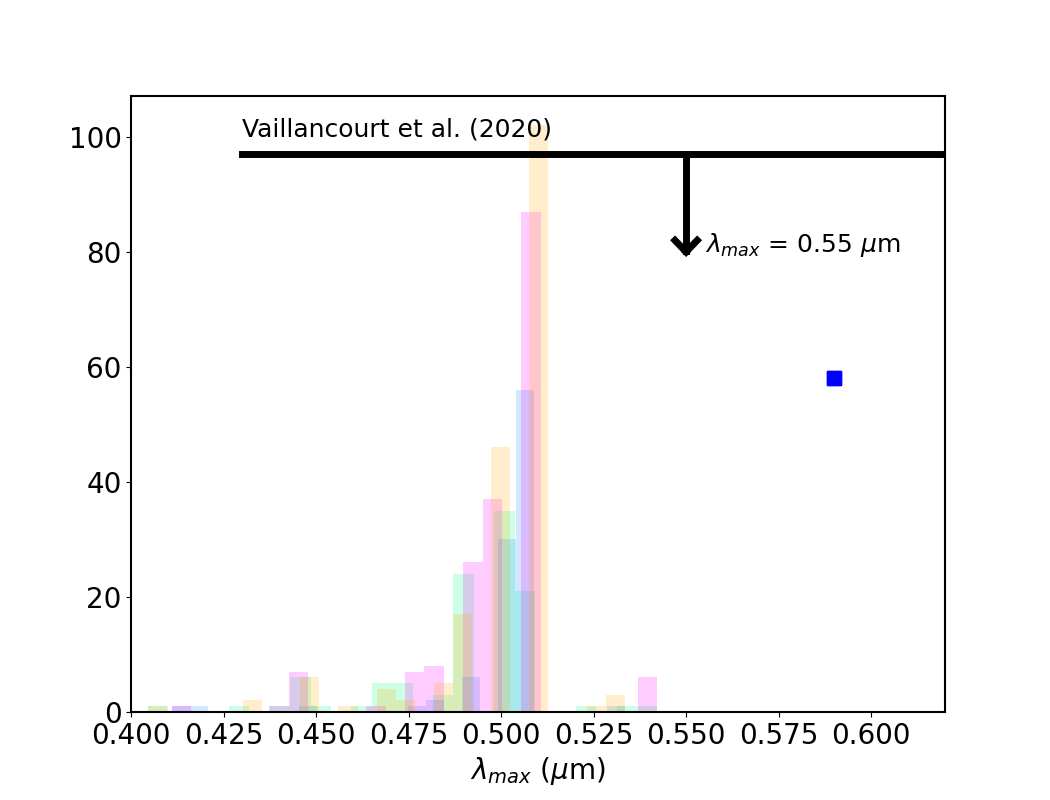} & \includegraphics[width=0.45\textwidth]{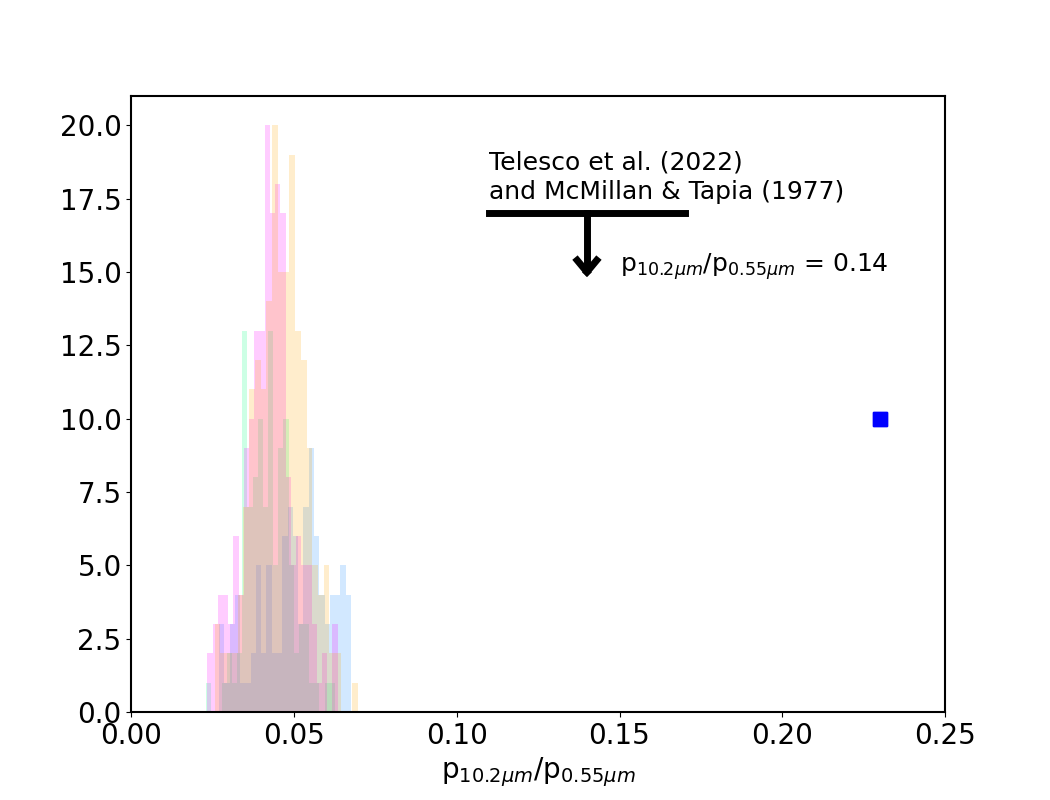} \\
\adjincludegraphics[width=0.45\textwidth,valign=T]{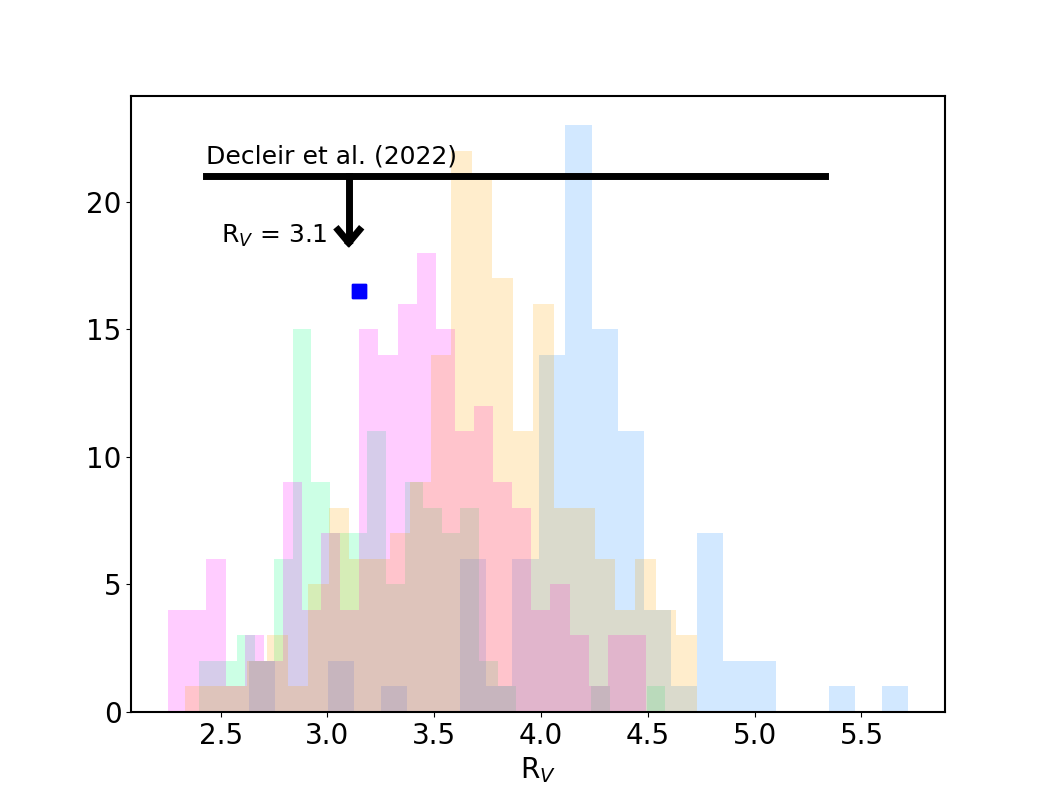} & \caption{Histograms of parameters derived from our best fits represented with the same colour scheme as in Fig.~\ref{fit_results}. For comparison, the parameters derived from the model of \citet{Hensley2023} are shown by the dark blue squares. The two figures in the top row show two characteristic ratios of the polarised observations made with the Planck satellite: $R_{P/p}$ and $R_{S/V}$. The measurements, including their uncertainties, made by \citet{Panopoulou2019} and \citet{PlanckXII2020}, are represented by the black horizontal lines. The two figures in the middle row show the dispersion of $\lambda_{max}$ (left) and of the polarised extinction at 10~$\mu$m normalised to the optical (right) of our best fits. The measurements, including their uncertainties, made by \citet{Vaillancourt2020} and \citet{Telesco2022} combined to \citet{McMillan1977} are shown by the black horizontal lines. The bottom row figure shows the histogram of the extinction parameter $R_V$. The measurements made by \citet{Decleir2022}, including their uncertainties, are shown by the black horizontal line.}
\end{tabular}}
\label{figure_ratios} 
\end{figure*}

The Planck data revealed a variability in the properties of dust in the diffuse medium \citep[$10^{19} \leqslant N_H \leqslant 2.5 \times 10^{20}$~H/cm$^2$,][]{PlanckXI2014}. Their results were based on a pixel-by-pixel modified blackbody $\chi^2$-fit between 353 and 3\,000~GHz:
\begin{equation}
\label{MBB}
I_{\nu} = \tau_{\nu_0} B_{\nu}(T) (\nu/\nu_0)^{\beta},
\end{equation}
where $\tau_{\nu_0}$ is the optical depth at $\nu_0 = 353$~GHz, $T$ is the dust colour temperature, and $\beta$ is the spectral index assumed to be constant from 100~$\mu$m to 353~GHz. Besides, \citet{PlanckXI2014} computed the dust luminosity $L_H = (\int_{\nu} I_{\nu} d\nu)/N_H$ and opacity $\sigma_{\nu_0} = \tau_{\nu_0}/N_H$. Using the Sloan Digital Sky Survey data \citep[SDSS,][]{Schneider2010}, they also derived the dust optical extinction $E(B-V)$.
Their results showed that (i) the dust luminosity does not vary with temperature, that (ii) the opacity and (iii) the spectral index are anti-correlated with the temperature, and (iv) presented $E(B-V)$ vs. $\tau_{353{\rm GHz}}$ for high-latitude diffuse lines of sight with $10^{19} \leqslant N_{\rm H} \leqslant 2.5 \times 10^{20}$~H/cm$^2$. None of these results can be explained for constant grain properties. In \citet{Ysard2015}, we explored how variations in the grain structures or abundances and in the radiation field intensity could explain these variations. Here we explore how the dispersion of models can explain the observations along the lines of sight with the highest polarisation fractions (Fig.~\ref{fit_results}) compared to the variations described in \citet{PlanckXI2014}.

To be able to compare our models with the dust parameters derived in \citet{PlanckXI2014}, we perform similar modified blackbody $\chi^2$-fit for all the models presented in Fig.~\ref{fit_results}. Figure~\ref{fit_Planck} shows that our collection of models can account for most of the parameters observed and that this depends neither on the $N_{\rm H}/E(B-V)$ normalisation choice nor on the maximum level of polarised extinction. If only the (in)homogeneity of the carbon mantles were taken into account, then the dispersion of the measurements could not be reproduced. The diversity of the chemical composition of silicates must be taken into account (see Fig.~\ref{fit_Planck_old} for instance, where only THEMIS I silicates are considered). Performing the fitting procedure with various polarisation fractions or with slightly different SEDs, would also help to fully populate the plots in Fig.~\ref{fit_Planck}. 

Figure~12 presents the comparison of our best fits with the $R_{P/p}$ and $R_{S/V}$ ratios, along with the distributions of the $\lambda_{max}$, $p_{10.2\mu{\rm m}}/p_{0.55\mu{\rm m}}$, and $R_V$ parameters. Our models calibrated on the Lenz's ratio and $p_V/E(B-V) = 9$\% are compatible with the values of $R_{P/p}$ measured by \citet{PlanckXII2020} whereas those calibrated at 13\% are rather compatible with the values measured by \citet{Panopoulou2019}. Those calibrated on the Bohlin's ratio and $p_V/E(B-V) = 9$\% are between the two, while those taking Bohlin's and 13\% are weaker overall. This illustrates once again that the variability of the silicate chemical compositions can explain the observed variations. Similarly, all our models are consistent with the measurements of $R_{S/V}$. For comparison, Fig.~12 also shows the ratios given by the model of \citet{Hensley2023} which are within the uncertainties given by \citet{PlanckXII2020}, both on the weak ratio side. Whatever their normalisation, most of our models have $\lambda_{max}$ values ranging between 0.50 and 0.51~$\mu$m. This is smaller than the usual 0.55~$\mu$m used to build the Serkowski curve but fully consistent with the measures made by \citet{Vaillancourt2020} towards low extinction lines of sight. For comparison, \citet{Hensley2023} give a higher value of $\lambda_{max} = 0.59~\mu$m, also compatible with the observations.
The polarised extinction 10~$\mu$m-silicate feature normalised to the optical of our models range from $0.03 \leqslant p_{10.2\mu{\rm m}}/p_{0.55\mu{\rm m}} \leqslant 0.07$, with most of them lying around 0.05. This is a factor of about 2 to 4.7 smaller than the value inferred from the observations towards Cyg OB2-12 of \citet{Telesco2022} and \citet{McMillan1977}. In contrast to our predictions, \citet{Hensley2023} predict $p_{10.2\mu{\rm m}}/p_{0.55\mu{\rm m}} = 0.23$, about 1.6 higher than the observations. This is probably linked to the fact that their model overpredicts the total extinction at this same wavelength (see their Fig.~5) whereas our predictions are quite lower in this same wavelength range. It is however difficult to conclude anything from that considering how much the mid-IR total extinction curve is different when moving from the Galactic Plane or Centre to higher latitudes \citep[see for instance][]{Gordon2021}. Observations of this polarised extinction feature at higher latitude would be beneficial for constraining dust models. Finally, apart from the models normalised to the Bohlin's ratio with $p_V/E(B-V) = 9$\%, most of our models have $R_V$ values that are larger than the average 3.1 but all remain consistent with the dispersion measured by \citet{Decleir2022} for lines of sight with $0.8 \lesssim A(V) \lesssim 3$.

\subsection{Final model}
\label{final_model}

The THEMIS model is public, accessible here: \url{https://www.ias.u-psud.fr/themis/}. It is fully integrated into the DustEM numerical code (\url{https://www.ias.u-psud.fr/DUSTEM/}), which is used to produce all the figures in Figs.~\ref{fit_results}, \ref{size_distribution}, and \ref{alignment}. All the optical properties calculated for the chemical compositions and mantle description presented in Sect.~\ref{grain_composition} can be accessed at these two addresses and used in the DustEM and DustEMWrap codes.

We also provide the best fits (thick lines in Figs.~\ref{fit_results}, \ref{size_distribution}, and \ref{alignment}) for the four normalisations considered here and which are obtained for:
\begin{itemize}
\item aSil-4/a-C$^{\rm 2.5nm}$ (oblates, $e = 1.3$) and a-CH/a-C$^{\rm 4nm}$ (prolates, $e = 2$) in the case of a normalisation to the Bohlin's ratio and for $p_V/E(B-V) = 9$\% ;
\item aSil-1/a-C$^{\rm 2.5nm}$ (prolates, $e = 2$) and a-CH/a-C$^{\rm 5nm}$ (prolates, $e = 2$) in the case of a normalisation to the Bohlin's ratio and for $p_V/E(B-V) = 13$\% ;
\item aSil4-/a-C$^{\rm 2.5nm}$ (prolates, $e = 2$) and a-CH/a-C$^{\rm 6nm}$ (prolates, $e = 1.3$) in the case of a normalisation to the Lenz's ratio and for $p_V/E(B-V) = 9$\% ;
\item aSil-2/a-C$^{\rm 2.5nm}$ (prolates, $e = 2$) and a-CH/a-C$^{\rm 5nm}$ (prolates, $e = 2$) in the case of a normalisation to the Lenz's ratio and for $p_V/E(B-V) = 13$\%.
\end{itemize}
From now on, this last model will be considered as the THEMIS 2.0 reference model (see Tab.~\ref{best_model} for the main model parameters). However, it should be borne in mind that all the models shown in Figs.~\ref{fit_results}, \ref{size_distribution}, and \ref{alignment} are compatible with all the observations available within $\pm 20$\%. When making a comparison with a particular dataset, it may therefore be useful to try to see which normalisation is the most relevant, and not to confine oneself to using a single combination of compositions, size distribution or alignment function. The authors of this study will be happy to discuss the relevance of the choice of properties should the need arise.

\begin{table}[t]
\centering
\caption{Main parameters defining the best dust model normalised to the Lenz's ratio and $p_V/E(B-V) = 13$\%.mag$^{-1}$ as in \citet{Siebenmorgen2022} and \citet{Hensley2023}. For each dust population, we indicate the grain shape and elongation ($e$), the dust-to-gas mass ratio ($M_{dust}/M_{\rm H}$), the minimum and maximum grain size ($a_{min}$, $a_{max}$), the parameters defining the power-law size distribution for small spherical a-C nanograins ($\propto a^{\alpha}$ and with an exponential cutoff of the form $\exp-[(a-a_t)/a_C]^{\gamma}$ for $a > a_t$), the parameters defining the log-normal size distribution of the bigger spheroidal grains (with $a_0$ the centre radius and $\sigma$ the width of the distribution) and the three parameters defining the grain alignment function ($a_{thresh}$, $p_{stiff}$, and $f_{max}$, see Eq.~\ref{alignment_fraction}). The total and polarised SEDs shown by the orange thick lines in Fig.~\ref{fit_results} are obtained for $G_0 = 1.4$.} 
\label{best_model}
\begin{tabular}{lccc}
\hline
Composition               & a-C                   & a-C:H/a-C$^{\rm 5nm}$     & aSil-2/a-C$^{\rm 2.5nm}$  \\
core/mantle               &                       &                           &                           \\
\hline
Shape                     & sphere                & prolate                   & prolate                   \\
Elongation                & ...                   & $e = 2$                   & $e = 2$                   \\
$M_{dust}/M_{\rm H}$      & $1.32 \times 10^{-3}$ & $8.00 \times 10^{-4}$     & $3.27 \times 10^{-3}$     \\
$a_{min}$ (nm)            & 0.4                   & 45                        & 11                        \\
$a_{max}$ (nm)            & 25                    & 700                       & 374                       \\
$\alpha$                  & -5                    & ...                       & ...                       \\
$a_C$, $a_t$ (nm)         & 50, 10                & ...                       & ...                       \\ 
$\gamma$                  & 1                     & ...                       & ...                       \\
$a_0$ (nm)                & ...                   & 6.2                       & 0.92                      \\
$\sigma$                  & ...                   & 1.32                      & 1.22                      \\
$a_{thresh}$ (nm)         & ...                   & 63                        & 63                        \\
$p_{stiff}$               & ...                   & 0.22                      & 0.22                      \\
$f_{max}$                 & ...                   & 0.72                      & 0.72                      \\
\hline
\end{tabular}
\end{table}

\section{Conclusion}
\label{conclusion}

THEMIS is a dust model that includes three compact dust populations: a population of (sub-)nanometric aromatic-rich amorphous carbon a-C grains, a population of larger core/mantle a-CH/a-C carbonaceous grains, and a population of core/mantle a-Sil/aC silicates. We have extended the THEMIS grain model, first described in \citet{Jones2013}, in three different ways. First, the most comprehensive laboratory data on silicates to date were used to calculate the optical properties of this grain population. Second, the homogeneity of the a-C mantles at the surface of the a-C:H/a-C grains was studied. Third, spheroidal grains were considered instead of spheres in order to account for polarised emission and extinction.

This new version of THEMIS, for the first time entirely based on laboratory measurements, reproduces the dust emission and extinction, both total and polarised, observed in the diffuse ISM at a high Galactic latitude, while respecting the constraints on the abundances of the various elements making up the grains. An interesting point is that the model has not been pushed to its limits yet since it does not require the perfect alignment of all grains to explain the observations ($f_{max} < 1$ in Eq.~\ref{alignment_fraction} and Fig.~\ref{alignment}) and it therefore has the potential to accommodate for the highest polarisation levels inferred from extinction measurements \citep[e.g.][]{Panopoulou2019, Angarita2023}. In addition, the dispersion of the optical properties of the different silicates measured in the laboratory, combined with the more or less homogeneous description of the a-CH/a-C mantles, explain the variations in both the total and polarised emission and extinction observed in the diffuse\ ISM. 

This study illustrates that a single, invariant model calibrated on one single set of observations is obsolete for explaining contemporary observations. Variations in shape, size, chemical composition, and alignment efficiency (Figs.~\ref{fit_Planck} and 12), eventually associated with variations in grain abundance and structure \citep{Ysard2015}, are able to fully reproduce the variations observed in the diffuse medium. Even if challenging, this is particularly relevant for future CMB missions that will aim to perform precise measurements of the CMB spectral distortions and polarisation. Detailed studies of particular lines of sight will also allow us to gain stronger knowledge about dust chemical composition and structure. In a new era in which observations are achieving unprecedented precision in terms of angular and spatial resolutions and ever greater sensitivity, we are proposing a completely flexible dust model based entirely on laboratory measurements that has the potential to make major advances in understanding the exact nature of interstellar grains and how they evolve as a function of their radiative and dynamic environment.

\begin{acknowledgements}
We would like to thank our referee, R. Siebenmorgen, for his insightful comments, which helped us to improve the content of the paper. This work was supported by the Programme National PCMI of CNRS/INSU with INC/INP co-funded by CEA and CNES. Finally, we thank K. Misselt for his careful proof reading.
\end{acknowledgements}

\bibliographystyle{aa}
\bibliography{biblio}


\appendix

\section{Contributions of the carbonaceous and silicate grains to the polarisation and extinction}
\label{appendix_contributions}

\begin{figure}[!t]
\centerline{\begin{tabular}{c}
\includegraphics[width=0.45\textwidth]{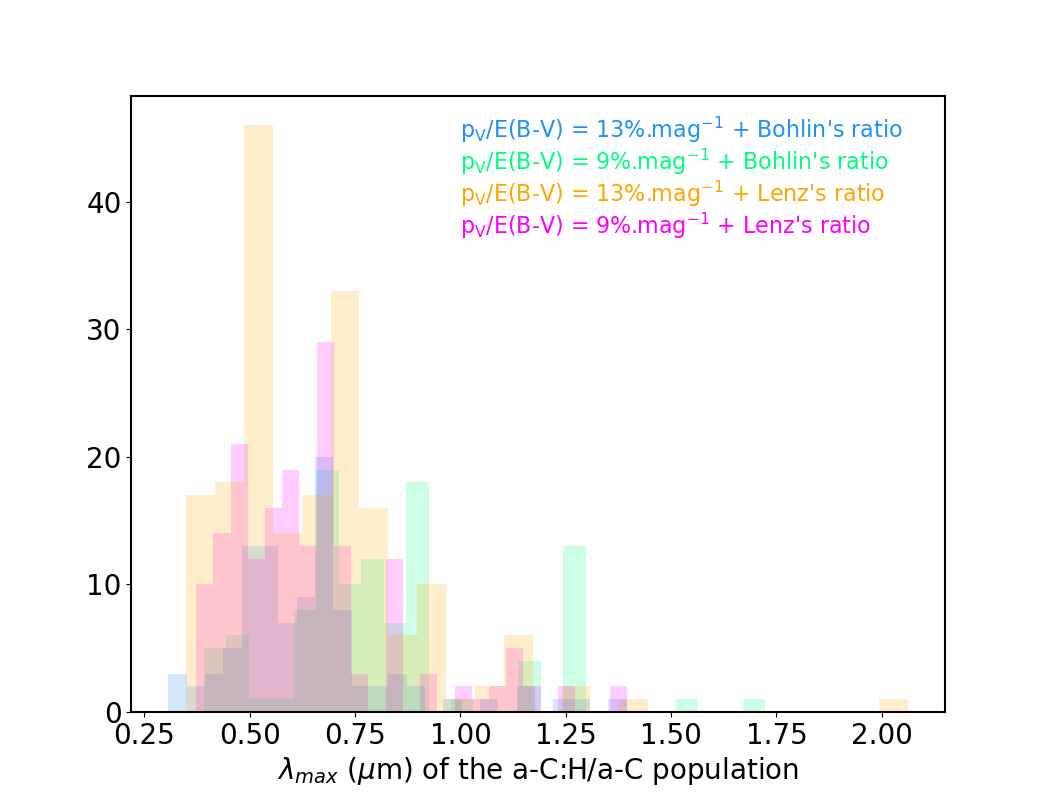} \\
\includegraphics[width=0.45\textwidth]{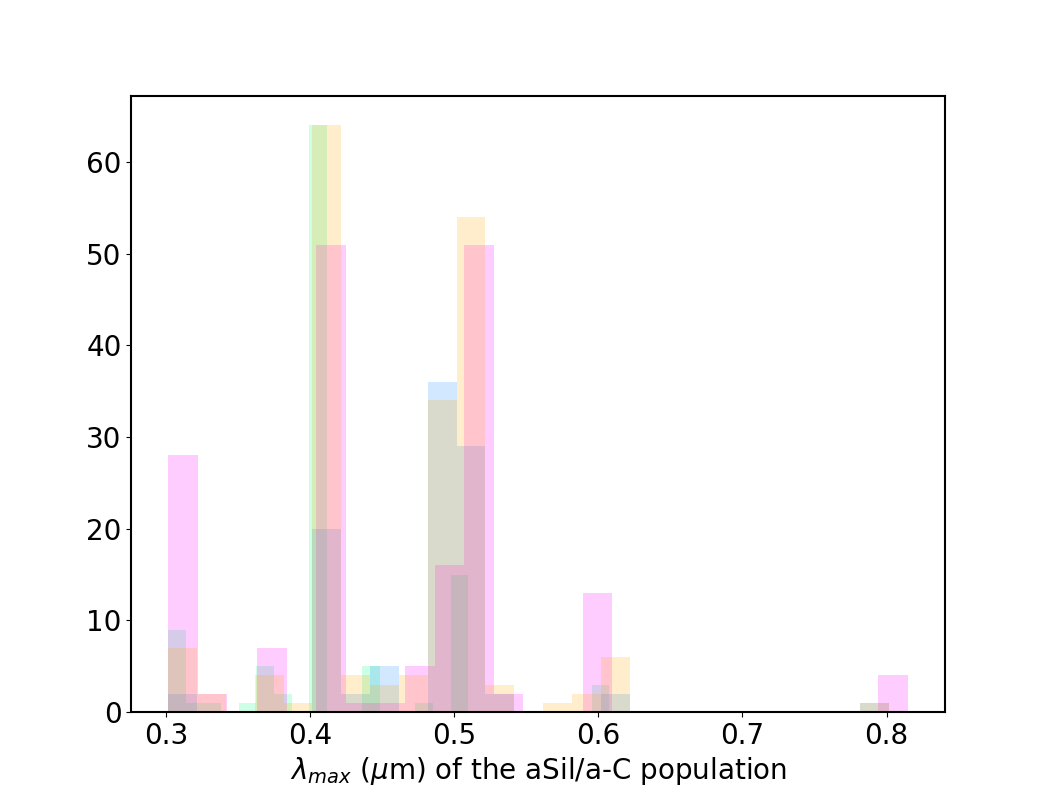} \\
\includegraphics[width=0.45\textwidth]{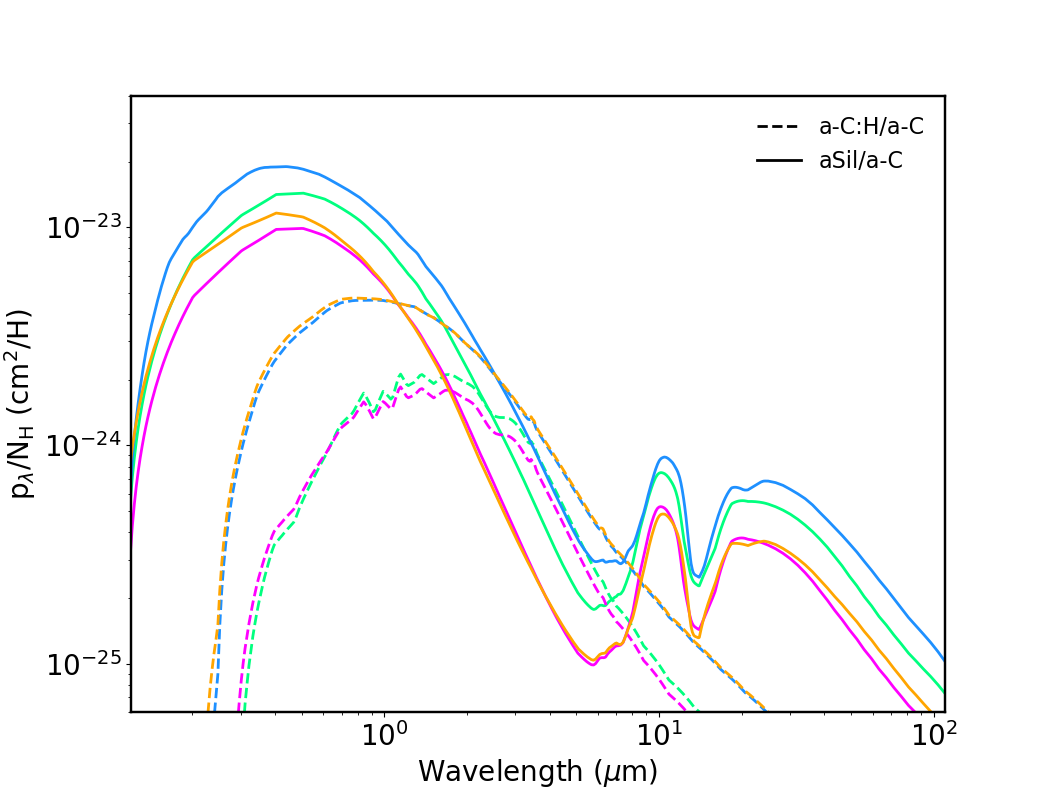} 
\end{tabular}}
\caption{Same colour scheme as in Fig.~\ref{fit_results}. {\it Top:} Histogram of the wavelengths at the maximum of the polarised extinction curve of carbonaceous grains. {\it Centre:} Same for the silicate grains. {\it Bottom:} Polarised extinction of the carbonaceous (dashed lines) and silicate grains (solid lines) for the best fits presented in Fig.~\ref{fit_results}.}
\label{appendix_pol_ext} 
\end{figure}

Figure~\ref{appendix_pol_ext} shows the $\lambda_{max}$ of each carbonaceous and silicate grain populations for the models in Fig.~\ref{fit_results}. The carbonaceous grains make a non-negligible contribution in the near- to mid-IR, before the rise of the 10~$\mu$m-silicate feature. This is also seen in total extinction (Fig.~\ref{appendix_extinction}).

\begin{figure*}[!t]
\centerline{\begin{tabular}{cc}
\includegraphics[width=0.45\textwidth]{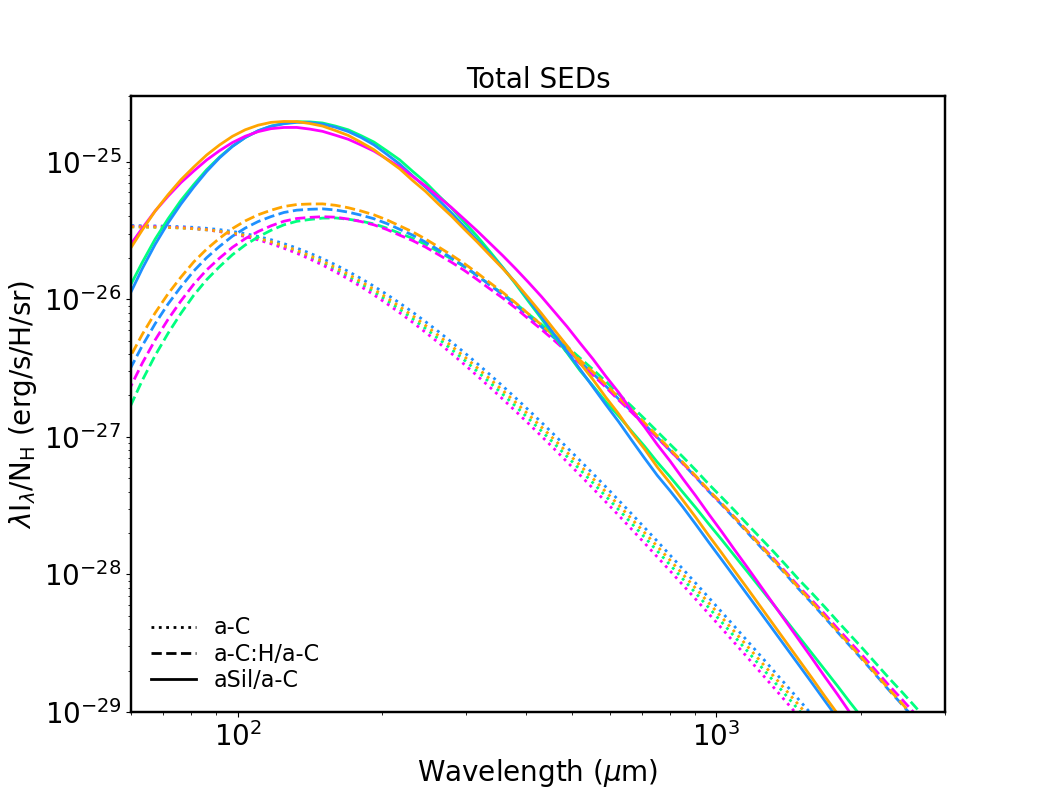} & \includegraphics[width=0.45\textwidth]{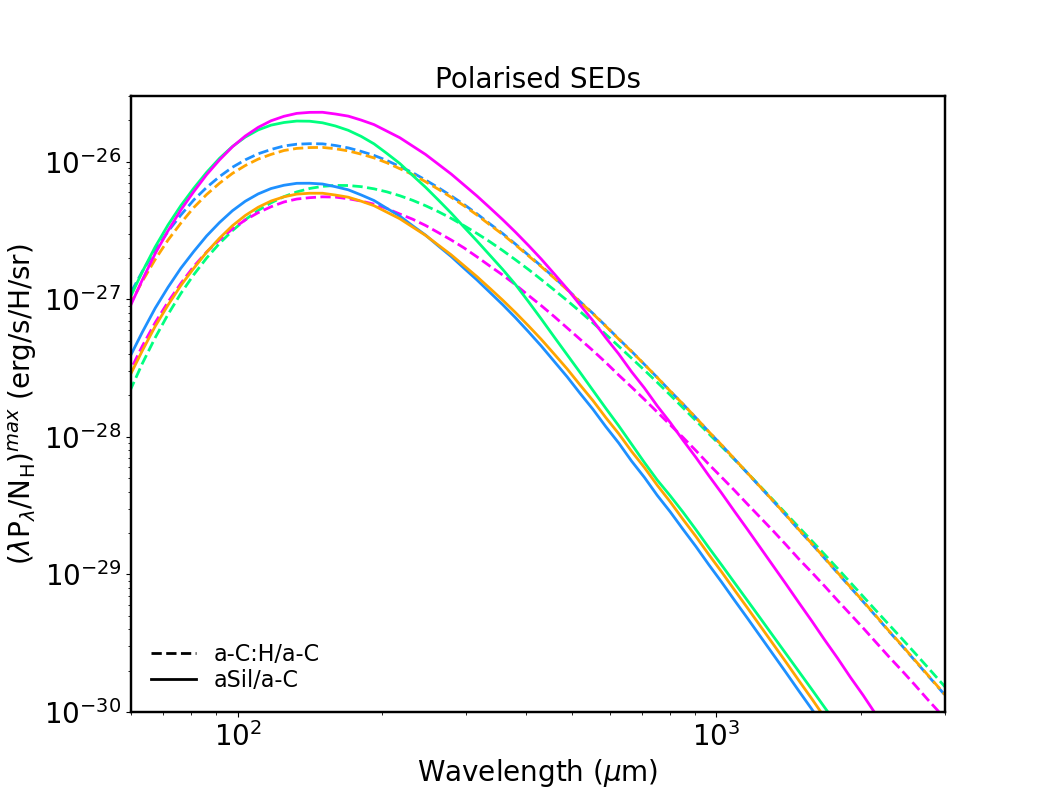}
\end{tabular}}
\caption{Same colour scheme as in Fig.~\ref{fit_results}. Contributions to the total and polarised SEDs of the nanometric hydrocarbons a-C shown by the dotted lines, of the a-C:H/a-C grains by the dashed lines and of the silicates by the solid lines. {\it Left:} Total SEDs of the best-fit models presented in Fig.~\ref{fit_results}. {\it Right:} Corresponding polarised SEDs.}
\label{appendix_seds} 
\end{figure*}

\begin{figure}[!h]
\centerline{\includegraphics[width=0.45\textwidth]{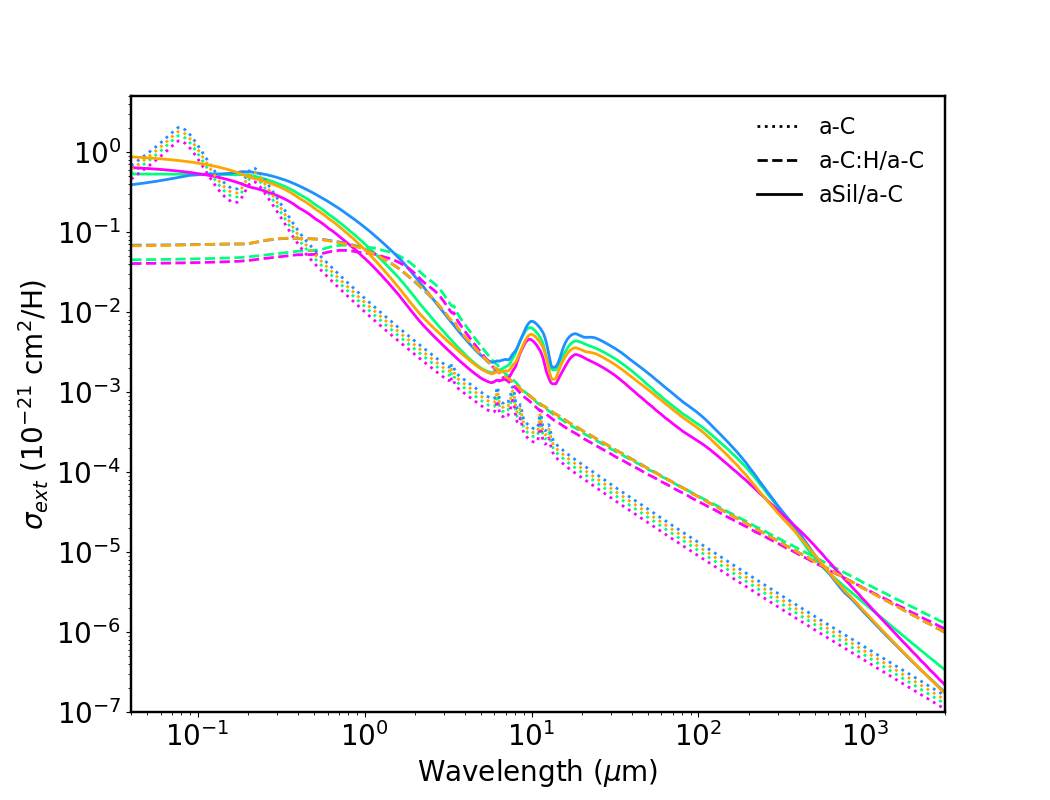}}
\caption{Same colour scheme as in Fig.~\ref{fit_results}. Contributions to the total extinction of the nanometric hydrocarbons a-C shown by the dotted lines, of the a-C:H/a-C grains by the dashed lines and of the silicates by the solid lines.}
\label{appendix_extinction} 
\end{figure}

Figure~\ref{appendix_seds} presents the total and polarised SEDs per population of the best-fit models shown in Fig.~\ref{fit_results}. In the case of the total SED, the silicates always dominate in the far-IR, but the carbonaceous grains dominate in the millimetre range. The wavelength at which the transition occurs between the two populations depends on the exact composition of the grains, since this determines their spectral index. The results for the polarised SED are more mixed. In the four examples presented, corresponding to the four best models shown in Fig.~\ref{fit_results}, we observe that (i) the two normalised to the Bohlin's ratio behave as for the total SED with a transition between silicates and carbonaceous grains from the far-IR to the millimetre range and that (ii) the two normalised to the Lenz's ratio are dominated by the carbonaceous grains for the entire wavelength range. This is just an illustration of what the collection of acceptable models presented in Fig.~\ref{fit_results} can produce. Indeed, for half of the models, the silicates dominate the polarised SED for both normalisations. The relative contributions of each population thus depend on two parameters: the choice made for the normalisation of the data and the exact composition of the grains.

\begin{figure}[!h]
\centerline{\includegraphics[width=0.45\textwidth]{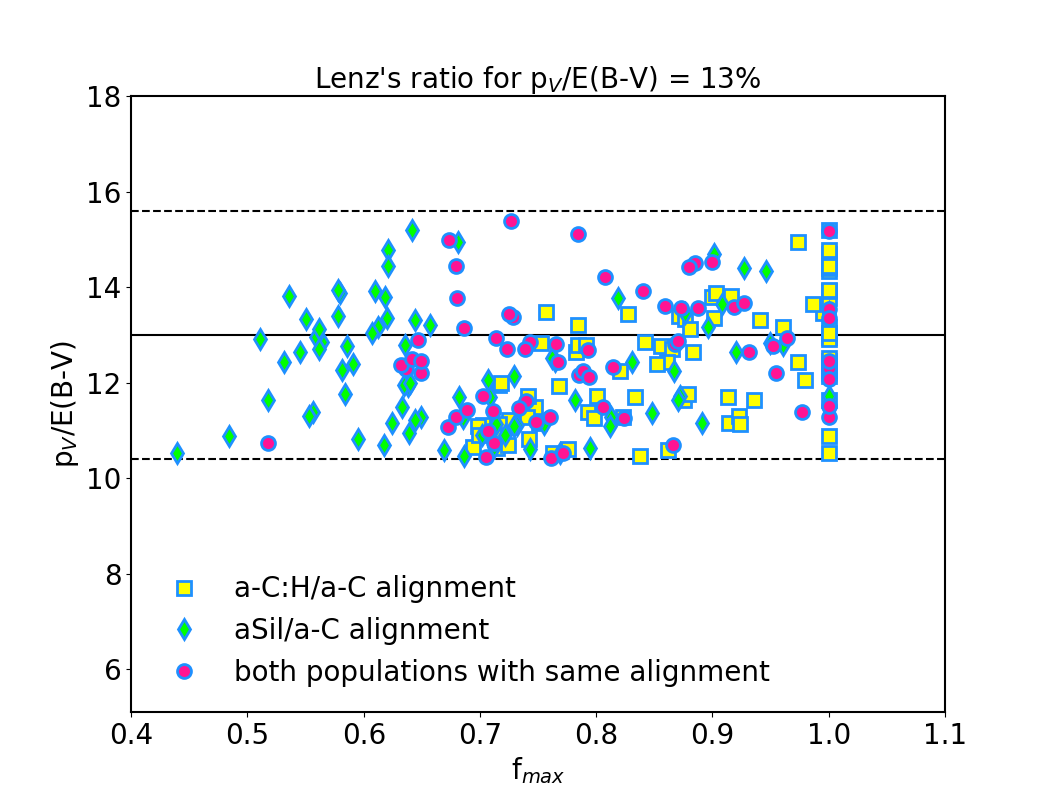}}
\caption{Maximum grain alignment fraction $f_{max}$ in the case of a normalisation to the Lenz's ratio for $p_V/E(B-V) = 13$\%.mag$^{-1}$ (black line and $\pm 20$\% uncertainty as dashed lines). Model results are shown for the case of a universal alignment function for both grain populations (pink circles) and for different alignment functions for carbonaceous and silicate grains shown as yellow squares and green diamonds, respectively.}
\label{appendix_fmax} 
\end{figure}

As an example, Fig.~\ref{appendix_fmax} shows the maximum grain alignment fraction $f_{max}$ in the case of a normalisation to the Lenz's ratio and $p_V / E(B-V) = 13$\%.mag$^{-1}$. There is no correlation between the two parameters, regardless of whether the two populations are aligned in the same way or whether the carbonaceous grains and silicates obey two different alignment functions. Indeed, in addition to $f_{max}$, $p_V/E(B-V)$ depends on the grain chemical composition as well as their size distribution. This result was already visible in \citet[][see their Table 3]{Guillet2018}.

\section{Deviation of the model from the total and polarised observed SEDs}
\label{pm20percent}

\begin{figure*}[!h]
\centerline{\begin{tabular}{cc}
\includegraphics[width=0.45\textwidth]{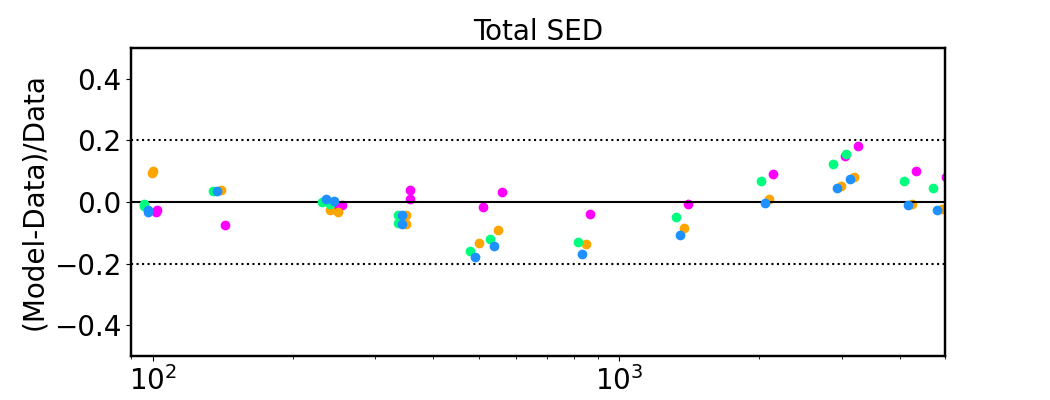} & \includegraphics[width=0.45\textwidth]{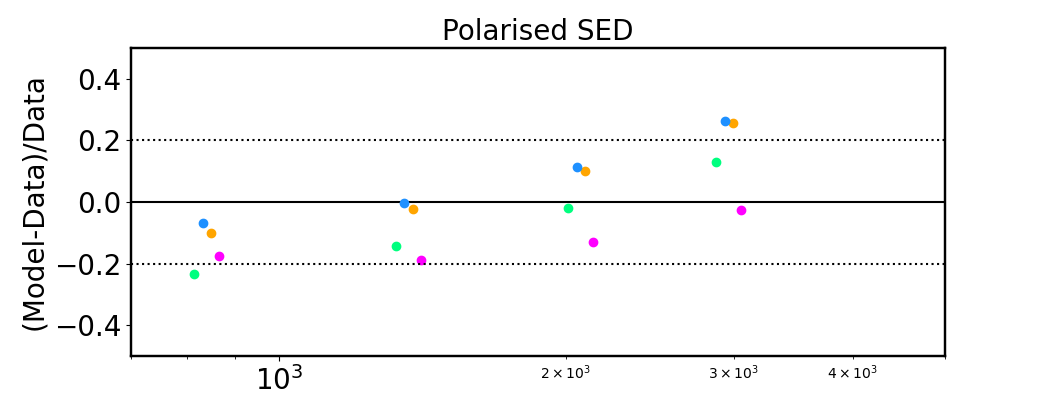}
\end{tabular}}
\caption{Deviation of the four best models presented in Fig.~\ref{fit_results} for the total (left) and polarised (right) SEDs. Same colour scheme as in Fig.~\ref{fit_results}. For clarity, the wavelengths are shifted by a few percents for the points not to overlap too much.}
\label{appendix_deviation} 
\end{figure*}

Figure~\ref{appendix_deviation} shows the deviation of the four best models presented in Fig.~\ref{fit_results} from the total and polarised SEDs. This shows an agreement within $\pm 20$\% at almost all wavelengths which is the precision requirement we adopted following \citet{Hensley2023}.

\section{Fits with THEMIS I silicates}
\label{appendix_THEMIS_I}

\begin{figure}[!t]
\centerline{\includegraphics[width=0.45\textwidth]{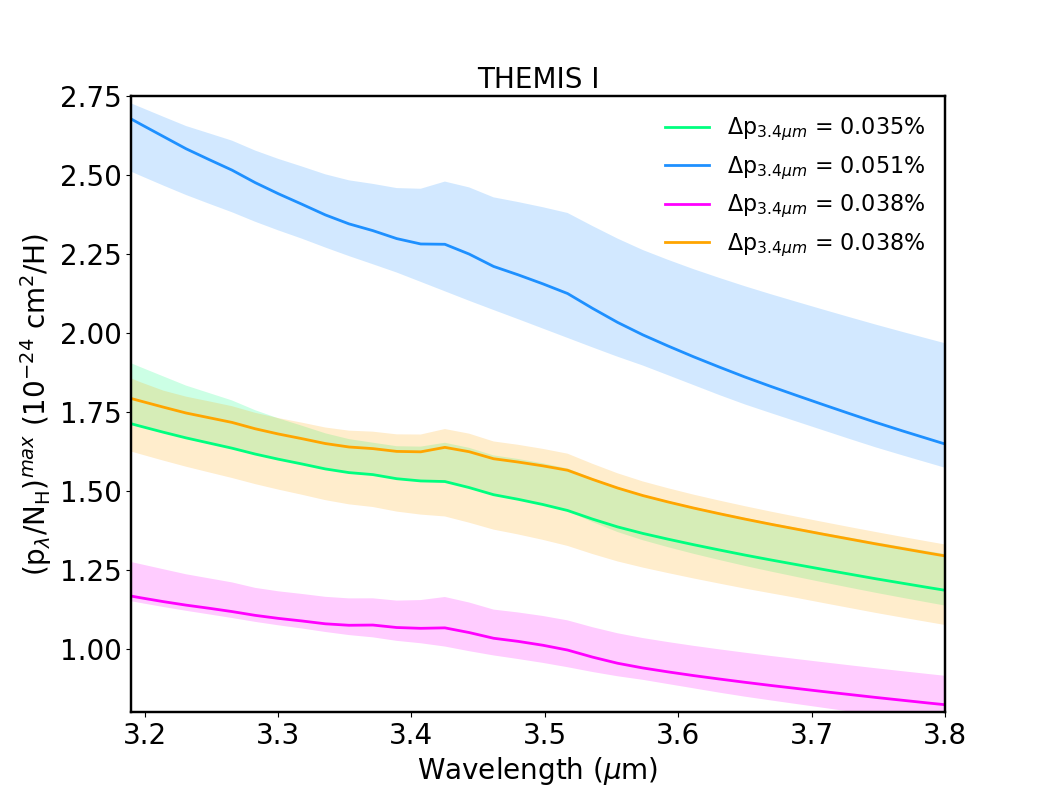}}
\caption{Same as in Fig.~\ref{3p4um} but for fits performed with THEMIS I silicates (see Table~\ref{table_composition}). Same colour scheme as in Fig.~\ref{fit_results_old}.}
\label{3p4um_old} 
\end{figure}

\begin{figure}[!t]
\centerline{\includegraphics[width=0.45\textwidth]{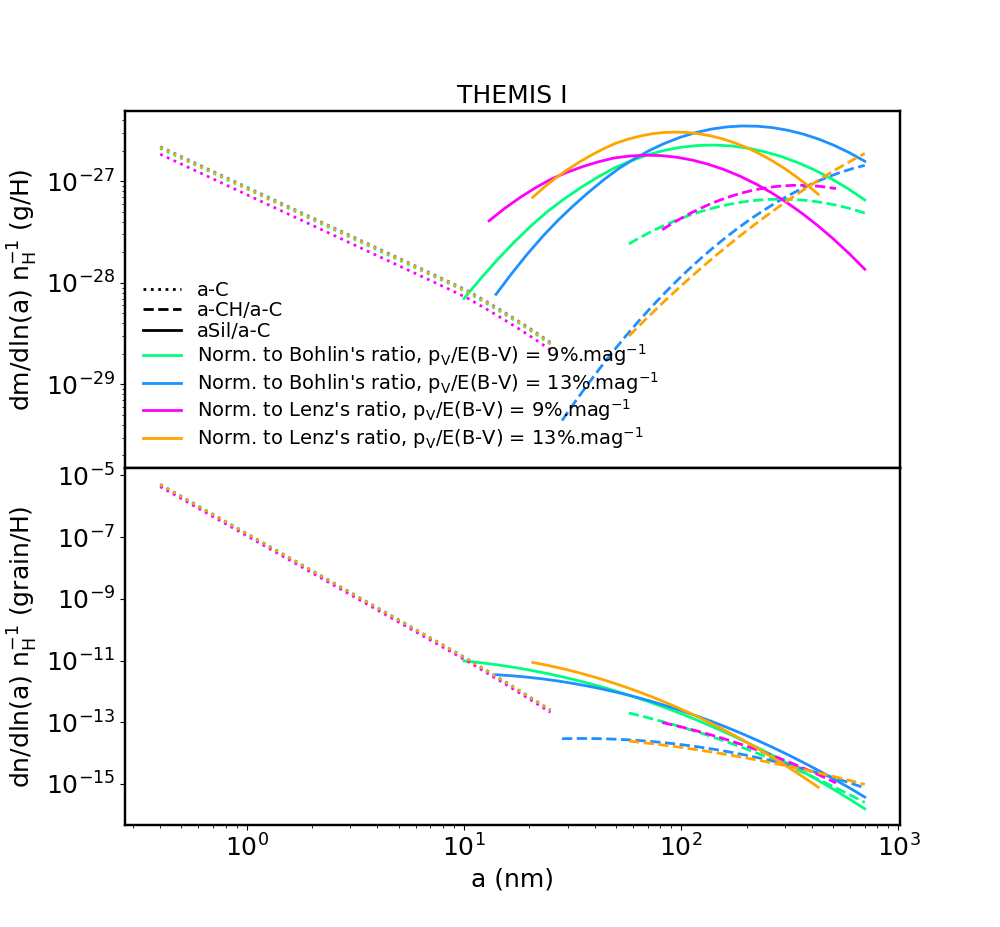}}
\caption{Same as Fig.~\ref{size_distribution} but for fits performed with THEMIS I silicates (see Table~\ref{table_composition}). Same colour scheme as in Fig.~\ref{fit_results_old}.}
\label{size_distribution_old} 
\end{figure}

\begin{figure}[!t]
\centerline{\includegraphics[width=0.45\textwidth]{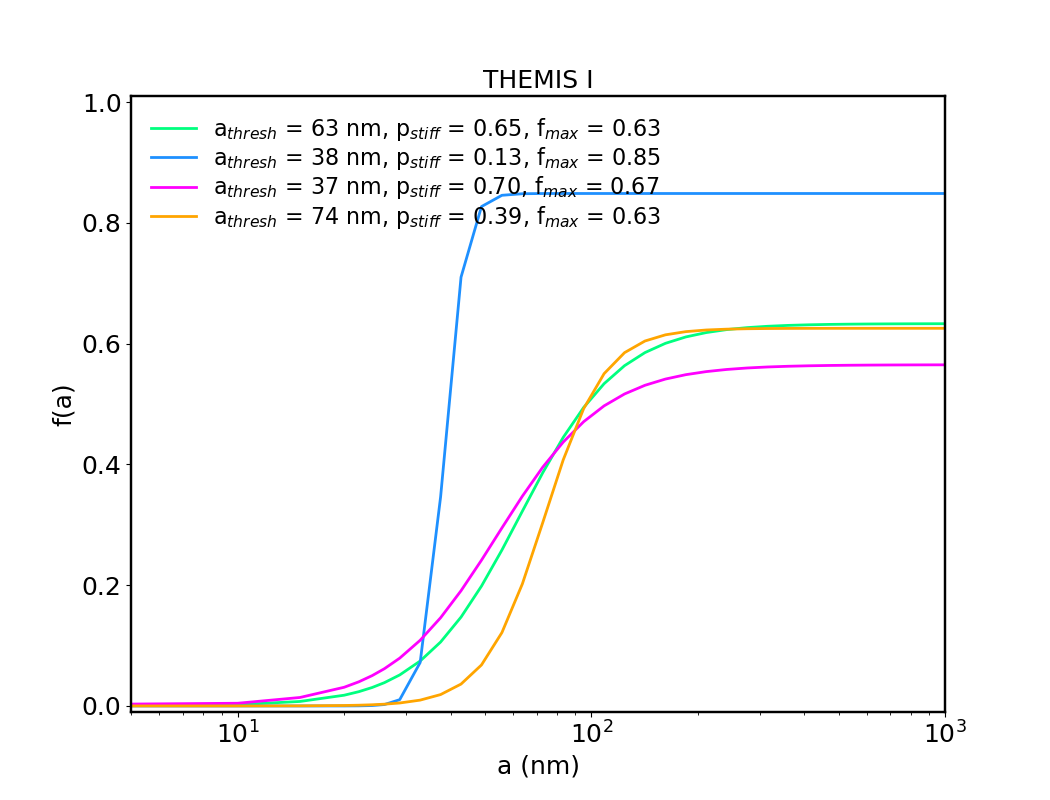}}
\caption{Same as Fig.~\ref{alignment} but for fits performed with THEMIS I silicates (see Table~\ref{table_composition}). Same colour scheme as in Fig.~\ref{fit_results_old}.}
\label{alignment_old} 
\end{figure}

Here we present the results obtained if instead of using the latest laboratory results for silicates \citep{Demyk2017, Demyk2022} we use the old silicates from the first version of THEMIS \citep{Jones2013, Koehler2014}. The grain shapes and elongations, as well as the compositions of the carbonaceous grains explored, are the same as those used in the rest of the paper.

\begin{figure*}[!h]
\centerline{\begin{tabular}{ccc}
\includegraphics[width=0.32\textwidth]{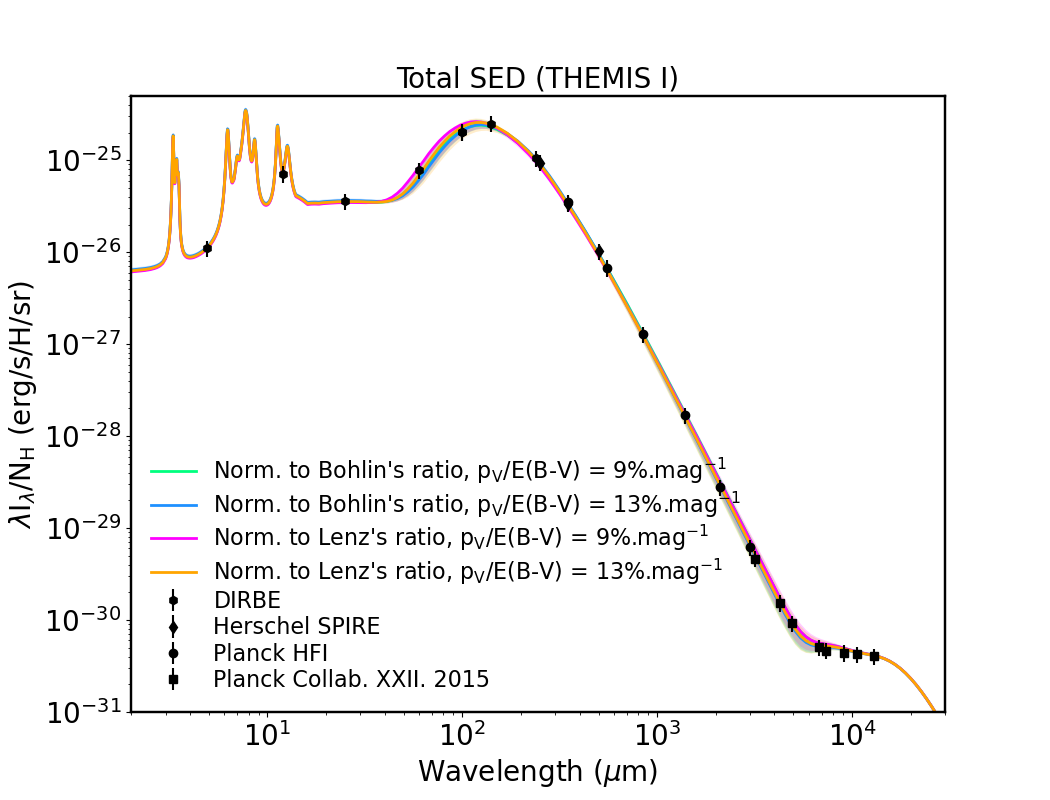} & \includegraphics[width=0.32\textwidth]{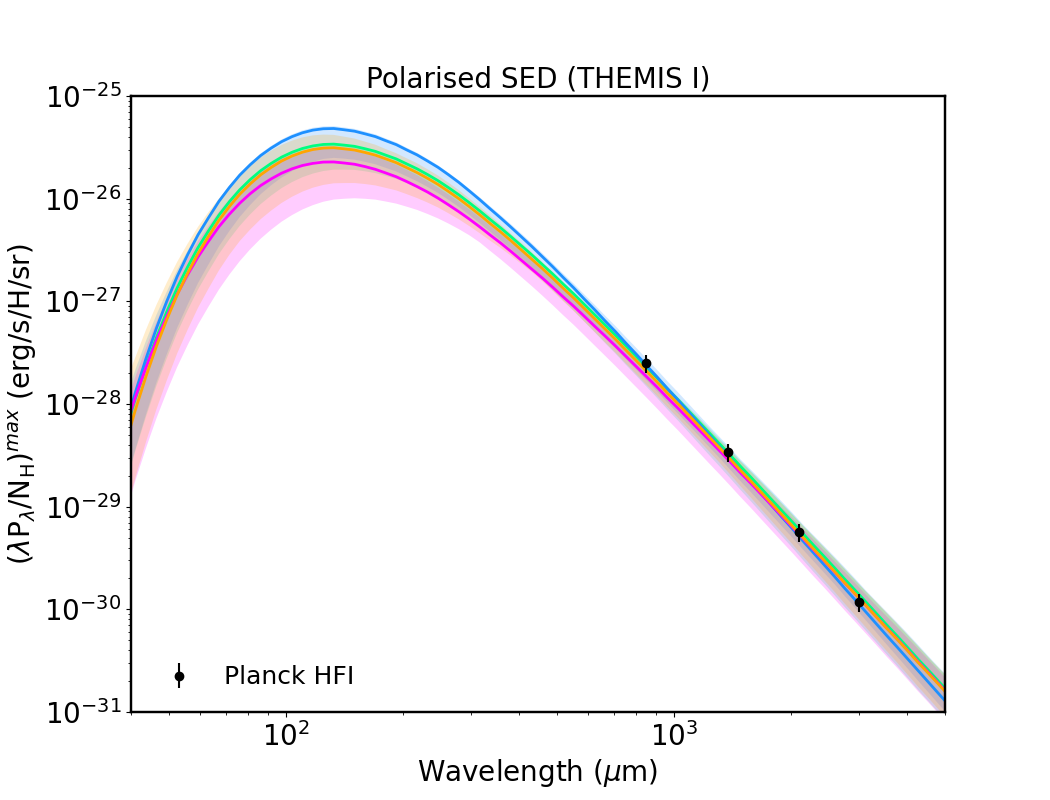} & \includegraphics[width=0.32\textwidth]{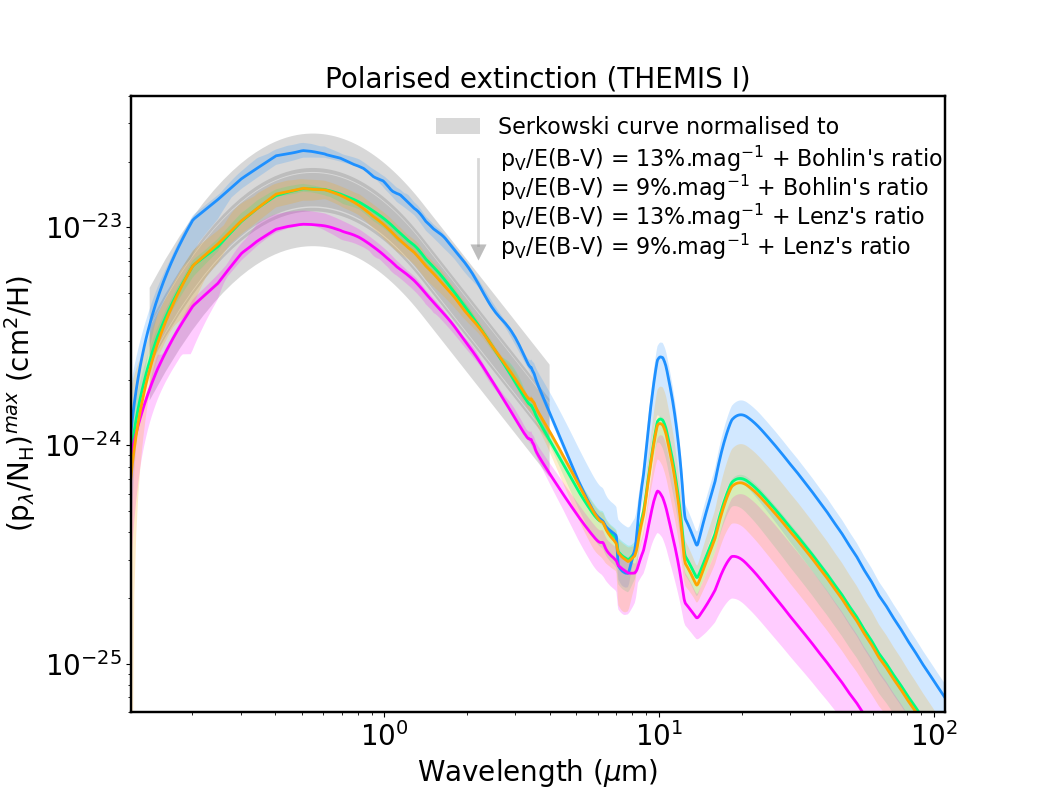} \\ 
\includegraphics[width=0.32\textwidth]{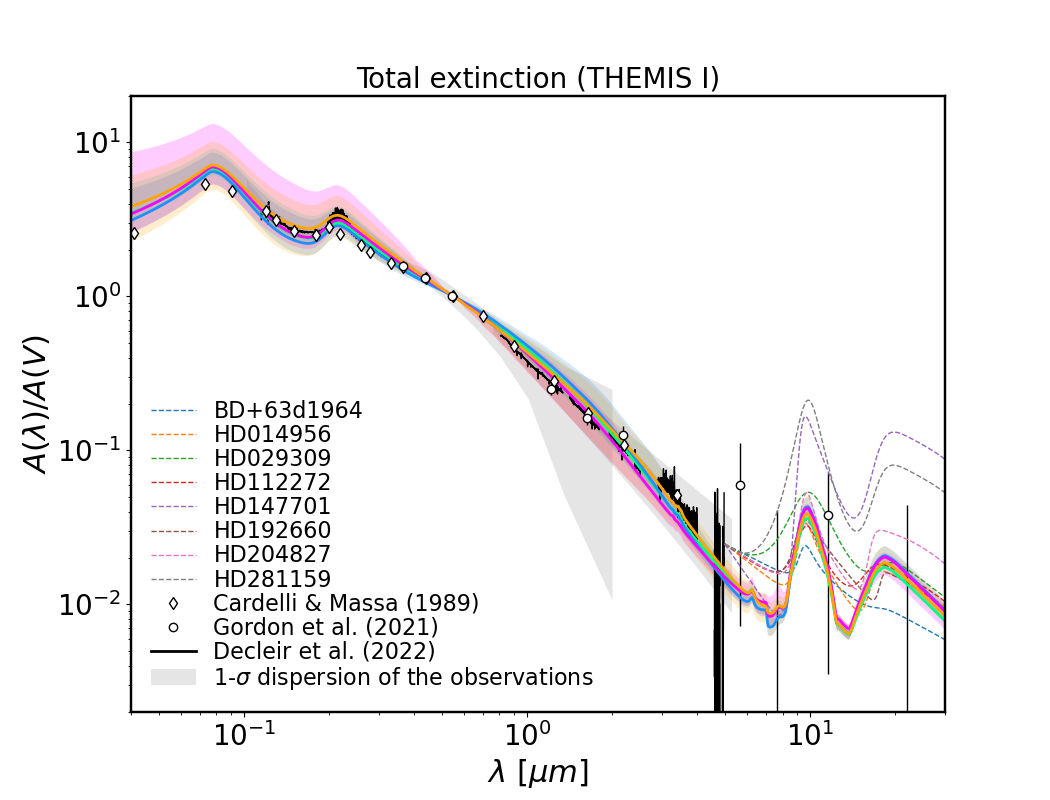} & \includegraphics[width=0.32\textwidth]{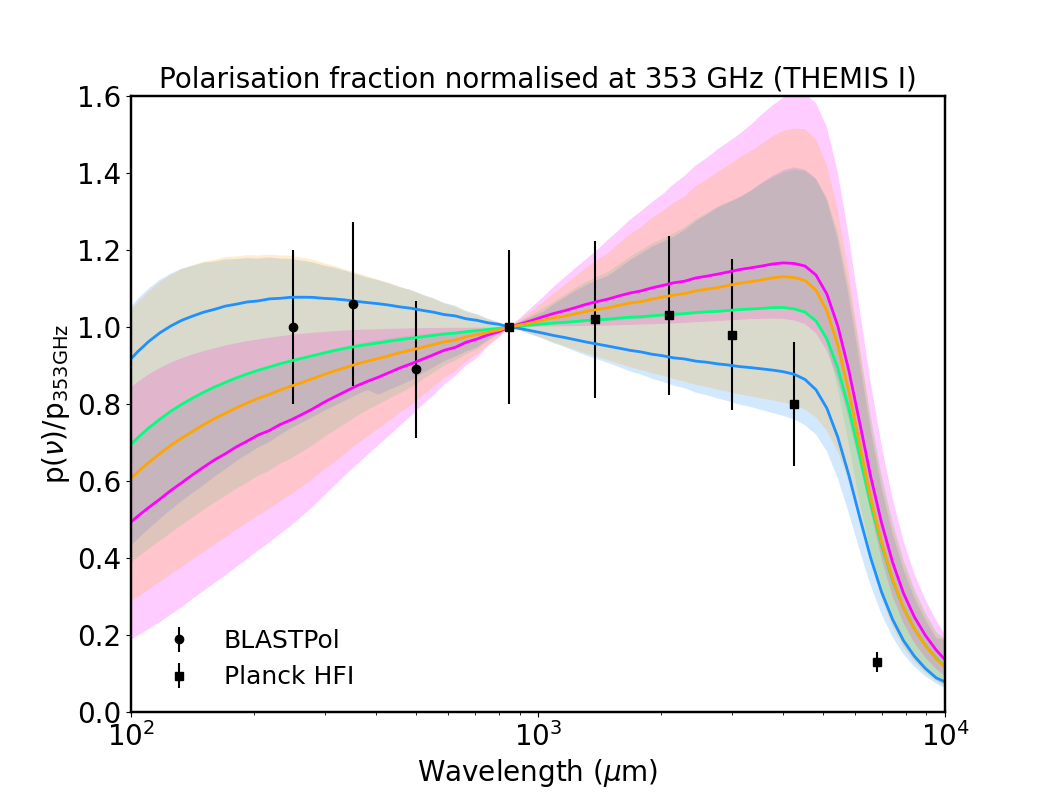} & \includegraphics[width=0.32\textwidth]{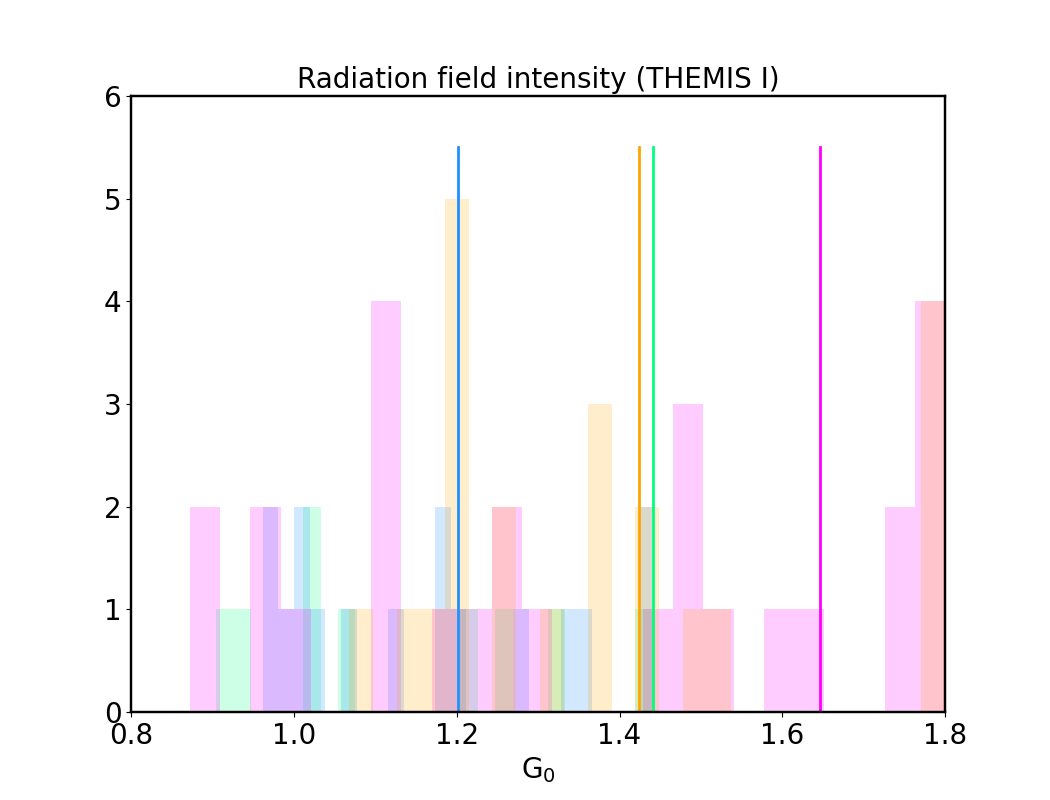} \\
\includegraphics[width=0.32\textwidth]{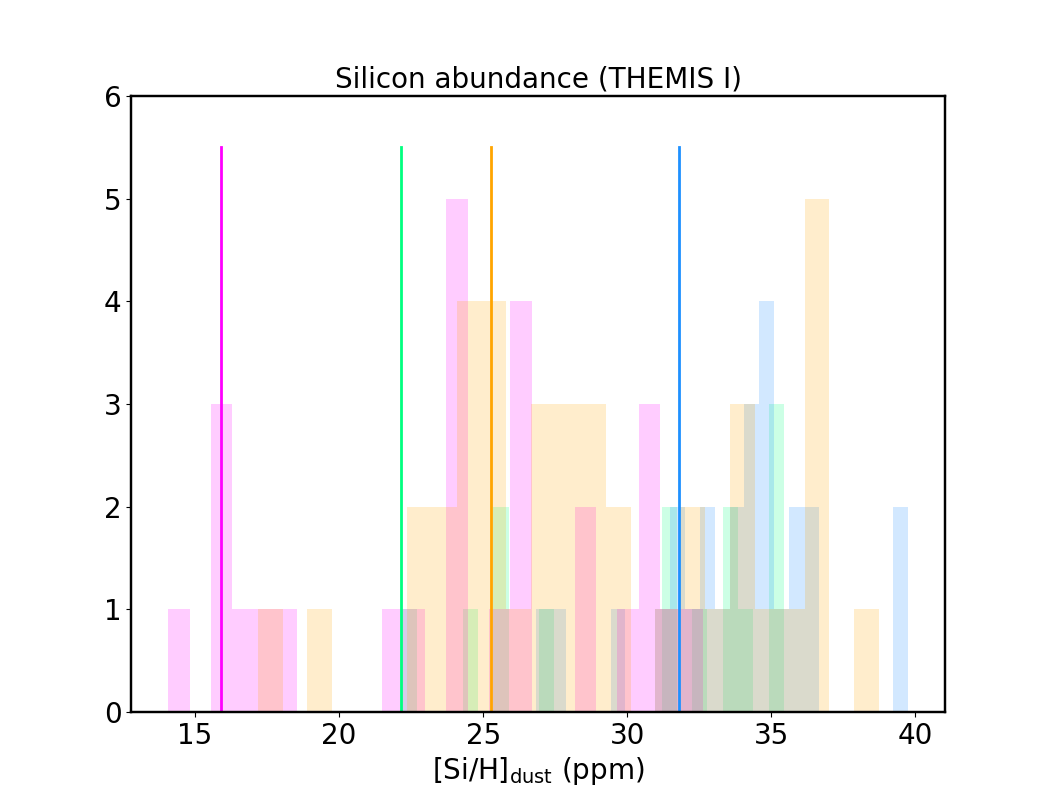} & \includegraphics[width=0.32\textwidth]{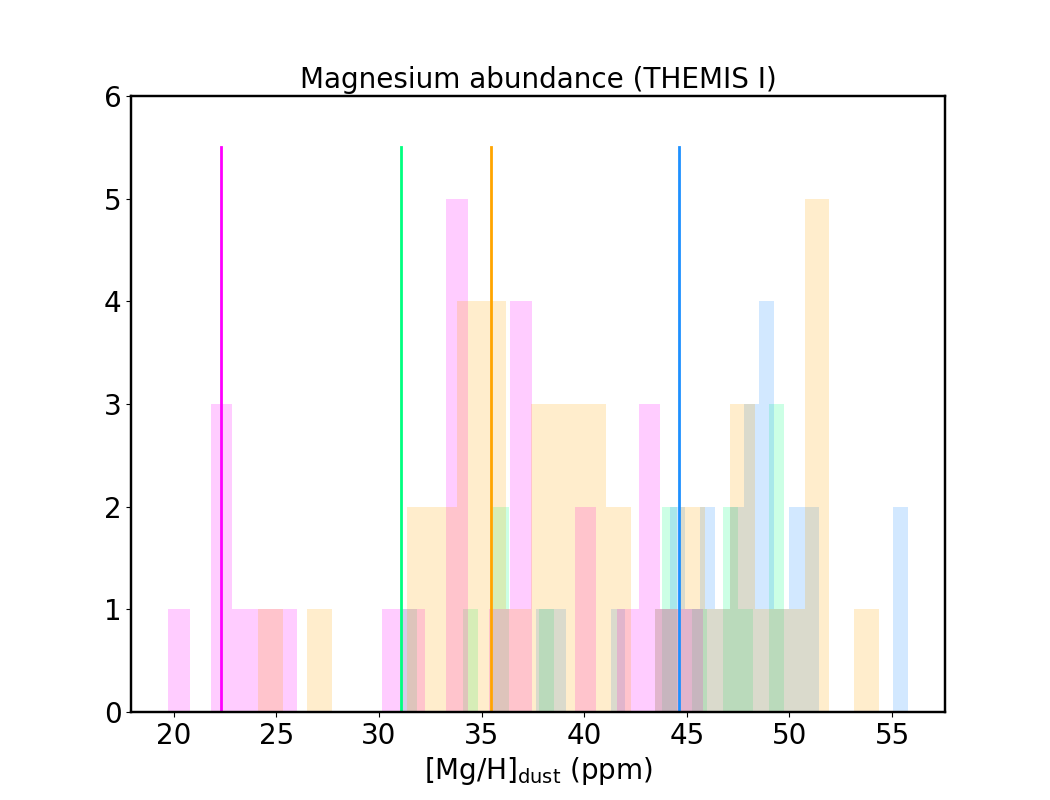} & \includegraphics[width=0.32\textwidth]{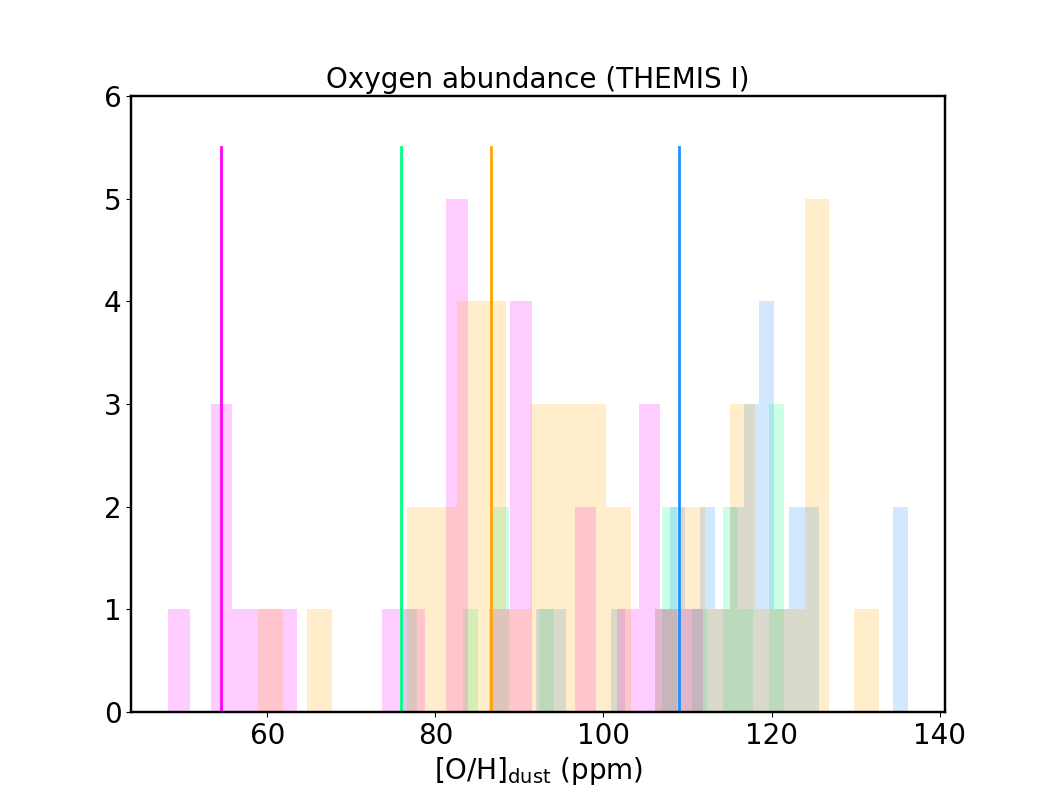} \\
\includegraphics[width=0.32\textwidth]{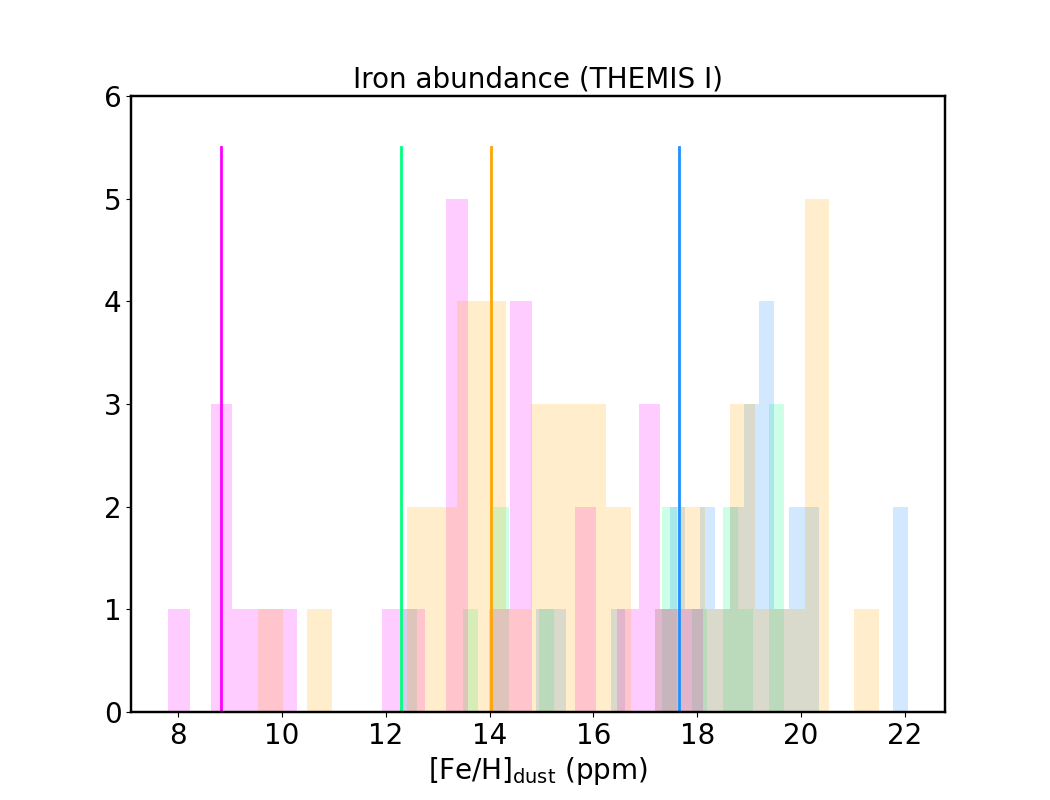} & \includegraphics[width=0.32\textwidth]{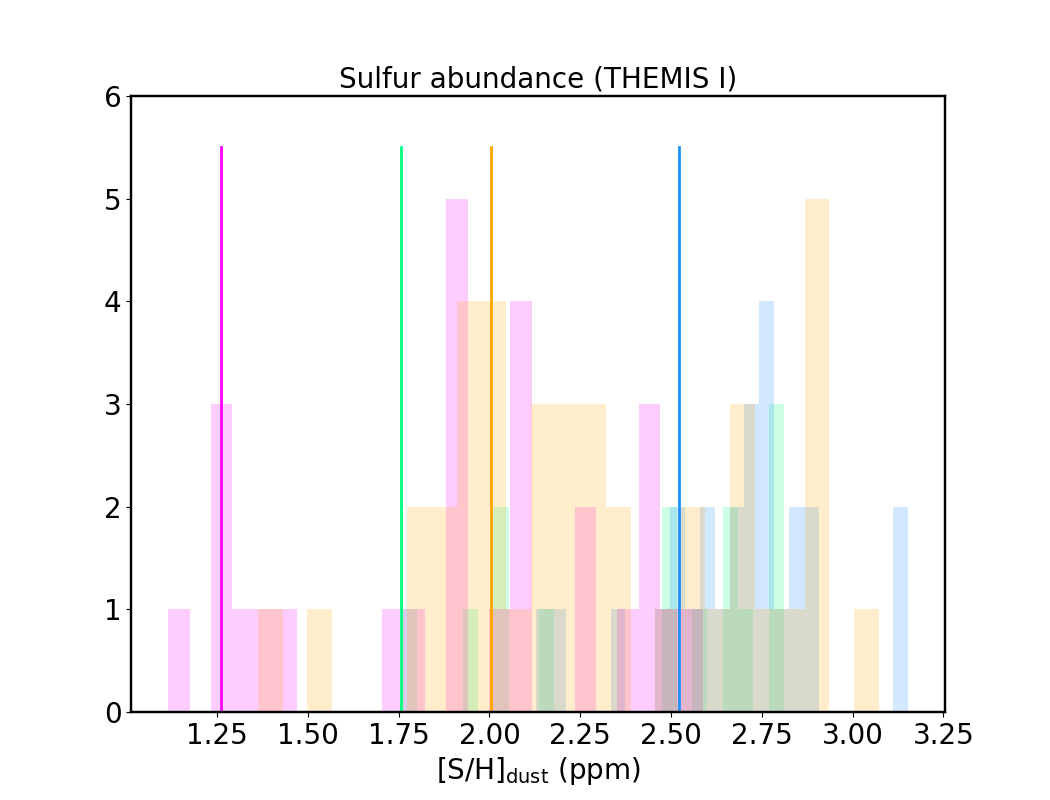} & \includegraphics[width=0.32\textwidth]{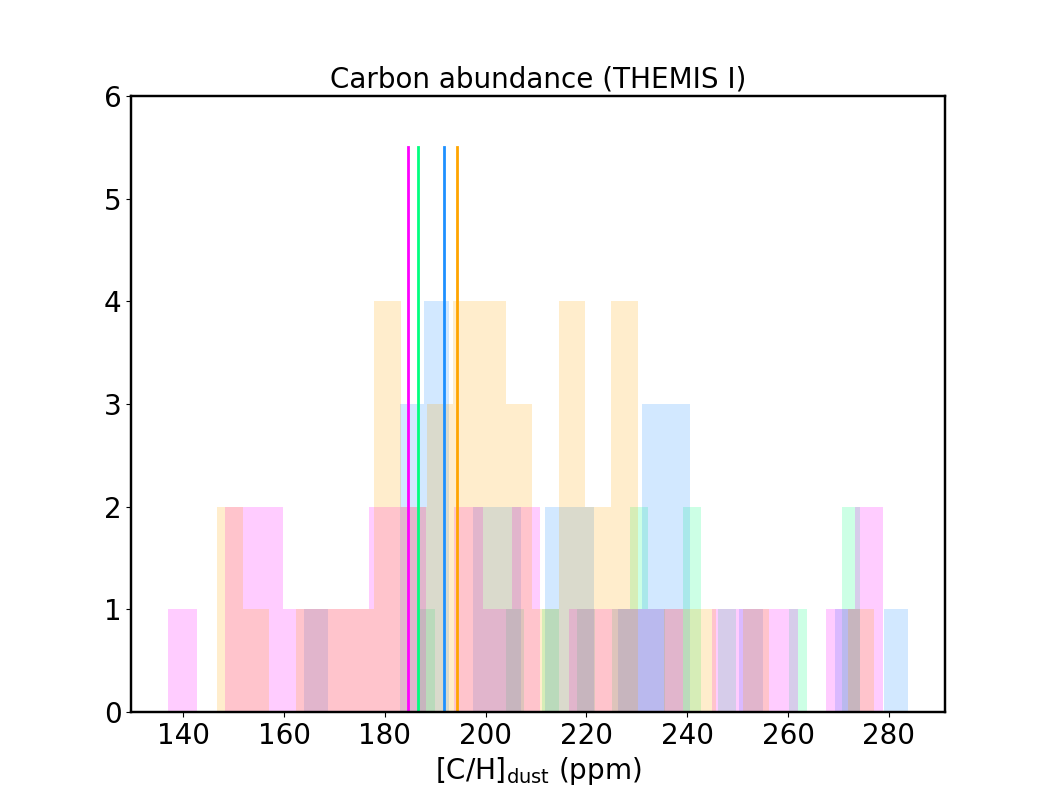} \\
\end{tabular}}
\caption{Same as Fig.~\ref{fit_results} but for fits performed with THEMIS I silicates (see Table~\ref{table_composition}). The data are those presented in Sect.~\ref{observations}. From top to bottom and left to right, the figures first show the models against the observations: total SED, polarised SED, polarised extinction, total extinction ; and then some model parameters: radiation field, silicium abundance, magnesium abundance, oxygen abundance, iron abundance, sulphur abundance, and carbon abundance. The light coloured areas represent the dispersion of the models in agrement with the observations and the matching thick solid lines show the best fits. As detailed in the text, two gas-to-dust mass ratios and two maximum optical polarisation values are considered to normalise the models. The models normalised to the ratio of \citet{Lenz2017} are shown in magenta and orange for $p_{\rm V}/E(B-V) = 9$\% and 13\%, respectively. Those normalised to the ratio of \citet{Bohlin1978} are shown in green and blue for $p_{\rm V}/E(B-V) = 9$\% and 13\%, respectively.}
\label{fit_results_old} 
\end{figure*}

\begin{figure*}[!t]
\centerline{\begin{tabular}{cc}
\includegraphics[width=0.45\textwidth]{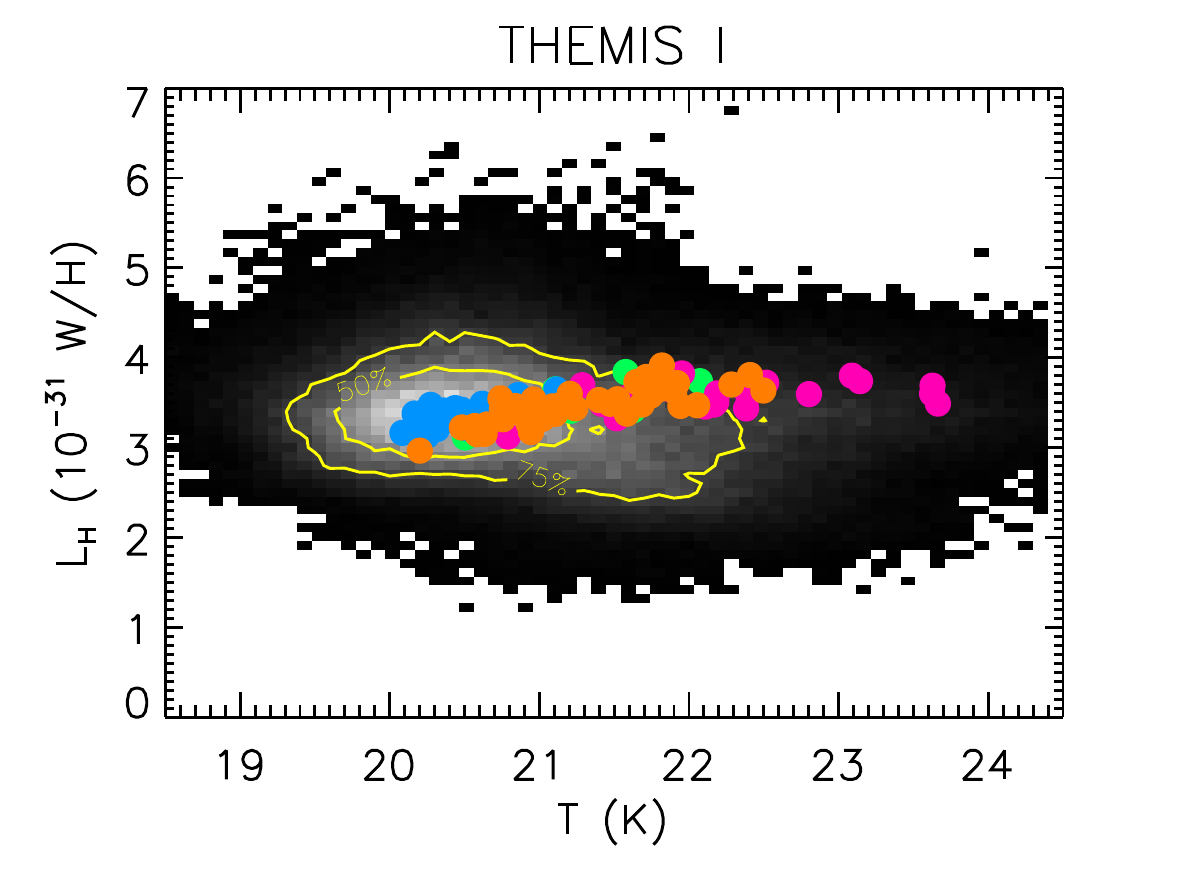} & \includegraphics[width=0.45\textwidth]{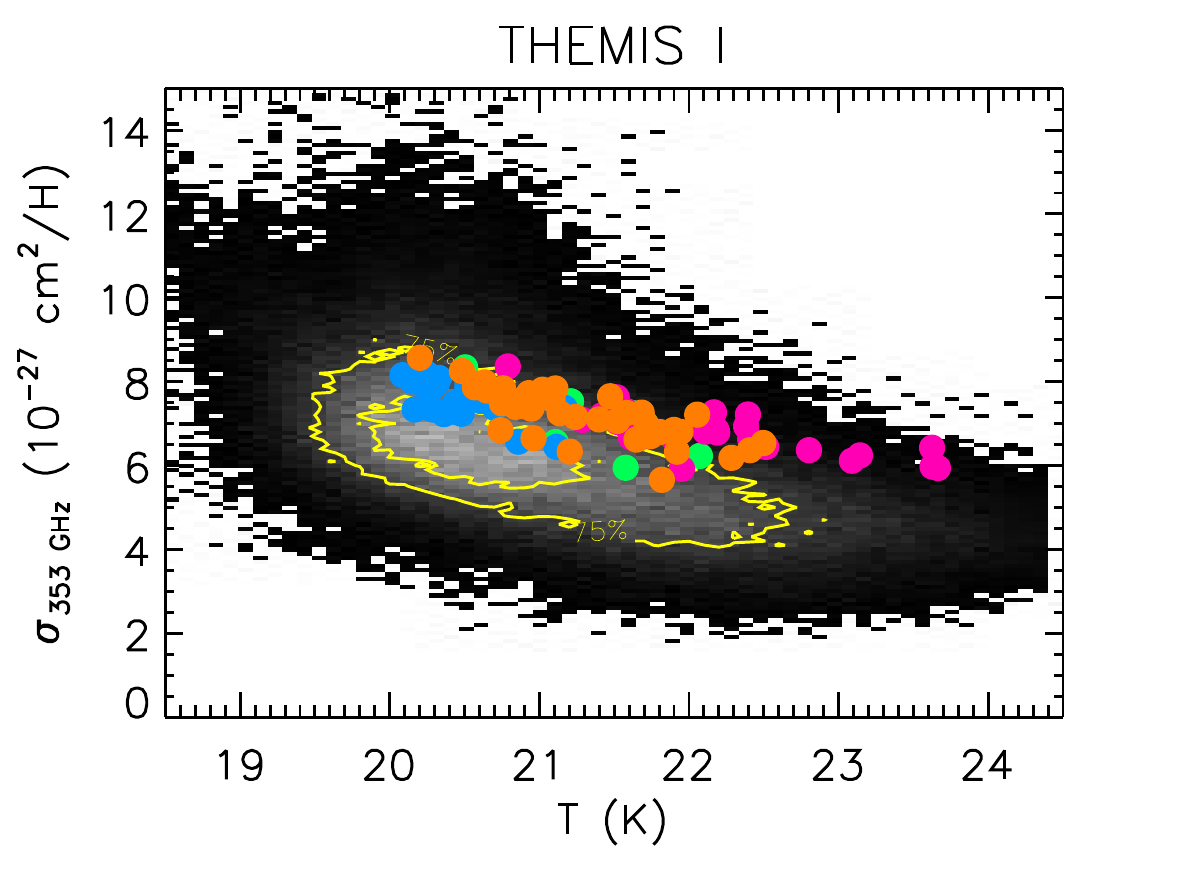} \\ 
\includegraphics[width=0.45\textwidth]{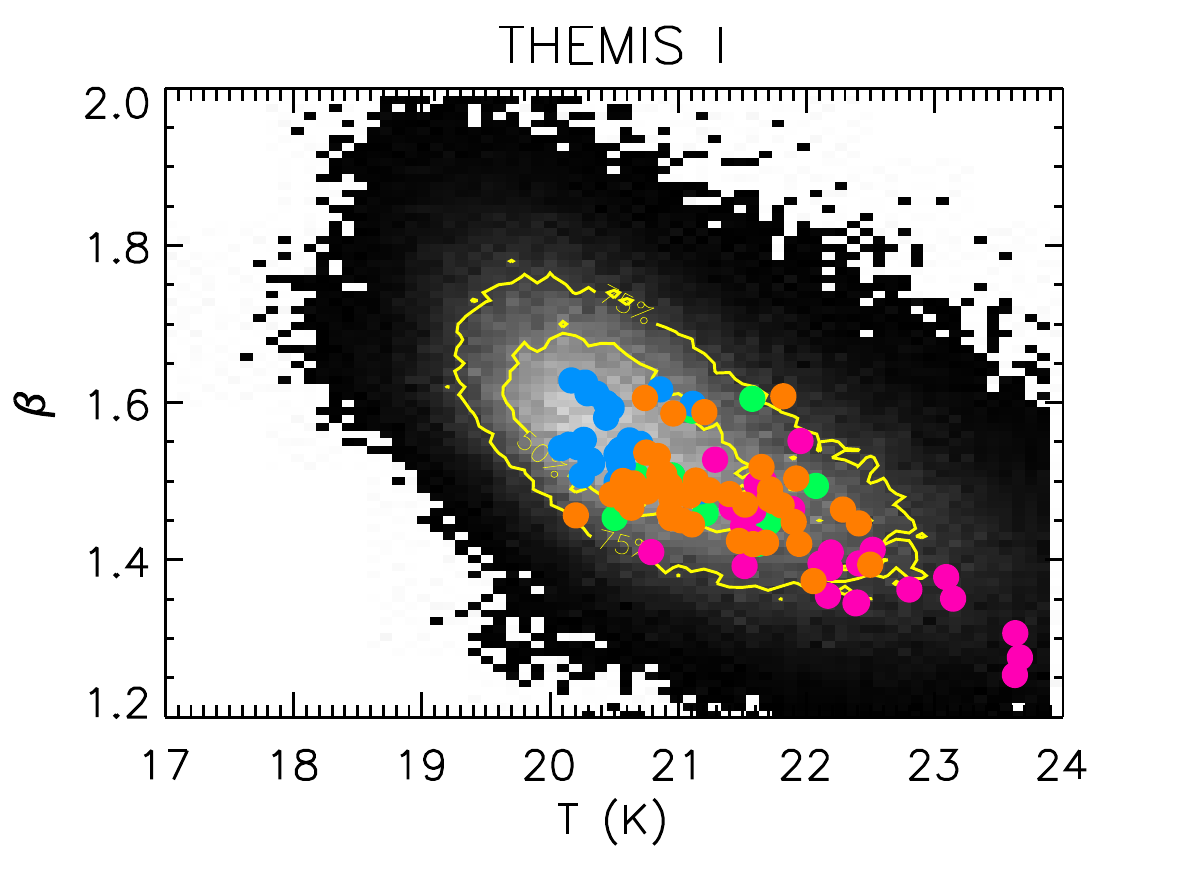} & \includegraphics[width=0.45\textwidth]{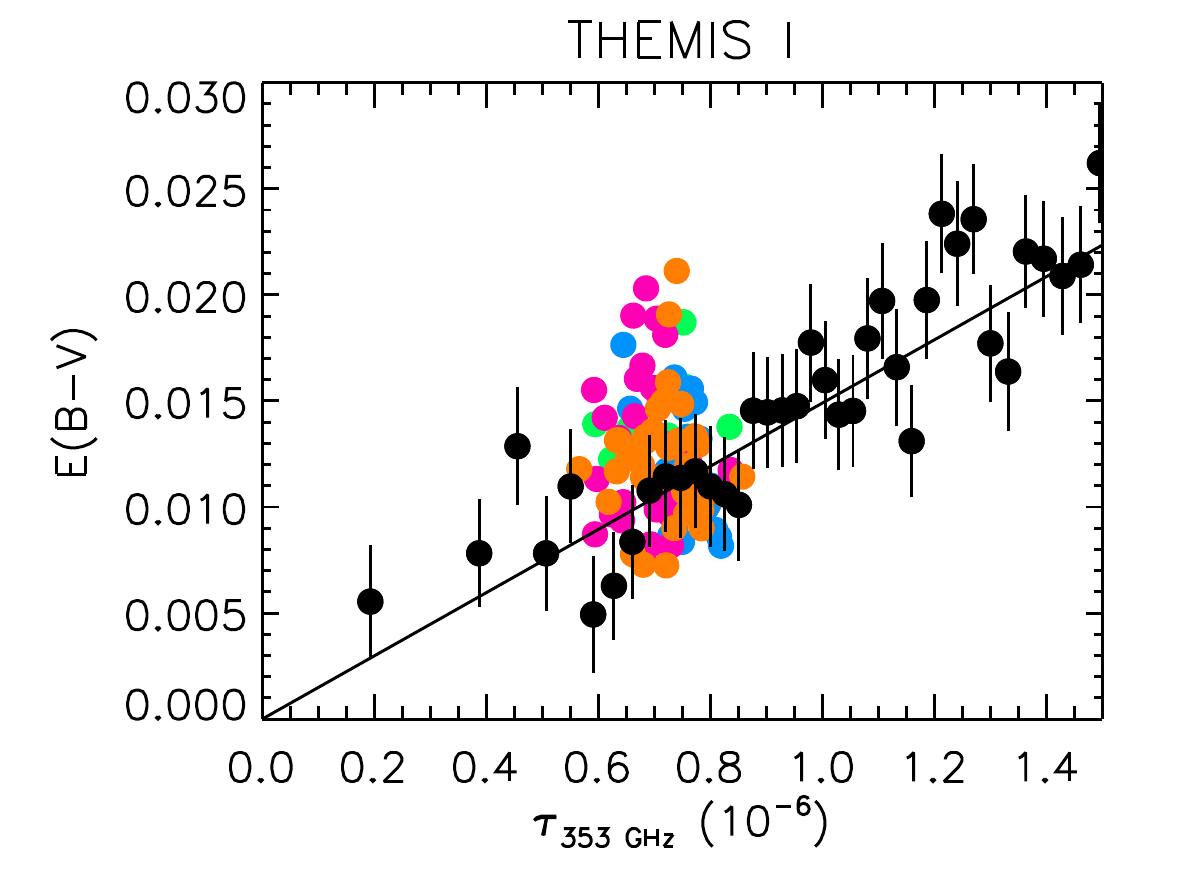}
\end{tabular}}
\caption{Same as Fig.~\ref{fit_Planck} but for fits performed with THEMIS I silicates (see Table~\ref{table_composition}). Variations in the dust parameters presented by \citet{PlanckXI2014}, based on a pixel-by-pixel modified blackbody $\chi^2$-fit. In the top left and right, and the bottom left figures, the observational results are the density of points maps, on which we overplot yellow contours: the central contour encloses 50\% of the observed pixels and the external contour 75\%. In the bottom right figure, the observational results are the black dots with error bars. In the four figures, the model results are the coloured dots, with the same colour scheme as in Fig.~\ref{fit_results_old}. Models normalised to the ratio of \citet{Lenz2017} are shown by magenta and orange dots for $p_{\rm V}/E(B-V) = 9$\% and 13\%, respectively. Those normalised to the ratio of \citet{Bohlin1978} are shown by green and blue dots for $p_{\rm V}/E(B-V) = 9$\% and 13\%, respectively. {\it Top left:} Luminosity vs temperature. {\it Top right:} Opacity at 353~GHz vs temperature. {\it Bottom left:} Spectral index vs temperature. {\it Bottom right:} E(B-V) vs optical depth at 353~GHz, models shown for $N_H = 10^{20}$~H/cm$^2$.}
\label{fit_Planck_old} 
\end{figure*}

Figure~\ref{fit_results_old} is the counterpart to Fig.~\ref{fit_results}, Fig.~\ref{3p4um_old} to Fig.~\ref{3p4um}, Fig.~\ref{size_distribution_old} to Fig.~\ref{size_distribution}, Fig.~\ref{alignment_old} to Fig.~\ref{alignment}, and Fig.~\ref{fit_Planck_old} to Fig.~\ref{fit_Planck}. Regardless of the normalisation chosen for the observations, it is always possible to find acceptable agreement between them and the model (Fig~\ref{fit_results_old}). The best fits are obtained for:
\begin{itemize}
\item THEMIS I silicates (oblates, $e = 1.3$) and a-CH/a-C$^{\rm 5nm}$ (oblates, $e = 2$) in the case of a normalisation to the Bohlin's ratio and for $p_V/E(B-V) = 9$\%
\item THEMIS I silicates (prolates, $e = 1.3$) and a-CH/a-C$^{\rm 5nm}$ (oblates, $e = 2$) in the case of a normalisation to the Bohlin's ratio and for $p_V/E(B-V) = 13$\%
\item THEMIS I silicates (oblates, $e = 1.3$) and a-CH/a-C$^{\rm 5nm}$ (prolates, $e = 2$) in the case of a normalisation to the Lenz's ratio and for $p_V/E(B-V) = 9$\%
\item THEMIS I silicates (prolates, $e = 2$) and a-CH/a-C$^{\rm 5nm}$ (prolates, $e = 2$) in the case of a normalisation to the Lenz's ratio and for $p_V/E(B-V) = 13$\%.
\end{itemize}
In the same way as with the silicates derived from the laboratory data, all shapes and elongations give acceptable fits with observations and most models require only imperfect grain alignment. The polarisation fraction at 3.4~$\mu$m is also compatible with the upper limits measured towards the Galactic Centre \citep[Fig.~\ref{3p4um_old}][]{Chiar2006}.

Finally, Fig.~\ref{fit_Planck_old} shows how models using the THEMIS I silicates compare with the dust parameters derived by \citet{PlanckXI2014} for the diffuse ISM. Even if the collection of acceptable models are compatible with the tendencies measured by \citet{PlanckXI2014}, it cannot explain the whole dispersion in the observations. This illustrates the need for diversity in the chemical composition of silicates to reconcile dust models and dust observations.

\end{document}